\documentclass[aps,prx,longbibliography,superscriptaddress,twocolumn,showpacs]{revtex4-2}
\usepackage[utf8]{inputenc}
\usepackage[american,british]{babel}
\usepackage[T1]{fontenc}

\usepackage{float}
\usepackage{amssymb}
\usepackage{wrapfig}
\usepackage{graphicx}
\usepackage{xcolor}
\usepackage{dcolumn}
\usepackage{bm}
\usepackage[normalem]{ulem}

\usepackage{amsmath,amsthm,amssymb,mathrsfs,mathtools,amsfonts,braket}
\usepackage[makeroom]{cancel}
\usepackage[caption=false]{subfig}

\usepackage[normalem]{ulem}
\usepackage[colorlinks=true, linktocpage=true]{hyperref}
\hypersetup{
    allcolors = {black}
}

\usepackage{braket}
\usepackage{todonotes}

\usepackage[Option]{overpic} 
\usepackage{mathdots}
\usepackage{dsfont}
\usepackage{mathrsfs}
\usepackage[version=4]{mhchem}
\DeclareMathAlphabet{\mathscrbf}{OMS}{mdugm}{b}{n}

 \newcommand{\RN}[1]{
  \textup{\expandafter{(\romannumeral#1)}}%
}
\usepackage{comment}

\usepackage{bigints}
\usepackage{float}
\usepackage{scalerel,stackengine,amsmath,amssymb}

\usepackage[most]{tcolorbox}
\newtcolorbox{textframe}[2][]{sharp corners, colback=white, boxrule=1pt}

\begin{document}
\title{
Quasiparticle cooling algorithms for quantum many-body state preparation 
}

\author{Jerome Lloyd}
\affiliation{Department of Theoretical Physics, University of Geneva, Geneva, Switzerland}

\author{Alexios Michailidis}
\affiliation{Department of Theoretical Physics, University of Geneva, Geneva, Switzerland}
\affiliation{PlanQC GmbH, Lichtenbergstr. 8, 85748 Garching, Germany}

\author{Xiao Mi}
\affiliation{Google Quantum AI, Santa Barbara CA, USA}

\author{Vadim Smelyanskiy}
\affiliation{Google Quantum AI, Santa Barbara CA, USA}

\author{Dmitry A. Abanin}
\affiliation{Google Quantum AI, Santa Barbara CA, USA}
\affiliation{Department of Physics, Princeton University, Princeton NJ 08544, USA}

\date{\today}

\begin{abstract}

Probing correlated states of many-body systems is one of the central tasks for quantum simulators and processors. A promising approach to state preparation is to realize desired correlated states as steady states of engineered dissipative evolution. 
A recent experiment with a Google superconducting quantum processor [X. Mi {\it et al.}, Science 383, 1332 (2024)] demonstrated a cooling algorithm utilizing auxiliary degrees of freedom that are periodically reset to remove quasiparticles from the system, thereby driving it towards its ground state. In this work, we develop a kinetic theory framework to describe quasiparticle cooling dynamics, and employ it to compare the efficiency of different cooling algorithms. In particular, we introduce a protocol where coupling to auxiliaries is modulated in time to minimize heating processes, and demonstrate that it allows a high-fidelity preparation of ground states in different quantum phases. We verify the validity of the kinetic theory description by an extensive comparison with numerical simulations for the examples of a 1d transverse-field Ising model, the transverse-field Ising model with an additional integrability-breaking field, and a non-integrable antiferromagnetic Heisenberg spin ladder. In all cases we are able to efficiently cool into the many-body ground state.
The effects of noise, which limits efficiency of variational quantum algorithms in near-term quantum processors, are investigated through the lens of the kinetic theory: we show how the steady state quasiparticle populations depend on the noise rate, and we establish maximum noise values for achieving high-fidelity ground states. This work establishes quasiparticle cooling algorithms as a practical, robust method for many-body state preparation on near-term quantum processors. 

\end{abstract}

\maketitle

\section{Introduction}

One of the central challenges for quantum simulation is to find reliable algorithms to prepare desired entangled quantum many-body states. In particular, preparing low-energy states of a given {many-body} Hamiltonian is an essential task for probing quantum dynamics, modeling correlated quantum materials, and for quantum chemistry applications of quantum computers~\cite{lanyon2010towards,kandala2017hardware,bauer2020quantum,google2020hartree}.
Classical tasks such as high-dimensional optimization problems can also be formulated in terms of finding ground states of appropriate Hamiltonians~\cite{finnila1994quantum, farhi2001quantum}. A number of quantum algorithms for preparing ground states of many-body Hamiltonians have been proposed~\cite{KitaevQPEArxiv1995,AbramsLloydPRL1999,farhi2000quantum,PoulinWocjanPRL2009,BoixoKnillSommaPhaseRandomization2009,GeCiracStatePreparationJMP2019}.

Two important classes of low-energy state-preparation protocols employed in experiments with noisy intermediate-scale quantum (NISQ)~\cite{preskill2018quantum} devices are: (i) {\it adiabatic} protocols~\cite{farhi2000quantum, albash2018adiabatic}, which rely on the adiabatic theorem and a slow variation of Hamiltonian parameters, and have been widely used in analogue quantum simulators such as ultracold atoms~\cite{Greiner2002Nature,Simon2011Nature} and Rydberg atom arrays~\cite{Semeghini2021Science,Scholl2021Nature}; (ii) {\it variational} methods~\cite{cerezo2021variational, TILLY2022PR, fedorov2022vqe}, suitable for digital quantum simulators, in which the target state is prepared by applying a variational quantum circuit to a simple initial state (e.g. a product state).

\begin{figure}[t!]
    \centering
    \includegraphics[width=0.8\columnwidth]{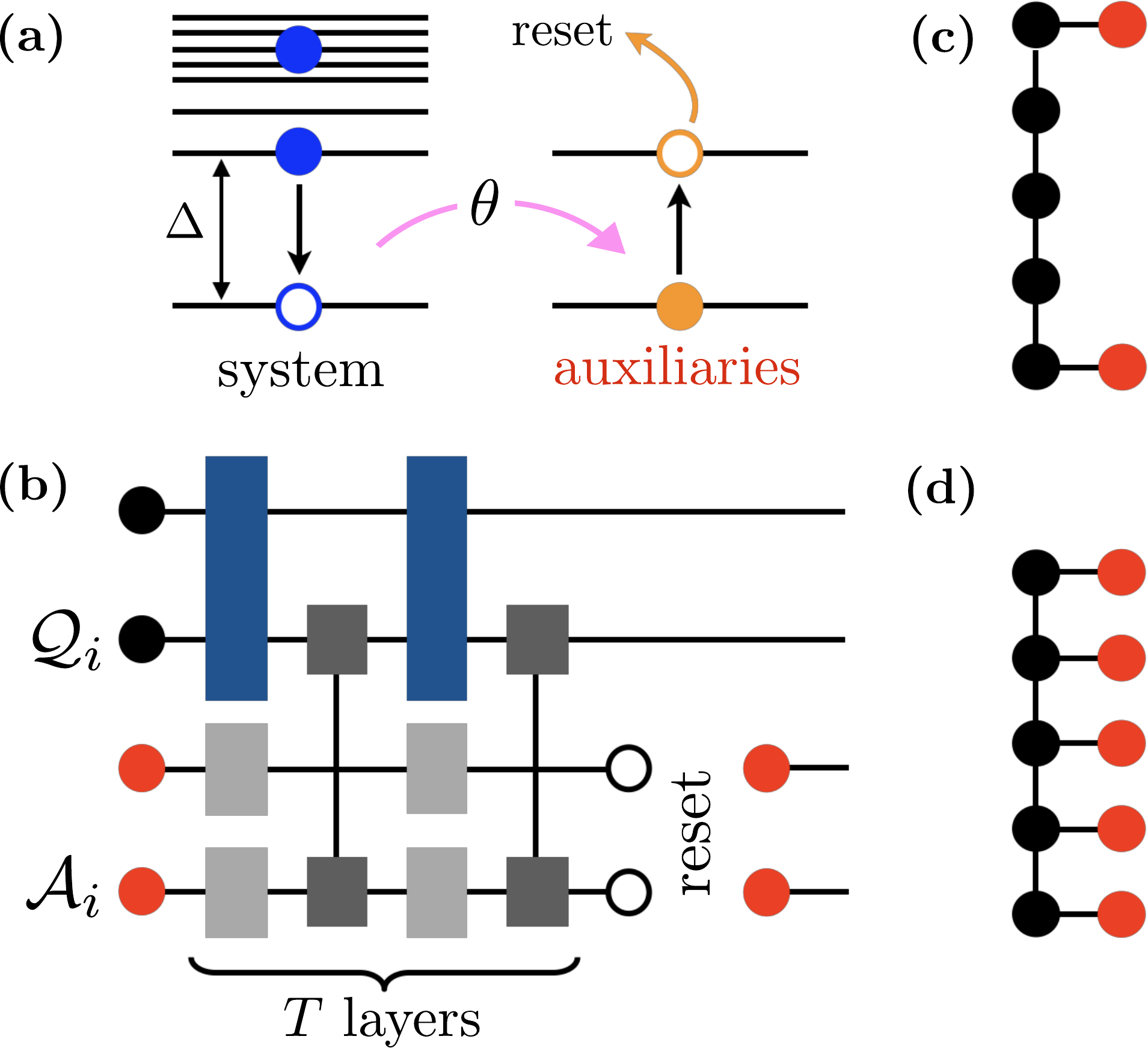}
    \caption{(a) Schematic mechanism of quasiparticle cooling: quasiparticles in the system, with energy above the system gap $\Delta$, are resonantly transferred to auxiliary qubits, which are then reset to ground state, removing entropy and energy from the system. (b) Digital quantum circuit used in reset protocol: a system of $N_S$ qubits $\mathcal{Q}_i$ is coupled to an auxiliary bath of $N_A$ qubits $\mathcal{A}_i$. Unitary evolution under system, auxiliary, and interaction gates is applied, and repeated $T$ times, followed by the reset of auxiliaries. This cooling cycle is repeated until the system converges to its steady state. (c) Edge cooling setup. (d) Bulk cooling setup, $N_A=N_S$.}
    \label{fig1}
\end{figure}

Unfortunately, unitary state-preparation protocols on NISQ processors are hindered by the presence of intrinsic decoherence. For example, superconducting qubits are subject to dephasing and decay processes, which introduce unwanted excitations to the system. Such protocols have no native mechanism to remove these excitations, so decoherence will eventually thermalize the system to a trivial state, e.g. a fully mixed state. This restricts the size of feasible circuits and limits the duration for which a quantum state can be stored on the device \cite{RevModPhys.94.015004, louvet2023gono}. Further, using variational circuits to prepare complex quantum states may be challenging due to flat optimization landscapes (`barren plateaus')~\cite{mcclean2018barren, Wang2021NatCom, Sack2022PRXQ}. 

A fundamentally different approach is to prepare many-body states as steady states of engineered dissipative evolution~\cite{TehralDiVincenzoPRA2000}. Because the dissipation directly competes with the decoherence effects, such state preparation methods can partially overcome the difficulties posed by noise. However, understanding which class of experimentally realisable dissipators can give rise to interesting quantum correlated steady states is a highly non-trivial question that has attracted much historic and recent interest \cite{davies1974markovian, Kraus2008, Diehl2008, Verstraete:2009vj, roy2020measurement, kastoryano2016quantum, chen2023quantum}. In particular, specific protocols that involve {\it local} engineered dissipation for frustration-free Hamiltonians (a restricted but important class of models) have been proposed and implemented experimentally in few-qubit systems of trapped ions and superconducting qubits~\cite{Barreiro2011,Wineland2013,Devoret:2013ve}. 

More recently, in an experiment~\cite{mi2024stable} on the digital Google Quantum Processor, a physics-inspired cooling algorithm was demonstrated for cooling 1d and 2d quantum spin systems. In this experiment, the system evolved under a periodic (Floquet) quantum circuit specified by a unitary Floquet operator $\hat{U}_S$, constructed as a shallow quantum circuit of gates with finite, but small rotation angles. In this regime, the time evolution of local observables is well-approximated by a prethermal Hamiltonian $H_{\rm eff}$~\cite{abanin17effective}. The approximate conservation of $H_{\rm eff}$ stems from the suppression of heating processes, which are exponentially slow in the inverse rotation angles of gates in $\hat{U}_S$. The key idea of the implemented algorithm (displayed schematically in Fig.~\ref{fig1}a,b --- see also ~\cite{metcalf2020engineered,Weimer_cooling_2020,Polla_PRA_2021,RudnerCoolingArxiv2022,puente2023quantum,MartiArxiv24} for related works) was to use a subset of qubits as auxiliaries, tuning their quasienergies and couplings to system qubits in order to resonantly transfer excitations from the system. The entropy of the system is removed by periodically resetting the auxiliary qubits. In the experiment, it was shown that the system reached a steady low-energy correlated state. Furthermore, the achieved steady states could be efficiently purified to yield high fidelities with respect to the true ground states, and it was argued that, asymptotically, the dissipative preparation is advantageous compared to variational unitary protocols. 

This progress calls for the development of practical, robust, dissipative many-body state preparation algorithms. In this work, we develop a theory of quasiparticle cooling on quantum processors, concentrating on efficient preparation of phases of matter with low-energy quasiparticle excitations, in realistic noisy environments. 
In such many-body systems, excited states at sufficiently low energy density can be described in terms of occupation numbers of non-interacting bosonic or fermionic degrees of freedom~\cite{Coleman_2015}. The  quasiparticles generally have a finite lifetime, due to the decay to other states; however, provided this lifetime is sufficiently long, a quasiparticle description can be employed. A celebrated example is the Landau Fermi-liquid theory, which provides a description of interacting electrons in metals in terms of fermionic quasiparticles~\cite{Coleman_2015}. More broadly,  quasiparticle descriptions apply to a variety of phases of matter including Fermi liquids, superconductors, and magnetically ordered states.

Focusing on 1d qubit -- or, equivalently, spin-$1/2$ -- systems, we consider generalised auxiliary-reset algorithms encompassing the one implemented in the experiment of Ref.~\cite{mi2024stable}. We derive a dissipative kinetic equation describing the quasiparticle population dynamics under the action of the protocol, which allows for a  transparent physical picture on how different parameters affect the state preparation fidelity. We show that, in particular, the success of the protocol relies on (i) the local cooling operators acting to remove quasiparticles from the system, (ii) the layout of auxiliary qubits, for example with auxiliaries only at the system boundaries or with a finite density of auxiliaries in the thermodynamic limit, as shown in Fig.~\ref{fig1}c,d, and (iii) the time dependence of the system-auxiliary couplings. The first two factors ensure that the matrix elements of the cooling operators between low-lying excited states and the ground state remain finite in the thermodynamic limit; we argue that only in topologically trivial systems is it possible to reach the ground state in constant time, while requiring the cooling operators to be local, few-qubit operators. The time-dependence of the couplings meanwhile ensures that heating transitions are suppressed relative to cooling transitions, and is crucial in reaching high fidelities.

Below, we characterize the protocol performance by log-fidelity per qubit, $-\log{\mathcal{F}}/N_S$, where $\mathcal{F}$ is the fidelity of the steady state with respect to the true ground state, and $N_S$ is the number of system qubits. Within our kinetic equation approach, we find $-\log{\mathcal{F}}/N_S \sim \exp(-\Omega(\Delta T))$ under the modulated coupling protocol, where $\Delta$ is the excitation gap and $T$ is the number of unitary layers before the auxiliaries are reset (see Fig.~\ref{fig1}b), which plays the role of an inverse temperature within the modulated coupling protocol. Thus, high fidelity can be achieved by increasing the unitary circuit depth. 


We assess the validity of our theory first with the example of an integrable 1d  transverse-field Ising model (TFIM) spin chain, where quasiparticles are infinitely long-lived and the relevant operators are known explicitly \cite{lieb1961two, sachdev1999quantum}. This model exhibits two phases, a paramagnetic (PM) and an antiferromagnetic (AFM) phase. The distinct nature of the quasiparticle types in the two phases (magnons in the paramagnet and domain walls in the antiferromagnet), as well as spontaneous symmetry breaking in the AFM  phase, lead to qualitative differences in the importance of different cooling processes. By performing extensive numerical simulations, we show that the kinetic equation correctly captures the system's evolution and its steady state even at a relatively strong auxiliary-system coupling --- the regime relevant for current experiments. We compare the experimental protocol of Ref.~\cite{mi2024stable} (where the couplings were kept constant) to our modulated coupling protocol, finding that the latter allows for much higher fidelities in the limit of large $T$. 


As a further test for the applicability of our protocol, we study two non-integrable and strongly interacting models: the TFIM with an integrability-breaking longitudinal field \cite{PhysRevB.68.214406}, and a Heisenberg spin ladder in a strongly interacting regime \cite{dagotto1996surprises}. Since many non-integrable systems have quasiparticle excitations which become weakly interacting at low energy densities, we expect our analysis to apply, at least at low energies (the most relevant regime for ground state preparation). In both cases we observe that the system converges to an energy below the excitation gap, thus demonstrating that ground state preparation is not restricted to integrable models.


While our results for the different models demonstrate the theoretical applicability of our protocol, to be practically useful the algorithm must also be able to compete against the unwanted effects of noise. We modify the kinetic equation by incorporating noise-induced heating on top of the cooling protocol, and investigate how the non-equilibrium distribution of quasiparticle level occupations in the steady state is affected by noise, for the Floquet TFIM example. We observe that the density of quasiparticles in the steady state scales differently with the Markovian noise strength in the two phases, such that achieving low quasiparticle densities in the AFM phase requires significantly weaker noise. This scaling behaviour was predicted in Ref.~\cite{RudnerCoolingArxiv2022}, and can be understood within the kinetic framework from the nature of different quasiparticle types in the two phases. Nevertheless, we find that the modulated coupling protocol achieves high fidelities when the noise rate is small compared to the auxiliary-system coupling strength.

The rest of the paper is organized as follows: in Section \ref{Sec:protocol} we describe the quasiparticle cooling protocol, and derive an equation for the evolution of the quasiparticle level populations, assuming weak auxiliary-system coupling. We discuss how the couplings can be tailored to remove quasiparticle excitations, and introduce the modulated coupling protocol (MCP) to suppress unwanted heating processes. In Section \ref{Sec:CoolingSpinChain}, we turn to cooling the integrable TFIM. We analyse cooling with boundary auxiliaries and with a finite density of auxiliaries, for different cooling protocols. We find that the MCP leads to effective cooling, which is accurately captured by the kinetic equation of Section \ref{Sec:protocol}. Non-integrable models are studied in Section \ref{Sec:Non-Integrable}, for the examples of the TFIM with an additional longitudinal field and a Heisenberg spin ladder. To check the protocol's robustness to decoherence, in  Section \ref{Sec:Noise} we generalise the kinetic theory to include noise, and examine how dephasing and decay processes affect the cooling of the TFIM. Finally, in Section \ref{sec:conclusions}, we provide concluding remarks and directions for future work.

\section{Quasiparticle cooling protocol}\label{Sec:protocol}

\subsection{Setup and protocol}

We consider the following general setup for a system of qubits, or spins-$1/2$, a particular version of which was realized in the recent experiment with the Google quantum processor \cite{mi2024stable}. The central idea follows Ref.~\cite{TehralDiVincenzoPRA2000}, and the main technical developments we present here are (i) the derivation of a simplified kinetic theory, based on the quasiparticle picture, and (ii) the introduction of time-dependent system-bath coupling, which, as we will see below, is crucial to reach high fidelities. 

The qubits are divided into two groups, $N_S$ qubits forming a system of interest, and $N_A$ auxiliary qubits acting as a bath. One cycle of the protocol (\emph{cooling cycle}), which effectively realizes a Floquet system coupled to an engineered bath, entails the following steps: 
\begin{itemize} 
\item[(i)] Auxiliary (bath) qubits are initialized in a state described by a density matrix $\hat\rho_B^0$. 
\item[(ii)] $T$ layers of unitary evolution are applied. A unitary applied at time step $1 \leq \tau \leq T$ is a combination of the Floquet operator $\hat U_S$ acting on system qubits, a unitary $\hat U_B$ acting on the bath qubits, and an interaction term $\hat U_{\theta, \tau}$ which couples system and bath qubits. The interaction term is generally time-dependent, and $\theta$ is the coupling strength (see Eq.~(\ref{eq:couplingops}) below).
\item[(iii)] The bath qubits are reset into the initial state $\hat \rho_B^0$, which we will assume to be a stationary state of the bath unitary $\hat U_B$. 
\end{itemize}
The main protocol control parameters are $T$, which we refer to as the \emph{reset time}, and the system-auxiliary coupling. Resetting auxiliary qubits at the end of the cycle defines a quantum channel acting on the system,  
\begin{equation}\label{eq:cooling_channel}
    \Phi (\hat \rho_S) = \text{Tr}_{B} \ \hat U_T (\hat \rho_S\otimes \hat \rho_{B}^0) \hat U_T^\dagger,
\end{equation}
where $\hat U_T$ is the combined cooling cycle unitary,
\begin{equation}\label{eq:unitarytrain}
    \hat U_T = \hat U_{\theta,T}\hat U_B\hat U_S\ldots \hat U_{\theta,1}\hat U_B\hat U_S,
\end{equation}
the operator $\hat \rho_S$ is the system density matrix at the start of the cycle, and $\text{Tr}_B$ is the trace over the bath degrees of freedom. The system density matrix after $\nu$ cooling cycles is given by
\begin{equation}
    \hat \rho_S^\nu = \Phi^\nu(\hat \rho_S^0).
\end{equation}
We will assume the steady state of the dynamics defined by this quantum channel is unique, and is reached independent of the initial state; we will not consider a scenario of initial states with anomalously slow relaxation to the steady state. In the analysis below we therefore take the initial state to be the maximally mixed state, $\hat \rho_S^0 \propto \hat I$, unless specified otherwise. 

The system Floquet $\hat U_S$ can in particular correspond to an exact or prethermal effective Hamiltonian $\hat H_{\rm eff}$, 
\begin{equation}
    \hat U_S \approx \exp\bigg(-\frac{i\pi}{2}\hat H_{\rm eff}\bigg),
\end{equation}
and our goal is to engineer dissipative evolution such that the steady state ($\nu \to \infty$) of the system is close to the ground state of $\hat H_{\rm eff}$. In order to have a concrete model in mind, we will later study cooling in the case of the 1d Floquet transverse-field Ising model, 
\begin{gather}\label{eq:TFIM}
    \hat U_S = \exp\bigg({-\frac{i\pi J}{2}\sum_{i=1}^{N_S-1} \hat X_i \hat X_{i+1}}\bigg) \exp\bigg(\frac{i\pi g}{2}\sum_{i=1}^{N_S} \hat Z_i\bigg).
\end{gather}
Here and below $\hat X_i,\hat Y_i, \hat Z_i$, denote the Pauli operators for qubit $i$, normalized as $\hat X_i^2=\hat Y_i^2=\hat Z_i^2=1$. The Floquet TFIM is exactly solvable via the Jordan-Wigner mapping to a model of non-interacting fermionic quasiparticles (see Appendix \ref{app:TFIM}), and so provides a convenient setting to study quasiparticle cooling.

We keep the bath evolution simple, assuming for the initial bath state the product state with all spins up,
\begin{equation}\label{eq:initialbath}
    \hat \rho_B^0 =\otimes_{j=1}^{N_A} |0\rangle_j \, _j\langle 0|
\end{equation}
--- note that in our convention the spin-up state is identified with the $\ket{0}$ state of the qubits ---  and for the bath unitary we choose
\begin{equation}\label{eq:UB}
    \hat U_B = \exp\bigg(\frac{i\pi h}{2}\sum_{j=1}^{N_A} \hat Z_{j}\bigg),
\end{equation}
so that each auxiliary qubit simply acquires a phase in the eigenbasis of the $\hat Z$ operator. Parameter $\pi h$ is referred to as the \emph{auxiliary quasienergy} and does not depend on the unitary step $\tau$. 

In contrast, the system-bath coupling will in general vary in time:
\begin{equation}\label{eq:couplingops}
    \hat U_{\theta,\tau} = \exp(-i \theta f_\tau \hat V),  \hspace{.5cm} \hat V =  \sum_{j=1}^{N_A}\hat V_{j}.
\end{equation}
We have chosen to separate the time-dependence into the function $f_\tau$, normalised according to $\sum_{\tau=1}^T f_\tau = 1$~\footnote{For protocols considered below, $f_\tau$ can be positive or negative, but for this choice the sum  $\sum_{\tau=1}^T |f_\tau|\gtrsim 1$.}, keeping  $\theta$ as a small parameter which controls the coupling strength. Both $f_\tau$ and $\theta$ are dimensionless, and the interaction operator, which has dimension of energy, can in general be written in the form
\begin{equation}\label{eq:AB}
    \hat V = \sum_a \hat A^{a\dagger}\hat B^a + \hat A^a\hat B^{a\dagger},
\end{equation}
where operators $\hat A$ ($\hat B$) acts on system (bath). We refer to the $\hat A^a$ operators as the (system) \emph{cooling operators}. 

In the following we will assume that the bath operators $\hat B$ act simply as annihilation operators on the non-interacting bath set by Eq.~(\ref{eq:UB}): 
\begin{equation}\label{eq:Bcreation}
    \hat U^\dagger_B \hat B^a \hat U_B = e^{-i\pi h} \hat B^a.
\end{equation}
This is simple to implement in the case of Eq.~(\ref{eq:UB}), by choosing $\hat B^a = \hat \sigma^+_{i_a}$, with $i_a$ labelling the auxiliary qubit, where $\hat \sigma^\pm_i = (\hat X_i\pm\hat Y_i)/2$ are the spin raising/lowering operators. While our choice of the bath parameters may seem rather simple to simulate realistic cooling, we will see below that the complexity is shifted into the coupling choice, Eq.~(\ref{eq:couplingops}).

\subsection{Weak-coupling analysis}

In the remaining theoretical analysis, we assume that the excitations of the many-body system are well-defined, long-lived quasiparticles, and that we are in the limit of weak system-bath coupling, $\theta |V_{j}|\ll 1$. The quasiparticle assumption holds exactly in the example of the TFIM, Eq.~(\ref{eq:TFIM}); we note, however, that interacting  systems away from criticality typically have quasiparticle excitations at low energies -- examples include Fermi-liquids, superconductors, and symmetry-broken magnetic phases~\cite{Coleman_2015}. We expect the quasiparticle cooling algorithm to remain  effective for such interacting systems, as we elaborate on in Subsection \ref{sec:coolingchoice}, and numerically demonstrate in Section \ref{Sec:Non-Integrable}. Our strategy will be to derive an equation describing evolution of the quasiparticle level occupations in the weak-coupling limit.

We define the quasiparticle occupation number operator $\hat{n}_k$, where index $k$ labels fermionic quasiparticle levels (e.g. momentum in a system with periodic boundary conditions). The quasiparticles are sufficiently long-lived, such that $\hat{n}_k$ is an integral of motion of the system's evolution operator, 
\begin{equation}\label{eq:nk_conserved}
    \hat{n}_k \hat U_S = \hat U_S \hat{n}_k. 
\end{equation}
In practice, it is sufficient to assume that the quasiparticle lifetime is much longer than the duration of one cooling cycle. 

It is convenient to work in the interaction picture with respect to the evolution operator $\hat U_0=\hat U_B \hat U_S$, such that an operator $\hat O_\tau$ (originally in the Schr\"odinger picture) is given by:
\begin{equation}
    \hat O_{I\tau}=\hat U_0^{-\tau} \hat O_\tau \hat U_0^\tau, \;\;\; 1\leq\tau\leq T. 
\end{equation}
The evolution operator can be rewritten as follows,
\begin{equation}
    \hat U_T=\hat U_0^T \mathcal{\hat U}_T, \;\;\; \mathcal{\hat U}_T=\mathcal{T} \exp\Big(-i\theta\sum_{\tau=1}^T  f_\tau \hat V_{I\tau}  \Big),
\end{equation}
where $\mathcal{T}$ denotes time-ordering. 
Taking into account Eqs.~(\ref{eq:cooling_channel},\ref{eq:nk_conserved}), the quasiparticle occupation number after one cooling cycle reads:
\begin{equation}\label{eq:}
   n_k' = {\rm Tr}_S ( \Phi(\hat\rho_S)\hat{n}_k)=\text{Tr}_{SB} \Big( \hat\rho_S\otimes\hat\rho_B^0 \Big[\mathcal{\hat U}^\dagger_T \hat n_k \mathcal{\hat U}_T \Big] \Big).
\end{equation}
\begin{widetext}
At weak coupling, the occupation number change over one cycle, $\delta n_k=n_k'-n_k$, is small, and we compute it by expanding the r.-h.s.~of the above equation to the second order in $\theta$. Noting that the linear order in $\theta$ vanishes, and defining
\begin{equation}
    \delta n_k={\rm Tr}_S (\hat\rho_S \delta\hat{n}_k), \hspace{0.5cm} \delta\hat n_k = \text{Tr}_B \Big(\hat\rho_B^0 \Big[\mathcal{\hat U}^\dagger_T \hat n_k \mathcal{\hat U}_T \Big]\Big) - \hat n_k , 
\end{equation}
this yields:
\begin{gather}\label{eq:delta_nk}
 \delta\hat n_k = \theta^2 \sum_{ab} \sum_{\tau_1=1}^T \sum_{\tau_2=1}^T  f_{\tau_1} f_{\tau_2}\Gamma_{ab}(\tau_1 -\tau_2) \Big(\hat A^{a\dagger}_{I\tau_1} \hat n_k \hat A^{b}_{I\tau_2} - \frac{1}{2} \{\hat n_k,\ \hat A^{a\dagger}_{I\tau_1}\hat A^{b}_{I\tau_2}\}\Big) + i\theta^2 [\hat H_{\rm LS}, \hat n_k].
 \end{gather}
 Here we defined the autocorrelators of the $\hat B$ operators,
\begin{equation}
     \Gamma_{ab}(\tau) = \text{Tr}_B \Big(\hat\rho_B^0 \hat B^{a}_{I\tau} \hat B^{b\dagger}_{I0} \Big),
\end{equation}
and used relations $\Gamma_{ab}(\tau) = \Gamma_{ba}^*(-\tau)$. The fact that the autocorrelator depends on a single time variable is due to the assumption that $\hat\rho_B^0$ is a stationary state of $\hat U_B$. For simplicity, in the following we will also assume that $\Gamma_{ab}$ is diagonal in the coupling operators, $\Gamma_{ab}(\tau) = \delta_{ab}\Gamma_{aa}(\tau)$; this will be the case in the examples we study. The second term in the r.-h.s. of Eq.~(\ref{eq:delta_nk}) can be viewed as arising from the ``Lamb shift'' renormalization:  
 \begin{gather}
 \hat H_{\rm LS} = \frac{i}{2}\sum_{ab}  \sum_{\tau_1=1}^T \sum_{\tau_2=1}^T f_{\tau_1} f_{\tau_2}(\Theta(\tau_1-\tau_2)-\Theta(\tau_2-\tau_1))\Gamma_{ab}(\tau_1-\tau_2) \hat A^{a\dagger}_{I\tau_1}\hat A^{b}_{I\tau_2}.
\end{gather}
where $\Theta(\tau)$ is the standard Heaviside step function, with the convention $\Theta(0)=1/2$. 
\end{widetext}

\subsection{Rate equation}

Note that the above derivation so far follows the steps of the textbook derivation of the Lindblad equation \cite{breuer2002theory, lidar2020lecture}, except here we focus on the Heisenberg-picture evolution of the occupation number operators, rather than on the system's density matrix. However, to derive the Lindblad equation one typically assumes a rapid decay of bath temporal correlations, compared to the system's dynamical timescales; in contrast, in our analysis the memory effects of the bath will play an essential role.

To simplify the analysis, we make an assumption that the system's density matrix at the beginning of a cooling cycle is diagonal in the quasiparticle basis, $[\hat\rho_S,\hat{n}_k]=0$. We further assume that the correlations between occupations of different levels can be neglected. The former assumption can be justified in the limit $\theta\to 0$, by coarse-graining over sufficiently many cooling cycles (secular approximation). The latter assumption is more difficult to prove, but is closely related to the molecular chaos hypothesis in kinetic theory. In the context of integrable quantum systems it is often referred to as the time-dependent Generalized Gibbs ensemble (GGE) ansatz for $\hat\rho_S$~\cite{RigolGGE2007, lange2018time, gerbino2024large}. As we will show below, these assumptions lead to a theory of quasiparticle cooling which is in quantiative agreement with numerical simulations. 

The diagonal assumption gives $\hat \rho_S = \sum_{\vec{\alpha}} \rho_{\vec{\alpha}\vec{\alpha}} \ket{\vec{\alpha}}\bra{\vec{\alpha}}$, where  $\ket{\vec{\alpha}}$ form a basis of states with occupation numbers $\{ \alpha_k \}$, $\alpha_k=0,1$ and quasienergy
\begin{equation}\label{eq:quasienergies}
\varepsilon(\vec{\alpha})=\sum_k \epsilon_k \alpha_k,
\end{equation}
where $\epsilon_k$ is quasienergy of quasiparticle level $k$. The diagonal assumption still leaves us with $2^{N_S}$ coupled equations to solve. The second assumption, however, implies that the probabilities $\rho_{\vec{\alpha}\vec{\alpha}}$ can be expressed as a product over the single-quasiparticle distribution functions: 
\begin{equation}
\rho_{\vec{\alpha}\vec{\alpha}}=\prod_q \frac{e^{-\alpha_q \lambda_q}}{1+e^{-\lambda_q}},\label{eq:GGE2}
\end{equation}
where the $\lambda_q$ serve to parameterise the single-particle distribution. Eq.~(\ref{eq:GGE2}) is the familiar form for a gas of non-interacting fermionic particles. Equivalently, the probabilities can be written in terms of the \emph{average} quasiparticle occupation numbers $n_q={\rm Tr}(\hat\rho_S \hat{n}_q)$, 
\begin{equation}
\rho_{\vec{\alpha}\vec{\alpha}}=\prod_q n_q^{\alpha_q} (1-n_q)^{1-\alpha_q}\label{eq:GGE},
\end{equation}
by using the relation $n_q = e^{-\lambda_q}(1+e^{-\lambda_q} )^{-1}$. 

Next, we compute the change in the occupation number over one cycle, $\delta n_k$, using the above form of the density matrix and Eq.~(\ref{eq:delta_nk}). Since density $\hat n_k$ commutes with $\hat \rho_S$, the Lamb shift contribution plays no role. Denoting the Bohr frequencies by 
\begin{equation}\label{eq:Bohr}
\Delta(\vec{\alpha},\vec{\beta}) = \varepsilon({\vec{\alpha}}) -\varepsilon({\vec{\beta}}),
\end{equation}
we obtain 
\begin{widetext}
\begin{equation}\label{eq:rate_general}
        \delta n_k = \theta^2 \sum_{\vec{\alpha},\vec{\beta}}\sum_{a,\tau_1,\tau_2} \rho_{\vec{\alpha}\vec{\alpha}} (\beta_k-\alpha_k)f_{\tau_1}f_{\tau_2} \Gamma_{aa}(\tau_1-\tau_2) e^{i(\tau_1-\tau_2)\Delta(\vec{\alpha},\vec{\beta})} |\bra{\vec{\beta}}\hat A^{a}\ket{\vec{\alpha}}|^2. 
\end{equation}
With our above choice of bath unitary (\ref{eq:UB}) and of the $\hat B$ operators (\ref{eq:Bcreation}), the bath spectral function reads
\begin{equation}
    \Gamma_{aa}(\tau) = e^{-i\pi h\tau},
\end{equation}
and Eq.~(\ref{eq:rate_general}) takes the following form: 
\begin{equation}\label{eq:rate_equation}
    \delta n_k = \theta^2 \sum_{a}\sum_{\vec{\alpha}\vec{\beta}} \rho_{\vec{\alpha}\vec{\alpha}} (\beta_k-\alpha_k)|F_{h,T}(\Delta(\vec{\alpha},\vec{\beta}))|^2 |\bra{\vec{\beta}} \hat A^{a}\ket{\vec{\alpha}}|^2.
\end{equation}
\end{widetext}
Here, $\rho_{\vec{\alpha}\vec{\alpha}}$ is defined via Eq.~(\ref{eq:GGE}) to yield a closed set of equations for the $N_S$ quasiparticle occupations, and we defined
\begin{equation}\label{eq:filterF}
    F_{h,T}(\epsilon) = \pi \sum_{\tau=1}^T f_\tau e^{i\tau(\epsilon-\pi h)}. 
\end{equation} 
We refer to $F_{h,T}(\epsilon)$ as the \emph{filter function} associated with $f_\tau$. In short, it acts to filter out transitions occuring at unwanted frequencies, such as heating processes. 

Eq.~(\ref{eq:rate_equation}) is a general rate equation for quasiparticle level occupation numbers. It shows that the quasiparticle evolution is controlled by (i) the filter function, which encodes spectral features of the bath and time-dependent coupling, and (ii) the matrix elements of the operators $\hat A^a$ entering the system-auxiliary coupling. In order for the cooling protocol to remove quasiparticles from the system and drive into the many-body ground state, both of these features should be properly engineered, as we discuss next.

\subsection{Choice of operators coupling to auxiliary qubits}\label{sec:coolingchoice}

First, we discuss the choice of the system cooling operators $\hat A^a$, entering the coupling Eq.~(\ref{eq:couplingops}). For simplicity, below we will consider 1d translationally invariant systems, with annihilation operator of a quasiparticle with quasimomentum $k$  given by
\begin{equation}
    \hat{Q}_k^-=\frac{1}{\sqrt{N_S}}\sum_{j=1}^{N_S} e^{ikj}\hat{Q}_j^-,
\end{equation}
where $\hat{Q}_j^-$ is an operator centered at site $j$. 

The auxiliary couplings $\hat{A}^a$ should be chosen such that the matrix elements for removing quasiparticle excitations in Eq.~(\ref{eq:rate_equation}) are maximized. In principle, this can be accomplished by choosing the operators $\hat{A}^a$ to be quasiparticle annihilation operators $\hat{Q}^-_j$. Then, the matrix element appearing in Eq.~(\ref{eq:rate_equation}) reads,
\begin{equation}
    |\bra{\vec{\beta}} \hat Q^{-}_j\ket{\vec{\alpha}}|^2 = \frac{1}{N_S} \sum_k (1-\beta_k)\alpha_k \prod_{q\neq k} \delta_{\alpha_q\beta_q},
\end{equation}
and the change of the quasiparticle occcupation numbers over one cooling period (assuming one auxiliary qubit per system site) is given by
\begin{equation}\label{eq:annihilationrate}
    \delta n_k = -\theta^2 |F_{h,T}(\epsilon_k)|^2 n_k,
\end{equation}
i.e.~the quasiparticle occupations decay exponentially with a rate set by the coupling strength and the filter function. The ground state is reached in time $O(\theta^{-2})$. 

In practice, the precise form of the $\hat{Q}_j^-$ operators for a generic many-body system is often not known. These operators may have support on multiple sites, with size of the order of the correlation length. Furthermore, in phases with topological excitations, the operators may not even be local in the qubit basis. A clear  example is the antiferromagnetic phase of the TFIM: while the excitations are local in a basis of non-interacting fermions, in the qubit basis, these excitations are domain walls, whose removal requires acting on a string of qubits (see  Sec.~\ref{Sec:CoolingSpinChain}). 

This complicated structure of the $\hat{Q}_j^-$ operators poses an obstacle to choosing the cooling operators to coincide with annihilation operators -- current hardware constraints dictate that the $\hat A^a$ operators should be relatively simple operators with support on at most a few neighbouring qubits. Despite this, the principle of adiabatic continuity \cite{Coleman_2015, hastings2005quasiadiabatic} implies that in gapped systems in which the quasiparticles are not topological, a choice of $\hat A^a$ operators can be made which is simultaneously hardware-efficient and which cools to the many-body ground state. 

To illustrate this, suppose that the Hamiltonian of a gapped system in a topologically trivial phase can be written as (extension to Floquet systems in the prethermal regime is straightforward):
\begin{equation}
    \hat H' = \hat H_0 + \lambda \hat V,
\end{equation}
where $H_0$ is a simple solvable Hamiltonian where quasiparticles are created by local operators with support strictly on a few nearby sites, and $\hat{V}$ is a perturbation away from the simple point. Decomposing a Hamiltonian $\hat H'$ into above form is not unique, and we choose one such representation which minimizes the perturbation strength $\lambda$. For the TFIM example we have $H_0 = -g\sum_i \hat Z_i$ and $\hat V = \sum_i \hat X_i \hat X_{i+1}$, with $\lambda = J$; the quasiparticle annihilation operators of $\hat H_0$ are the single-spin operators $\hat{Q}_{0,j}^-=\hat\sigma^+_j$ in this case. According to Eq.~(\ref{eq:annihilationrate}), the non-interacting ground state at $\lambda=0$ can be prepared exactly by choosing the cooling operators $\hat A^j = \hat Q^-_{0,j}$, and the operators are strictly local in the qubit basis. 

When the interactions are switched on ($\lambda \neq 0$), then as long as $\hat H'$ and $\hat H_0$ remain in the same quantum phase, the principle of adiabatic continuity implies that the two Hamiltonians are related by a quasi-local unitary transformation \cite{hastings2005quasiadiabatic}. Both the ground state and the quasiparticle operators become `dressed' by the transformation, while retaining their quantum numbers. This dressing of quasiparticles by the interactions can be characterised by the spectral function
\begin{equation}
    S_{\hat Q^-_0}(k,\omega) = \sum_{\gamma'} |\bra{\Omega'} \hat Q_{0,k}^- \ket{\gamma'}|^2\delta(\omega-(\varepsilon_{\gamma'} -\varepsilon_0)),
\end{equation}
where the sum runs over all excited states of the Hamiltonian $\hat H'$ and $\Omega'$ is now the interacting ground state with energy $\varepsilon_0$. In the $\lambda \to 0$ limit the spectral function exhibits a delta-function pole at the quasiparticle energy, $S(k,\omega) = \delta(\omega - \epsilon_k)$, while for finite $\lambda$, as long as the quasiparticle lifetime remains long, the spectral function retains a sharp feature centred around $\epsilon_k$ and with a width proportional to the inverse quasiparticle lifetime~\footnote{In practice it is the spectral function that is measured in experiment and the validity of the quasiparticle description requires the existence of a sharp pole.}. The weight (quasiparticle residue) $Z_k$ of the pole is directly related to the matrix elements between the interacting ground state and single-quasiparticle excited states, $Z_k = |\bra{\Omega'} \hat Q^-_{0,k} \ket{k'}|^2$ \cite{Coleman_2015}. Thus, as long as the quasiparticle picture holds, $Z_k$ remains non-zero and choosing the cooling operators according to $\hat A^j = \hat Q^-_{0,j}$ should lead to effective cooling of the low-energy excitations, albeit with cooling rate reduced by the factor of the quasiparticle residue.

This argument relies on the phase being adiabatically connected to a fixed-point Hamiltonian with simple, local quasiparticles, which is the case for gapped, topologically-trivial phases. In contrast, in systems where the underlying excitations are fractionalised, e.g.~in the TFIM antiferromagnet or topologically ordered system, it is not possible to adiabatically connect the quasiparticle operators to local operators. Typically however, in such systems pairs or higher multiplets of quasiparticles may be annihilated locally. Similarly, as a system is tuned toward a critical point, the quasiparticle description may break down, leading to a vanishing quasiparticle residue, and a continuum of low-energy excitations emerges. In both of these cases we do not expect that the ground state can be prepared in constant time (scaling in the size of the system). We will see this explicitly in the case of the antiferromagnetic TFIM in Section \ref{Sec:CoolingSpinChain}.

To summarise our discussion, a possible strategy for choosing {\it local} cooling operators is to adiabatically connect the system's (effective) Hamiltonian to the nearest simple solvable Hamiltonian having local quasiparticle operators, and choose the coupling to auxiliary qubits to annihilate these quasiparticles. We note that, although above we viewed $\hat V$ as a perturbation, our argument is not restricted to weak-coupling: in Sec.~\ref{Sec:Non-Integrable}, we will show that this method is effective in cooling an interacting spin ladder in the regime where the perturbation $\lambda$ becomes of order one.

\subsection{Choice of the filter function and two cooling protocols}

We now turn to the second essential ingredient of efficient cooling, the engineering of the filter function in  Eq.~(\ref{eq:filterF}). When the cooling operators are chosen according to the argument in the previous Subsection, the operators will still generally have small matrix elements for transitions that increase the system energy, i.e.~those corresponding to $\Delta(\vec{\alpha},\vec{\beta})<0$ in Eq.~(\ref{eq:rate_equation}). As we now discuss, a proper choice of the filter function corrects for this deficiency by suppressing negative frequency (heating) transitions.

The filter function can be seen as a Fourier summation of the time-varying function $f_\tau$, truncated to a finite number of terms, and so acts as a spectral filter on allowed transitions. Since in experiment the reset time is finite, spectral features of the filter function are broadened, with width of order $\delta\epsilon = 1/T$; however, assuming that $T$ is sufficiently large compared to the inverse many-body gap, $T\Delta \gg 1$, it should be possible to choose the modulated function $f_\tau$ so that the filter function strongly suppresses heating processes, 
\begin{equation}\label{eq:filtercondition}
    \bigg|\frac{F_{h,T}(+\epsilon)}{F_{h,T}(-\epsilon)}\bigg| \gg 1, \hspace{.5cm} \epsilon \gtrsim \Delta.
\end{equation}

For Floquet systems, the quasienergy is $2\pi$-periodic and this statement needs clarification. It is natural to consider the prethermal limit where quasiparticle quasienergies are close to the energies of quasiparticles of the corresponding prethermal Hamiltonian~\cite{abanin17effective}. For example, the transverse-field Ising model with arbitrary integrability-breaking local terms is in the prethermal regime as long as $|J|, |g|\ll 1$ and a similar condition is satisfied by the additional, integrability-breaking terms. In the prethermal regime, generally $|\epsilon_k|\ll \pi$, and heating processes due to Floquet evolution, and the associated energy non-conservation, are suppressed exponentially in $1/|g|,1/|J|$. The prethermalization assumption then allows us to neglect processes in Eq.~(\ref{eq:rate_equation}) where quasienergy changes by $\pm 2\pi, \pm 4\pi$ etc, since the corresponding matrix elements are strongly suppressed. 

Next, we introduce two protocols which realise Eq.~(\ref{eq:filtercondition}) to different degrees. In the first strategy, one aims to resonantly remove single quasiparticles from the system, by approximately matching auxiliary excitation quasienergy $\pi h$ to a value $\epsilon$ within the quasiparticle band, $\Delta<\epsilon<\Delta + W$, where $\Delta$ is the single-particle quasienergy gap and $W$ is the bandwidth. The function $f_\tau$ can be chosen to have a sufficiently broad profile, with width of order $1/W$, so that the filter function is centered at $\pi h$, with width of order $W$. A simple example is to use a constant pulse:

\vspace{.5cm}
\begin{textframe}{}
\subsubsection{Step-wise cooling protocol (SCP)}

The auxiliary quasienergy is matched to the quasiparticle band, $\Delta<\pi h<\Delta + W$, and the function 
\begin{equation}
    f_\tau=1/T 
\end{equation} is kept constant, therefore giving a `step' pulse for $1\leq \tau \leq T$. We will refer to this approach as the {\it step-wise cooling protocol (SCP)}.

\end{textframe}
\vspace{.3cm}

The SCP was the strategy used in the experiment of Ref.~\cite{mi2024stable}. The filter function corresponding to the SCP gives, up to a phase factor,
\begin{equation}\label{eq:Fstep}
    F^{\rm SCP}_{h,T}(\epsilon)=\frac{\pi}{T}\frac{\sin[(\epsilon-\pi h)T/2]}{\sin[(\epsilon-\pi h)/2]}, 
\end{equation}
which for $T\gg 1$ has a sharp peak at $\epsilon=\pi h$, with a width of order $\delta \epsilon= \frac{2\pi}{T}$. Thus, in the SCP, the processes of quasiparticle removal in Eq.~(\ref{eq:rate_equation}) with quasienergies $\epsilon\approx \pi h$ are favored approximately according to 
\begin{equation}
    \bigg|\frac{F^{\rm SCP}_{h,T}(+\pi h)}{F^{\rm SCP}_{h,T}(-\pi h)}\bigg| \approx h T.
\end{equation}

The SCP is efficient in systems with a sufficiently large quasiparticle gap and small bandwidth, as demonstrated in \cite{mi2024stable}; in this case the resonance peak of the filter function can be made broad enough to cover the entire quasiparticle band, while only having small weight at negative frequencies, suppressing heating processes. However, as the gap is decreased, in particular when a quantum critical point is approached, the heating processes become more important. Even in a gapped phase, the suppression of heating is only algebraic in quasiparticle energy, which leads to non-zero quasiparticle density in the steady state. Therefore, SCP does not allow for high fidelity state preparation in most cases. A second disadvantage is the need to tune the auxiliary quasienergy $\pi h$ to overlap with the possibly unknown quasiparticle band.

Instead of transitions occurring only at the natural resonant frequencies of the auxiliary qubits, transitions that lead to cooling may be engineered by strongly modulating the coupling $f_\tau$ in the time domain. In the task of cooling the system toward thermodynamic equilibrium (in particular the ground state), it is natural to consider those functions which satisfy the \emph{detailed balance} condition \cite{lidar2020lecture}:
\begin{equation}\label{eq:detailedbalance}
    \bigg|\frac{F_{h,T}(+\epsilon)}{F_{h,T}(-\epsilon)}\bigg| = e^{\epsilon\beta/2},
\end{equation} 
where $\beta$ is the inverse temperature of the target equilibrium. The detailed balance condition ensures that heating processes are thermally suppressed, with the relative ratio of cooling to heating rates matching that in thermal equilibrium. Note that the SCP explicitly violates this condition, and thus drives the system into a non-equilibrium state with small quasiparticle density.

In order to construct a filter function which satisfies the detailed balance condition, we start by observing that the `Fermi' function, $1+\tanh(\epsilon\beta/4)$, satisfies  Eq.~(\ref{eq:detailedbalance}). However, the filter function must satisfy periodicity in the quasienergy argument, and so a correct choice is given by the $2\pi$-periodic `symmetrised' Fermi function 
\begin{equation}\label{eq:symmFermi}
    F_\beta(\epsilon) = \frac{\pi}{2A}\sum_{m=-\infty}^\infty  e^{im\pi}\tanh([\epsilon - m\pi]\beta/4),
\end{equation}
where $A$ is a normalisation constant, and the function is symmetric about $\epsilon = \pi/2$. For $|\epsilon| \ll \pi$, this function looks like the Fermi function with `thermal' broadening $\beta$, and satisfies the detailed balance condition. By decomposing Eq.~(\ref{eq:symmFermi}) as a Fourier series (see Appendix \ref{app:MCP}), we are led to the following coupling time-dependence:

\vspace{.5cm}
\begin{textframe}{}
\subsubsection{Modulated cooling protocol (MCP)}

The coupling strength $f_\tau$ is modulated in time according to 
    \begin{equation}\label{eq:ftau_step}
    f_\tau = \frac{2}{A\beta} \frac{\sin [\pi (\tau-\tau_0)/2]}{\sinh[2\pi(\tau-\tau_0)/\beta]},
    \end{equation}
where $\tau_0 = T/2$, $A$ is the normalization factor~\footnote{We note that the normalization factor is chosen according to the condition $\sum_{\tau=1}^T f_\tau=1$. The modulated coupling in Eq.~(\ref{eq:ftau_step}) is sign-changing, however, $\sum_{\tau=1}^T |f_\tau|$ is of order 1 for all cases considered below.}, and the auxiliary quasienergy is fixed at $\pi h = \pi/2$. We call this approach the {\it modulated cooling protocol (MCP) }. 

\end{textframe}
\vspace{.3cm}

For finite $T$ and $1\ll \beta \ll T$, the filter function associated to the MCP gives a good approximation to Eq.~(\ref{eq:symmFermi}), $F^{ \rm MCP}_{h,T}(\epsilon) \approx F_\beta(\epsilon)$, with the Fourier series being effectively truncated at order $T$. Hence, the MCP satisfies detailed balance and leads to \emph{exponential} suppression of heating processes in the limit given above. We have verified that $F^{\rm MCP}_{h,T}(\epsilon)$ in fact closely approximates $F^\beta(\epsilon)$ even when $\beta/T \approx 1$ (see Fig.~\ref{fig:MCP}, Appendix \ref{app:MCP}). Using the experimental values of Ref.~\cite{mi2024stable}, and setting $\gamma T/\theta^2 \approx 1$, where $\gamma$ is the noise rate and $\theta$ the coupling strength, we find that the current practical regime in the presence of noise is around $T \approx 10$. Reduction of the noise levels in future quantum devices will allow the reset time to be further increased, which as we will see leads directly to higher state preparation fidelities.

In the rest of this paper, we focus on cooling \emph{gapped} quasiparticle systems, where the many-body gap is $\Delta > 0$. Then for high-fidelity ground state preparation, it is sufficient to ensure $\beta \Delta \gg 1$ for the MCP. Due to the detailed balance condition, it is also possible to use the MCP to prepare finite-temperature thermal states, as we review in Appendix \ref{app:thermal}, simply by choosing $\beta$ accordingly. We note that the strategy given above, of first writing down the filter function in the frequency domain and Fourier transforming to obtain the coupling modulation, is general and can be used to target other steady states which are not necessarily in thermal equilibrium.

\section{Theory of quasiparticle cooling in a spin chain}\label{Sec:CoolingSpinChain}

In this Section, we study the performance of the cooling algorithms using the example of the TFIM, specified by the Floquet operator in Eq.~(\ref{eq:TFIM}). We assume the absence of noise (which we include in Section \ref{Sec:Noise}). We first consider the edge auxiliary setup (Fig.~\ref{fig1}c). We will see that this leads to solvable dynamics in the sense that the density matrix of the system remains Gaussian. We show that the predictions of the rate equation derived above are in excellent agreement with the exact numerical solution. Then, we study the bulk auxiliary setup (Fig.~\ref{fig1}d), where integrability is broken by the coupling to auxiliaries. In this case we compare the rate equation predictions to matrix-product states simulations of the dissipative evolution, finding again a good agreement. We investigate the dominant cooling and heating processes in the two phases, paramagnetic (PM) and antiferromagnetic (AFM), of the TFIM, finding qualitative differences due to the different nature of quasiparticles. By comparing the performance of the SCP and MCP, we find  that the latter allows ground state preparation with high fidelity, in the noiseless case, by increasing the parameter $\beta$ ($T$). 

\subsection{Transverse-field Ising model}

We start by briefly summarizing the properties of the Floquet TFIM. This model can be mapped onto a $p$-wave superconducting chain by the Jordan-Wigner transformation: 
\begin{equation}\label{eq:JW}
    \hat{\sigma}_j^-=\prod_{i=1}^{j-1} e^{i\pi \hat c_i^\dagger \hat c_i} \hat c_j^\dagger, \hspace{.5cm} \hat Z_j=1-2\hat c_j^\dagger \hat c_j, 
\end{equation}
where $\hat c_j^\dagger, \hat c_j$ are fermionic creation/annihilation operators. The quasiparticle bands have quasienergies $\epsilon_k$ given by:
\begin{equation}\label{eq:dispersion}
    \cos \epsilon_k=\cos(\pi J) \cos (\pi g)-\sin(\pi J)\sin(\pi g)\cos k,  
\end{equation}
where $k\in [-\pi;\pi]$ is a quasimomentum. As a function of parameters $J$ and $g$, the model has four phases, separated by phase transition lines at $J=g$, $J=1-g$, where the quasiparticle gap $\Delta$ closes. The four phases are distinguished by the $0$ and $\pi$ Majorana edge modes~\cite{ThakurathiFloquetPRB2013,BauerFloquetPRB2019,LerosePRB2021}. Here we will be interested in the part of the phase diagram where $0< (J,g)<1/2$, and two phases exist: a paramagnetic (PM) phase at $g>J$, and an antiferromagnetic (AFM) phase at $g<J$. For  $g,J\to 0$ these are the phases of the Hamiltonian TFIM; by analogy to  this case, we define the Floquet vacuum projector 
\begin{equation}
\Omega = \sum_{g} |\Omega_g\rangle\langle{\Omega_g}|
\end{equation}
to project onto states where the valence quasiparticle band is filled, and the  conduction band is empty i.e.~$\hat \eta_k|\Omega_g\rangle = 0$, $\forall k$, with the quasiparticle annihilation operators $\hat \eta_k$ specified below. The index $g$ runs over the possible ground states: in the PM phase the vacuum state is unique, while in the AFM phase in the limit $N_S\to\infty$, the vacuum state is doubly degenerate, $g=\pm1$, and the $\mathbb{Z}_2$ symmetry is spontaneously broken. In the fermionic counterpart of the TFIM, this degeneracy stems from the existence of zero-(quasi)energy Majorana edge modes, and the two vacuum states of the AFM differ by occupation of the Majorana level. The quasienergy gap between the ground and first excited state is given by $\Delta = 2\pi|J-g|$. In the rest of the paper, unless otherwise stated, results for the AFM (PM) phase correspond to the parameter points $J=0.2\ (0.1),\ g=0.1\ (0.2)$.

For the open-boundary chain we consider, the allowed quasimomenta are quantized, $k_m \approx \pi (m-1)/N_S$, $m=1,2,.., N_S$, and the fermionic eigenmodes, defined by creation (annihilation) operators $\hat \eta_k^\dagger$ ($\hat \eta_k$), are superpositions of plane waves with wave-vectors $\pm k$ (standing waves). The relation between on-site fermionic operators and eigenmodes is 
\begin{equation}\label{eq:site-eigenmodes}
    c_j=\sum_k u_{jk}\eta_k +v_{jk} \eta_k^\dagger,
\end{equation}
with the coefficients $u$, $v$ referred to as the `Bogolioubov coefficients'. We refer to Appendix \ref{app:TFIM} for a more detailed discussion and explicit relations. 

In the paramagnetic limit $J=0$, the quasiparticle annihilation operators are given by $\hat \eta_{k} = \frac{1}{\sqrt{N_S}} \sum_j e^{ijk}\hat\sigma_j^+$. Following the discussion in Subsection \ref{sec:coolingchoice} regarding the optimal choice of system-auxiliary coupling in Eq.~(\ref{eq:couplingops}), for the rest of this section we set 
\begin{equation}\label{eq:SWAP}
    \hat V = \pi \sum_{j} (\hat\sigma^+_{\mathcal{Q}_j}\hat\sigma^-_{\mathcal{A}_j}+\hat\sigma^-_{\mathcal{Q}_j}\hat\sigma^+_{\mathcal{A}_j}), 
\end{equation}
where $\mathcal{Q}_j$ ($\mathcal{A}_j$) represent system (auxiliary) position associated with $j$-th coupling.

\subsection{A solvable model: cooling at the edge}

We now study a solvable example of cooling in the edge auxiliary setup of Fig~\ref{fig1}c. In this case, the system cooling operator $\hat\sigma_1^+$ ($\hat\sigma_1^-$) is free of the Jordan-Wigner string in the fermionic language, being  proportional to $\hat c_1$ ($\hat c_1^\dagger$) (see Eq.~(\ref{eq:JW})). Therefore, in the rate equation (\ref{eq:rate_equation}), only processes where occupation of one of the quasiparticle levels changes by $\pm 1$ are allowed. Due to reflection symmetry, the auxiliary at the other edge contributes equally to the cooling process, and Eq.~(\ref{eq:rate_equation}) takes the form:
\begin{gather}\label{eq:bryrateeqn}
    \delta n_k= -n_k W_k^- +(1-n_k) W^+_k ,
\end{gather}
with the single-particle cooling and heating rates 
\begin{gather}
    W^-_k = 2\theta^2|F_{h,T}(+\epsilon_k)|^2 |u_{1k}|^2,  \\
    W^+_k = 2\theta^2 |F_{h,T}(-\epsilon_k)|^2 |v_{1k}|^2.
\end{gather}
The coefficients $|u_{1k}|^2, |v_{1k}|^2$ naturally represent the probabilities for the quasiparticle (quasihole) to be found at the boundary. Assuming that the system is away from the critical point, the probabilities scale as $O(N_S^{-1})$ (from the normalization condition) away from the band edge, but are $O(k^2 N_S^{-1}) \approx O(N_S^{-3})$ near the band edge $k \approx 0, \pi$. This leads to a pronounced dependence of the cooling rates on the quasienergy $\epsilon_k$, with the late time dynamics dominated by the cooling of modes near the band edges. Eq.~(\ref{eq:bryrateeqn}) can be solved for the steady state populations
\begin{equation}\label{eq:steady_state_edge}
    n_k^{\infty} = \frac{W^+_k}{W^+_k+W^-_k},
\end{equation}
showing that a finite heating rate will inevitably lead to non-zero quasiparticle population.

As a first characterisation of the cooling performance, we study the dynamics of the quasiparticle density $n(t)=N_S^{-1}\sum_k n_k(t)$. We compare the rate equation predictions to exact numerical simulations,  which can be performed efficiently in the case of edge cooling due to the problem's free-fermion nature (see Appendix \ref{app:numerics}). In Fig.~\ref{fig2}a,b we plot $n(t)$ for both protocols, SCP and MCP, in the two phases, starting from the maximally mixed state $n(0) = 1/2$. We take $T=4$, $h = 0.3$ in the SCP, and $T= 3\beta/2 = 28$, $h=0.5$ in the MCP; the smaller reset time in the SCP case is approximately optimal (see Fig.~\ref{fig2}c) while the MCP choice corresponds to $\Delta\beta \approx 4\pi$. In both cases $\theta = 0.02$ and $N_S=20$. We observe a very good agreement between the rate equation theory (dashed lines) and the exact numerical simulation (solid lines). In  the considered time window, for both PM and AFM phases under the MCP, the density decays predominantly exponentially in time, with a rate proportional to $\theta^2$. In accordance with the arguments regarding heating rates in Section \ref{Sec:protocol}, the MCP exhibits a saturation of $n(t)$ at a much smaller value than the SCP (though in that case, the saturation is still to a relatively small value on account of the large  quasiparticle gap). We also note the long time-scales associated with cooling to states of low quasiparticle density, with $\nu \approx 300\theta^{-2}$ for the PM phase under the MCP and even greater for the AFM. This is a consequence of the edge cooling setup, and the scaling of the Bogoliubov coefficients near the band edge discussed above.

As a second test, we examine the fidelity of the steady state distribution with respect to the Floquet vacuum, $\mathcal{F} = \text{Tr}(\Omega\rho).$ In the case where Eq.~(\ref{eq:GGE}) holds, the fidelity can be rewritten
\begin{equation}
    \mathcal{F} = \prod_{k} (1-n_k) \approx \exp(-\sum_k n_k).
\end{equation}
The second line holds in the limit of sufficiently low total quasiparticle density. In this regime, the log-fidelity per qubit and quasiparticle density are approximately synonymous. In Fig.~\ref{fig2}c we plot the steady state fidelity for the two pulses as the parameter $\beta$ is increased, with $T=3\beta/2$ fixed (for the SCP this is just a variation of parameter $T$). 
For parameters considered, the SCP allows to reach fidelity of order $0.99$ per qubit\, while the MCP fidelity per qubit is better than one part in $10^6$ by the point $\beta\Delta \approx 8\pi$. From Eq.~(\ref{eq:detailedbalance}), we see that the density of quasiparticles is bounded approximately as $n \sim e^{-\beta\Delta}$ for large $\beta$. The apparent saturation of MCP around $\beta\Delta=8\pi$ is due to small ringing errors, which can be removed by increasing the ratio $T/\beta$ --- we refer the reader to Appendix \ref{app:MCP} for more details. In practice, dominant heating processes arise due to noise and it pays to keep the ratio $T/\beta\gtrsim O(1)$. 

From the above results, we conclude that the MCP is able to successfully prepare the ground state of the TFIM, with fidelity approaching 1 in the limit of large $\beta$. However, the associated time scale diverges with the system size, naturally a weakness of the edge cooling setup. The rate equation accurately captures the quasiparticle dynamics for the case of edge cooling. Next, we will consider the bulk cooling setup with a finite density of auxiliaries.

\begin{figure*}[t]
    \centering
    \includegraphics[width=0.66\columnwidth]{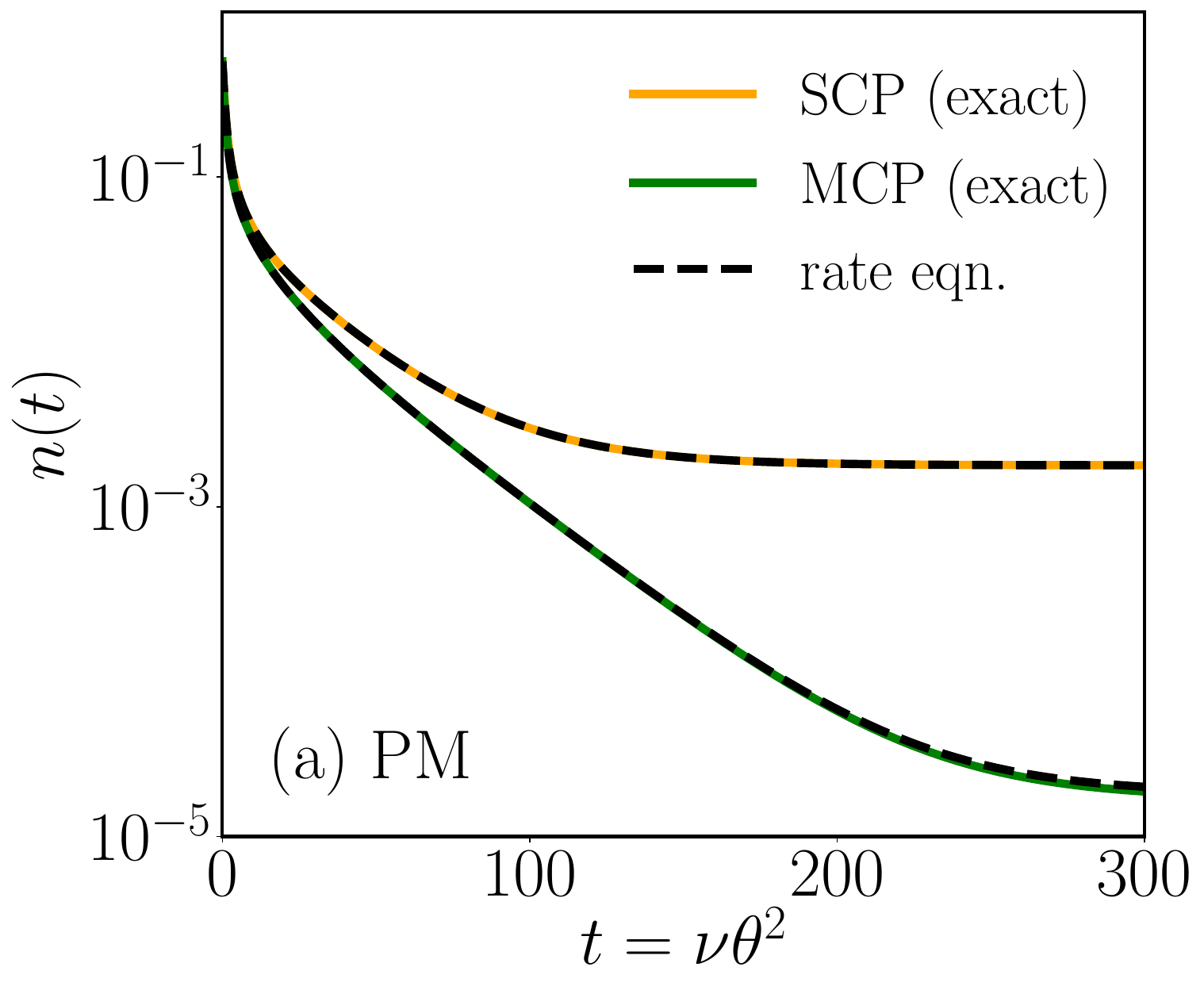}
       \includegraphics[width=0.66\columnwidth]{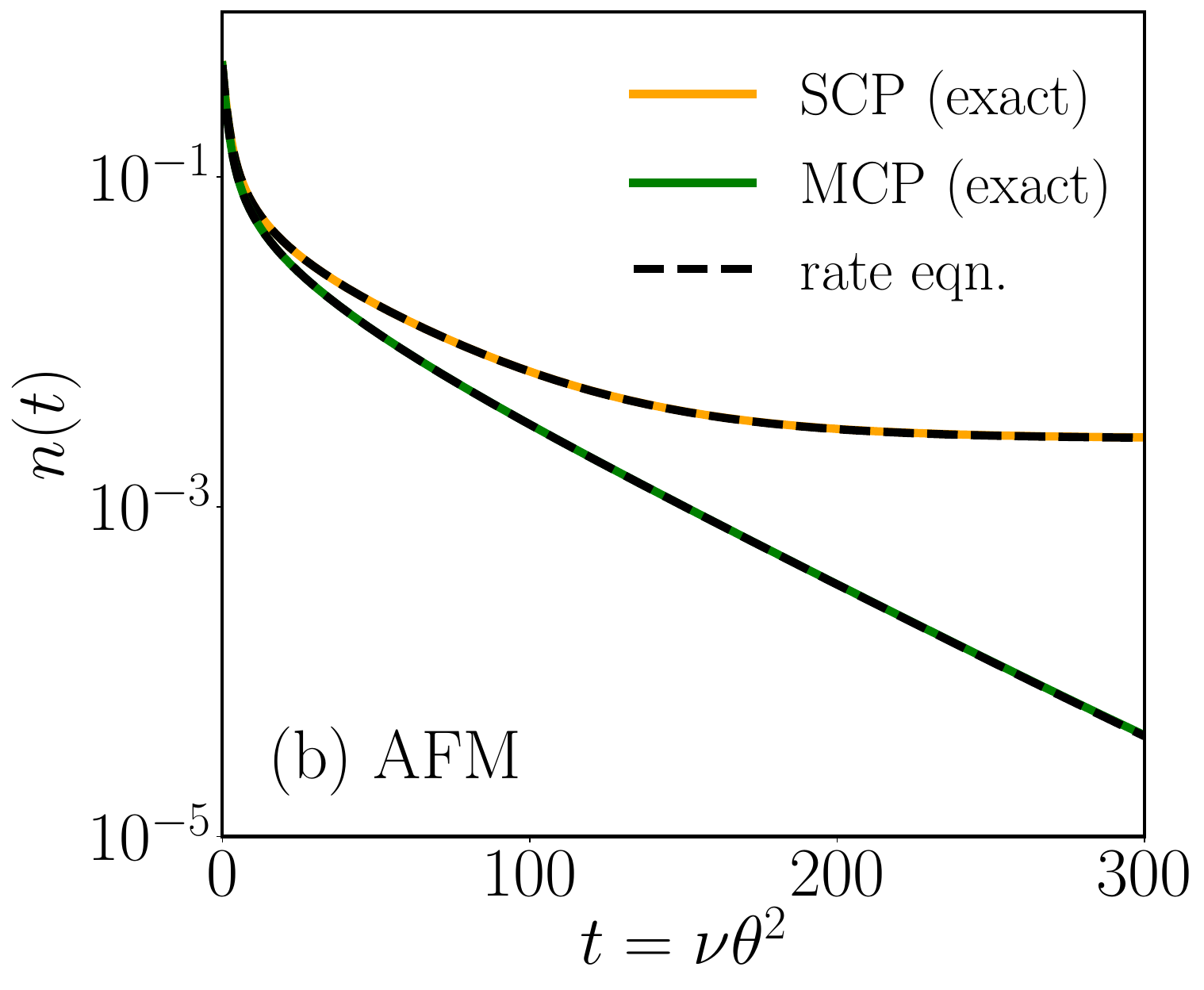}
          \includegraphics[width=0.66\columnwidth]{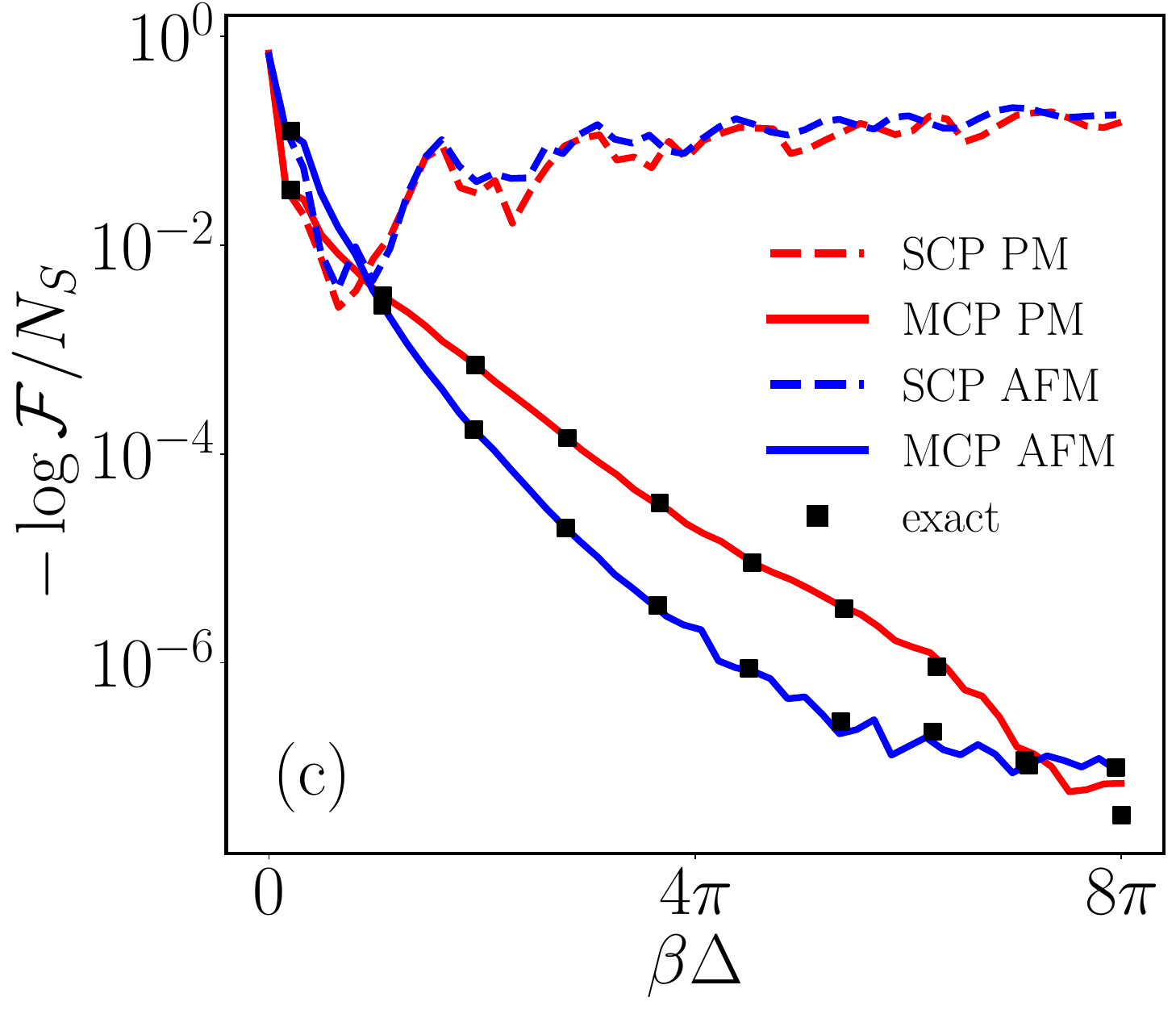}
    \caption{Performance of edge cooling protocol, for the parameter choice described in the main text. (a) Evolution of quasiparticle density $n(t)$ as a function of rescaled timescale $t = \nu\theta^2$, where $\nu$ is total number of cooling cycles, for the PM phase. We compare exact numerical results with $\theta=0.02$ (solid lines) and the rate equation (dashed lines), for both SCP and MCP; (b) Same as in (a) but for AFM phase; (c) Log-fidelity per qubit of the steady state with quasiparticle vacuum, obtained from kinetic equation, as a function of the MCP parameter $T = 3\beta/2$. Black squares are exact numerical data, with $\theta=0.001$, showing that kinetic theory accurately captures the steady state properties.} 
    \label{fig2}
\end{figure*}

\subsection{Finite density of auxiliaries}

\begin{figure*}[t]
    \centering
    \includegraphics[width=0.65\columnwidth]{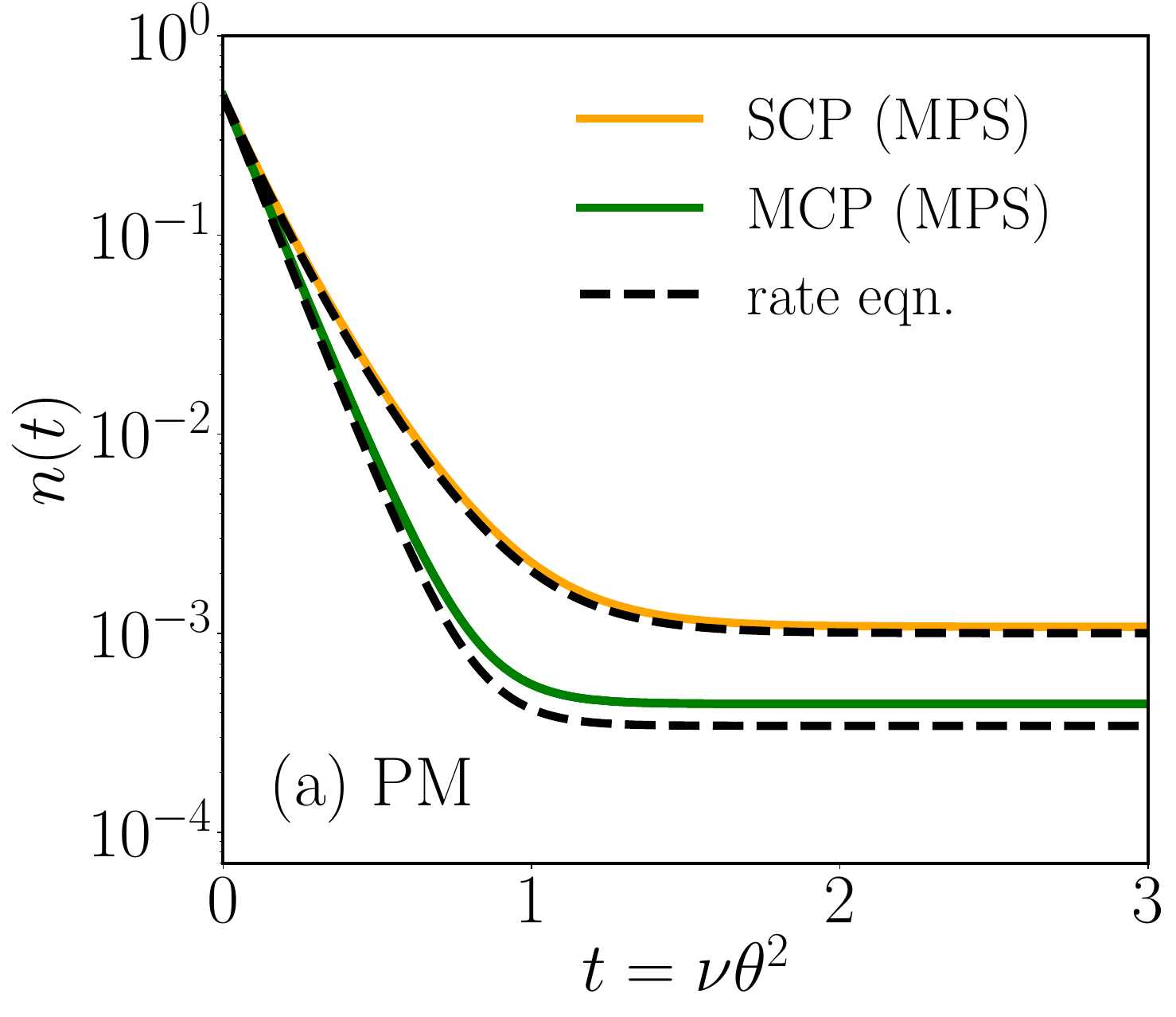}
       \includegraphics[width=0.65\columnwidth]{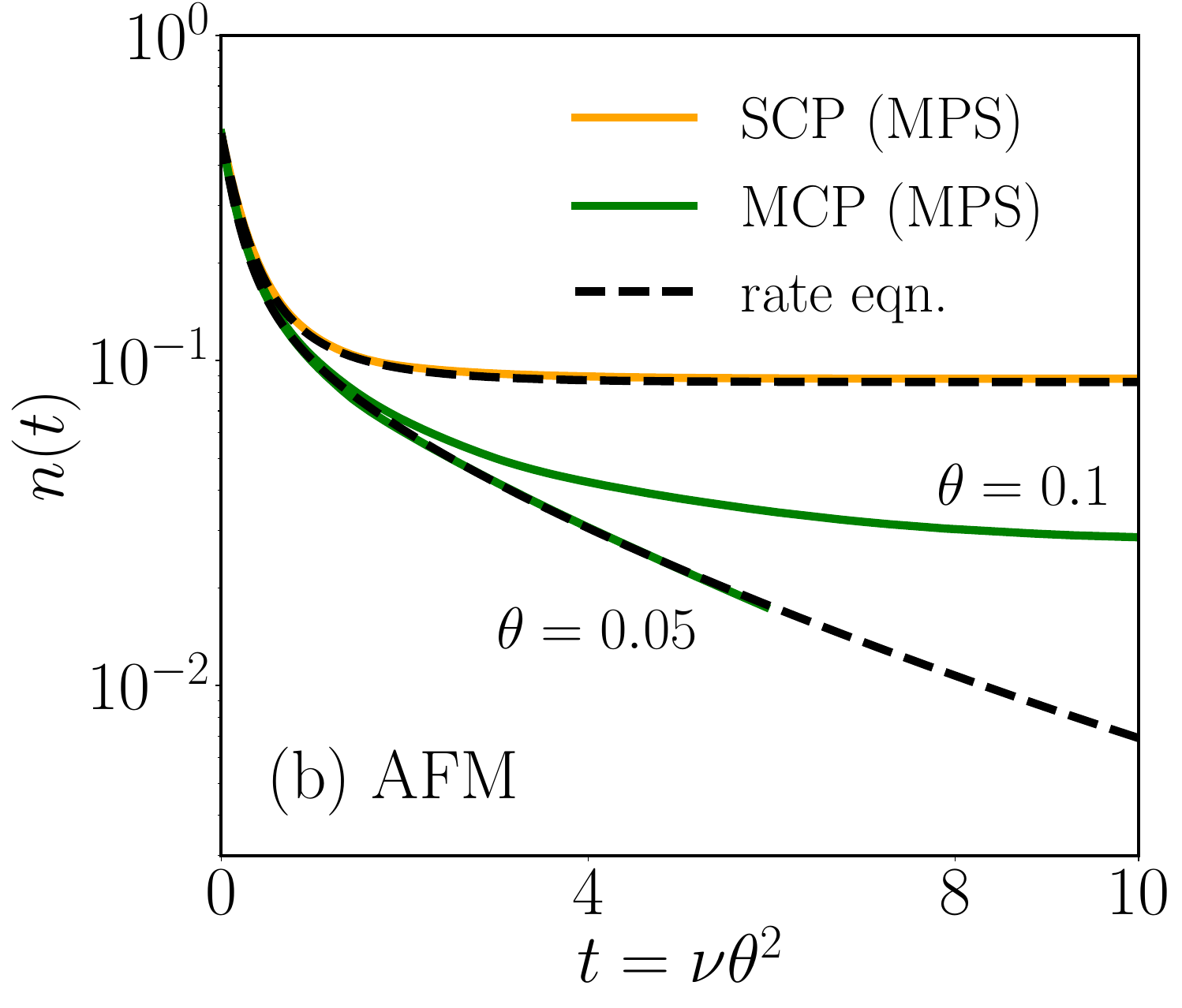}
          \includegraphics[width=0.65\columnwidth]{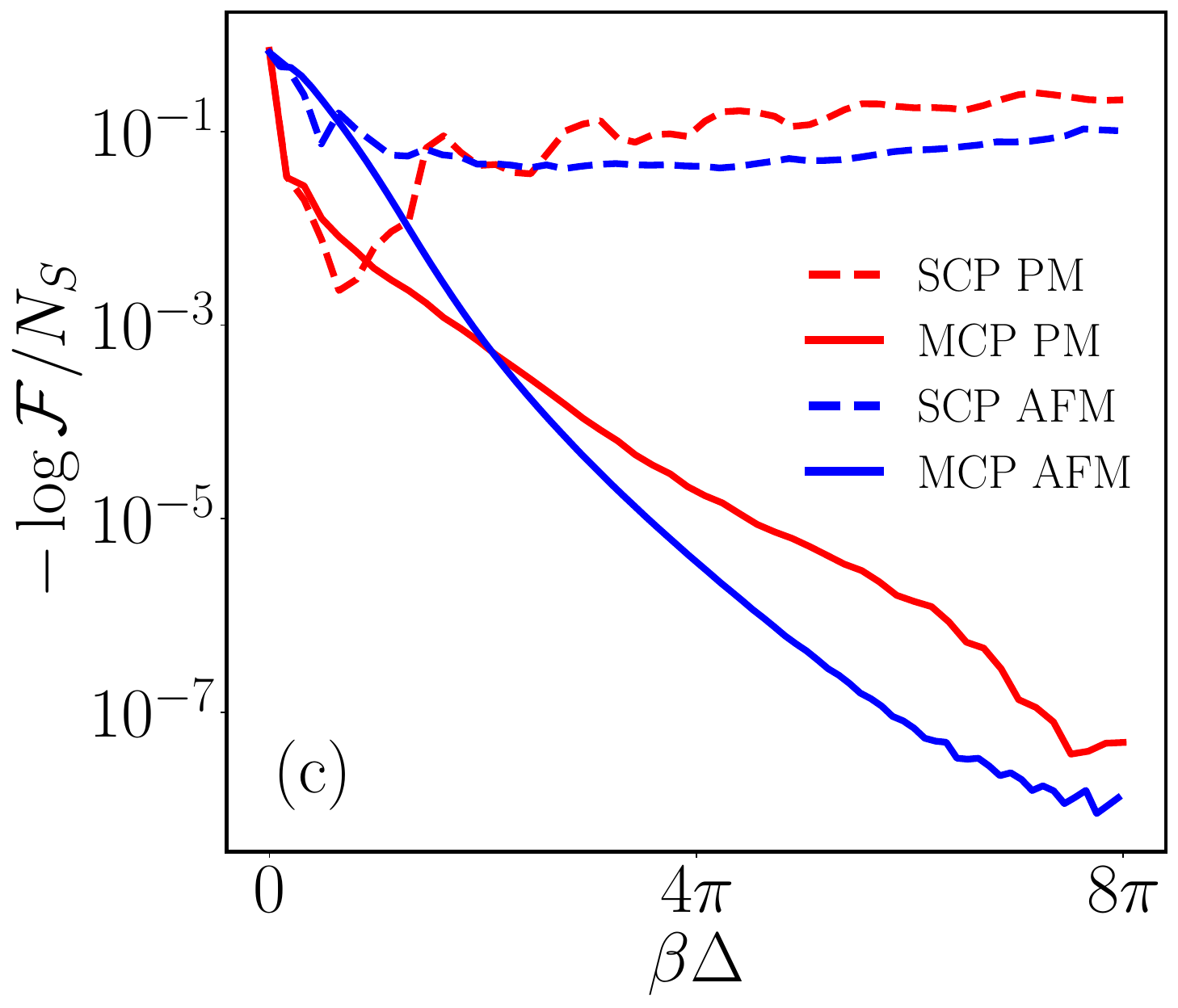}
            
    \caption{Performance of the bulk cooling protocol, with one auxiliary per system qubit and $N_S=20$. The parameter choice is described in the main text. (a) Quasiparticle density $n(t)$ vs.~rescaled timescale $t = \nu\theta^2$ in the PM phase. The MPS results (solid lines) are compared to the kinetic theory predictions (dashed lines), for both SCP and MCP. Parameter $\theta=0.1$. (b) Same as in (a) but for the AFM phase. We show MPS results for two coupling strengths, $\theta = 0.1$ and $0.05$, with the larger showing corrections to the rate equation. 
    The AFM cools more slowly than the PM, as expected due to the topological nature of the quasiparticles. (c) Log-fidelity per qubit of the steady state with quasiparticle vacuum, calculated from the kinetic equation, as a function of the MCP parameter $T = 5\beta/2$.} 
    \label{fig3}
\end{figure*}

While providing an instructive example, the edge cooling setup is not practical for large systems. We argued above that the cooling timescale diverges with the system size in that case, which holds generally for models with local boundary dissipation \cite{vznidarivc2015relaxation}. Furthermore, the cooling rate remains constant while the heating rate due to noise scales with system size. Therefore, we now consider the setup in Fig.~\ref{fig1}d, with a finite density of auxiliaries. For simplicity, we will assume that there is one auxiliary per system qubit.

In this setup, the dynamics is no longer integrable, and the system's density matrix develops non-Gaussian correlations due to the fact that the matrix elements for bulk $\hat\sigma_j^\pm$ operators carry Jordan-Wigner strings in the fermionic formulation (see Eq.~(\ref{eq:JW})). 
This allows for transitions between many-body eigenstates where the occupation numbers of several quasiparticle levels change at once. 

To make progress, we therefore assume that the processes involving a small number quasiparticles are dominant. For cooling processes, this assumption is justified in the low-density limit $n_k \ll 1$, which describes the system's dynamics close to the ground state~\footnote{Heating processes, in particular due to noise, may in principle involve a large number of particles; however, we will find that few-quasiparticle processes are dominant, and considering them gives an accurate approximation of heating rates.}. Restricting to processes involving at most two quasiparticles, we derive the following kinetic equation, containing cooling and heating terms, as well as terms that describe scattering between quasiparticles:

\begin{widetext}

\begin{align}\label{eq:bulk_rate_equation}
\begin{split}
     \delta n_k & =-n_k W_k^-+(1-n_k)W_k^+ + \sum_q \Big[-n_k W_{k,q}^- n_q+(1-n_k) W_{k,q}^+ (1-n_q)-n_kV_{k,q}^{-}(1-n_q)+(1-n_k) V_{k,q}^+ n_q\Big].
\end{split}
\end{align}
\end{widetext}

We compute the rates in Eq.~(\ref{eq:bulk_rate_equation}) assuming the system is near the vacuum state $\Omega$, which yields:
\begin{equation}\label{eq:bulk_rates1}
        W_k^{\mp} = \theta^2 \sum_{j=1}^{N_S}  |F_{h,T}(\pm \epsilon_k)|^2 |\sigma^\pm_{j;0;k}|^2,
\end{equation}
\begin{equation}\label{eq:bulk_rates2}
    W_{k,q}^{\mp} =\theta^2 \sum_{j=1}^{N_S}  
    |F_{h,T}(\pm (\epsilon_k+\epsilon_q))|^2 |\sigma^\pm_{j;0;k,q}|^2,
\end{equation}
\begin{equation}\label{eq:bulk_rates3}
    V_{k,q}^{\mp} = \theta^2  \sum_{j=1}^{N_S} |F_{h,T}(\pm (\epsilon_k-\epsilon_q))|^2 |\sigma^\pm_{j;q;k}|^2,
\end{equation}
where we introduced the shorthand matrix-elements
\begin{equation}
    \sigma^\pm_{j;k_1,\ldots,k_m;k'_1,\ldots,k'_n} = \bra{k_1,\ldots,k_m}\hat\sigma^\pm_{j}\ket{k'_1,\ldots,k'_n}.
\end{equation}
Above, the terms $W^\mp_k$ correspond to cooling/heating processes involving one quasiparticle, while the terms $W^\mp_{k,q}$ describe removing/adding a pair of quasiparticles, and $V^\mp_{k,q}$ describe quasiparticle scattering. Note that while the scattering does not directly change the total quasiparticle occupation number, it will tend to move quasiparticles toward lower quasienergies due to the presence of the filter function $F_{h,T}$. We expect that these rates will be weakly affected by a small finite background density of quasiparticles. The matrix elements can be evaluated using Wick's theorem (see Appendix \ref{app:matrixelements} and a review~\cite{IsingModelReviewSantoro2020}).

\begin{figure}[t]
    \centering
    \includegraphics[width=0.85\columnwidth]{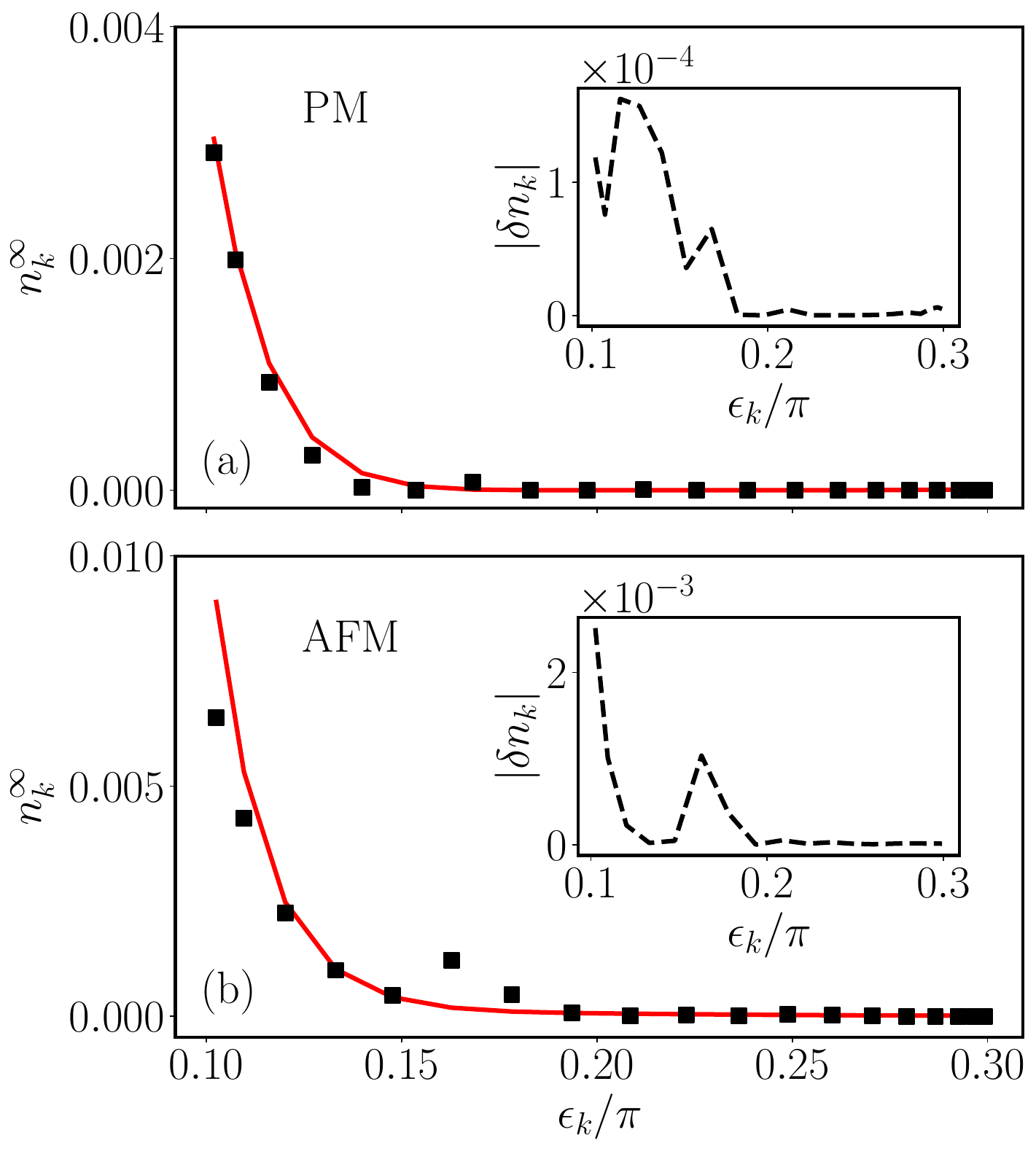}
    \caption{Quasiparticle occupation numbers $n^\infty_k$ in the steady state, as a function of quasienergy $\epsilon_k$. We compare MPS simulations (scatter points) and kinetic equation results (red line), for a system with $N_S = N_A = 20$ qubits in (a) PM and (b) AFM phases. The reset time is $T=\beta=12$, and the coupling $\theta=0.05$. The MPS simulations use a maximum bond dimension of $d= 100$. Inset: deviation $|\delta n_k|$ between rate equation and MPS values shown in main figure.}
    \label{fig4}
\end{figure}

Crucially, the matrix elements entering in the above rates obey different selection rules in the PM and AFM phases. In the PM phase, there is a unique symmetric ground state. Operators $\hat \sigma_j^\pm$ anticommute with the parity operator $\prod_j \hat{Z}_j$, so in the fermionic language they must change the fermion parity. Thus, two particle processes are forbidden in the PM phase, $W^\mp_{k,q} = V^\mp_{k,q} = 0$. In contrast, in the AFM phase, there are two vacuum states, $|\Omega_\pm\rangle$, which have parity $\pm 1$. The cooling process can induce transitions between them by changing the occupation of the Majorana edge fermion. $W_{k,q}^\mp$ and $V_{k,q}^\mp$ matrix elements are non-vanishing in this case. The bra and ket states should be understood as quasiparticle states on top of different vacua. Note that the Majorana edge mode has a vanishing energy in the limit of large system size; hence, this mode cannot by cooled by our protocol and will remain effectively at an infinite temperature, corresponding to an equal weight mixture of the two vacuum states.

In the spin language, these selection rules follow from the phenomenological discussion in Subsection \ref{sec:coolingchoice}: since we chose the system-auxiliary couplings in Eq.~(\ref{eq:SWAP}) to match the local magnon excitations of the paramagnetic fixed point, we expect that also within the PM phase single excitations can be removed locally; in contrast, in the AFM phase, quasiparticles are dressed domain walls  and can only be removed/created locally in pairs. This intuitive argument was pointed out in Ref.~\cite{RudnerCoolingArxiv2022}. In Appendix \ref{app:matrixelements}, we study the distinct behavior of matrix elements in the two phases in detail. In particular, the two-particle matrix elements in the AFM phase are homogeneous in the bulk, but single-particle matrix elements decay exponentially with the distance of the auxiliary qubit from the boundary (since single domain walls can be removed at the edges of the chain), with a decay length set by the correlation length of the model. We will see below that this difference in quasiparticle nature leads to different efficiencies of the cooling in the two phases. 

We now compare the predictions of the above kinetic theory to matrix product state (MPS) simulations of the system's density operator dynamics (more precisely, density matrix product operator simulations~\cite{VerstraetePRL04,ZwolakVidalPRL2004}). The dynamics are performed using the standard time-evolving block decimation (TEBD) algorithm~\cite{vidal2003efficient, vidal2004efficient}, implemented via the ITensor library~\cite{itensor}. Implementation details and numerical verifications of asymptotic convergence are presented in Appendix \ref{app:numerics}.

As illustrated in Fig.~\ref{fig3}a,b, the theory continues to be accurate for cooling with a finite density of auxiliaries, for both the SCP and MCP protocols. We fix system size $N_S=20$ with one auxiliary per site. We use cooling parameters $T=5$, $h=0.3$ in the SCP and $T=\beta=12$ in the MCP. The coupling is chosen to be $\theta = 0.1$ in the PM phase. In the AFM phase, this relatively large coupling leads to visible deviations between the MPS results and the kinetic equation prediction at late times; hence we include simulation results for a weaker coupling, $\theta=0.05$, which leads to a good agreement. We use bond dimension $d=100$ for both the SCP and MCP ($\theta=0.05$) data, and $d=150$ for the MCP ($\theta=0.1$) data. These choices are converged in bond dimension (see Appendix \ref{app:numerics}). Our choice of larger $\theta$ and smaller MCP reset time (compared to the edge cooling case) helps to avoid high entanglement at intermediate times, which affects the accuracy of the MPS simulations. The MCP outperforms the SCP, and in Fig.~\ref{fig3}c we provide further confirmation that the MCP can reach much higher fidelities, plotting the increase in the log-fidelity per qubit as the parameter $T=5\beta/2$ is increased (we choose a large value of $T/\beta$ to remove the ringing artifacts for the range of fidelities shown, see Appendix \ref{app:MCP}). 

The difference in cooling distinct quasiparticle types in the PM and AFM phases is also evident in the timescales shown in Fig.~\ref{fig3}a,b: the PM reaches a quasiparticle density below $10^{-3}$ after a small number of cooling cycles $O(\theta^{-2})$, with a clear exponential decay to the steady state value. In contrast, in  the AFM phase, cooling slows down once quasiparticle density is below $n(t)\lesssim 10^{-1}$. This separation of timescales in the AFM is due to the fact that initial cooling proceeds quickly through pair annihilation processes, but this mechanism is suppressed $\propto n^2$ in the low density regime. Late-time cooling is therefore dominated by single-particle processes, which may only occur around the system edge. 

As a final comparison between the kinetic equation and the MPS numerics, we look at the momentum-resolved quasiparticle occupations in the steady state, in Fig.~\ref{fig4}, focusing on the MCP. To reach the steady state in our MPS simulations, we perform the simulations starting from the exact ground state. As we expect the (unique) steady state to be close to the true ground state, this allows for faster real-time simulations compared to starting from a high energy-density state. The figure illustrates that occupation numbers of all levels are below $10^{-2}$, and the remaining population is mostly due to quasiparticles at the band edge. The finite weight on low energy modes stems from the relatively small value of the parameter $\beta\Delta \approx 2.4\pi$ chosen to aid the convergence of the MPS simulations.

To sum up, the analysis above demonstrates efficiency of MCP with a finite density  of auxiliaries for removing quasiparticles after $O(\theta^{-2})$ cooling cycles, in the PM phase. In agreement with the arguments of Ref.~\cite{RudnerCoolingArxiv2022}, we also find that topological, domain-wall type quasiparticles are more challenging to remove. Nevertheless, in an idealized system without noise, our cooling protocol allows preparation of the ground state with very high fidelity, limited only by the choice of reset time $T$.

\section{Cooling non-integrable systems}\label{Sec:Non-Integrable}

Having established the validity of our theory for the case of the TFIM, in this Section, we investigate the efficiency of the MCP for preparing ground states of non-integrable models. While these models are generally thermalizing, they typically have long-lived quasiparticles at low energy density. Therefore, the theory of dissipative cooling described in the previous sections for integrable systems, is expected to hold, once the system has been cooled to a state with sufficiently low energy density. 

We study two examples of non-integrable systems using tensor network and quantum trajectory simulations: the TFIM in the paramagnetic regime with an additional longitudinal field \cite{PhysRevB.68.214406}, and an antiferromagnetic spin ladder with Heisenberg couplings between neighbouring sites~\cite{barnes1993excitation}. In the longitudinal-field TFIM the physics is similar to the integrable case at low energies, except scattering now occurs between quasiparticles and multi-particle bound states emerge. Two-leg antiferromagnetic spin ladders model strongly-correlated low-dimensional magnetic systems \cite{azuma1994observation, notbohm2007one}, with suggested connections to high-temperature superconductivity \cite{dagotto1996surprises}. The low-energy spectrum features a band of spin-triplet excitations, as well as multi-particle bound states for intermediate coupling \cite{barnes1993excitation, vanderstraeten2015scattering}: thus, this relatively complex low-energy structure represents a good test for our cooling protocol.

\subsection{Longitudinal field TFIM}

The longitudinal field TFIM is defined by the Floquet unitary

\begin{gather}\label{eq:Ug}
    \hat U'_S =e^{-i\hat G_0} e^{-i\hat G_1}, \\
    \hat G_0 = \frac{\pi J}{2}\sum_{i=1}^{N_S-1} \hat X_i\hat X_{i+1}, \quad \hat G_1 = -\frac{\pi}{2}\sum_{i=1}^{N_S} g \hat{Z}_i+ g_X \hat{X}_i. \nonumber
\end{gather}
We fix $J = 0.1,\ g = g_X = 0.15$, for which the model is in a paramagnetic phase~\cite{PhysRevB.68.214406}. As discussed above, since quasiparticles in the PM phase are simple magnons, we expect this to be the fastest regime to cool, and simulation accuracy is higher since the system reaches the low-density regime much faster.

Our parameter choice is in the Floquet regime, and not in the strict Trotter limit ($J,g_x,g_z \ll O(1)$). Thus, to characterize the cooling efficiency, we consider effective Hamiltonians obtained in the first two orders of the Baker–Campbell–Hausdorff (BCH) expansion of Eq.~(\ref{eq:Ug}), $\hat U_S'\approx e^{-i \hat H_l}$, $l=0,1$. The BCH formula gives
\begin{equation}\label{eq:Hams}
    \hat H_0 = \hat G_0+\hat G_1,\quad  \hat H_1 = \hat G_0+\hat G_1 - \frac{i}{2}[\hat G_0,\hat G_1].
\end{equation}
The cooling efficiency can then be characterized by measuring the energy and the fidelity of the system's density matrix with respect to the ground states of the above Hamiltonians:
\begin{equation}\label{eq:EFNI}
  E_{l} = \text{tr} ( \hat H_l \hat \rho), \qquad \mathcal{F}_l =\langle \Omega_l|\hat\rho |\Omega_l\rangle, 
\end{equation}
where $l\in \{0,1\}$ denotes the BCH expansion order, and $\Omega$ is the corresponding ground state. The system is in the prethermal regime, due to the fact that $J,g,g_X \sim O(0.1)$, and is expected to cool towards  the ground state of an effective Hamiltonian $\hat H_{l_*}$, with an optimal truncation order $l_*\geq 1$ of the BCH series. Thus, we expect the ground state of $\hat H_1$ to provide a better approximation than that of $\hat H_0$. 

\begin{figure}[t]
    \centering
    \includegraphics[width=0.85\columnwidth]{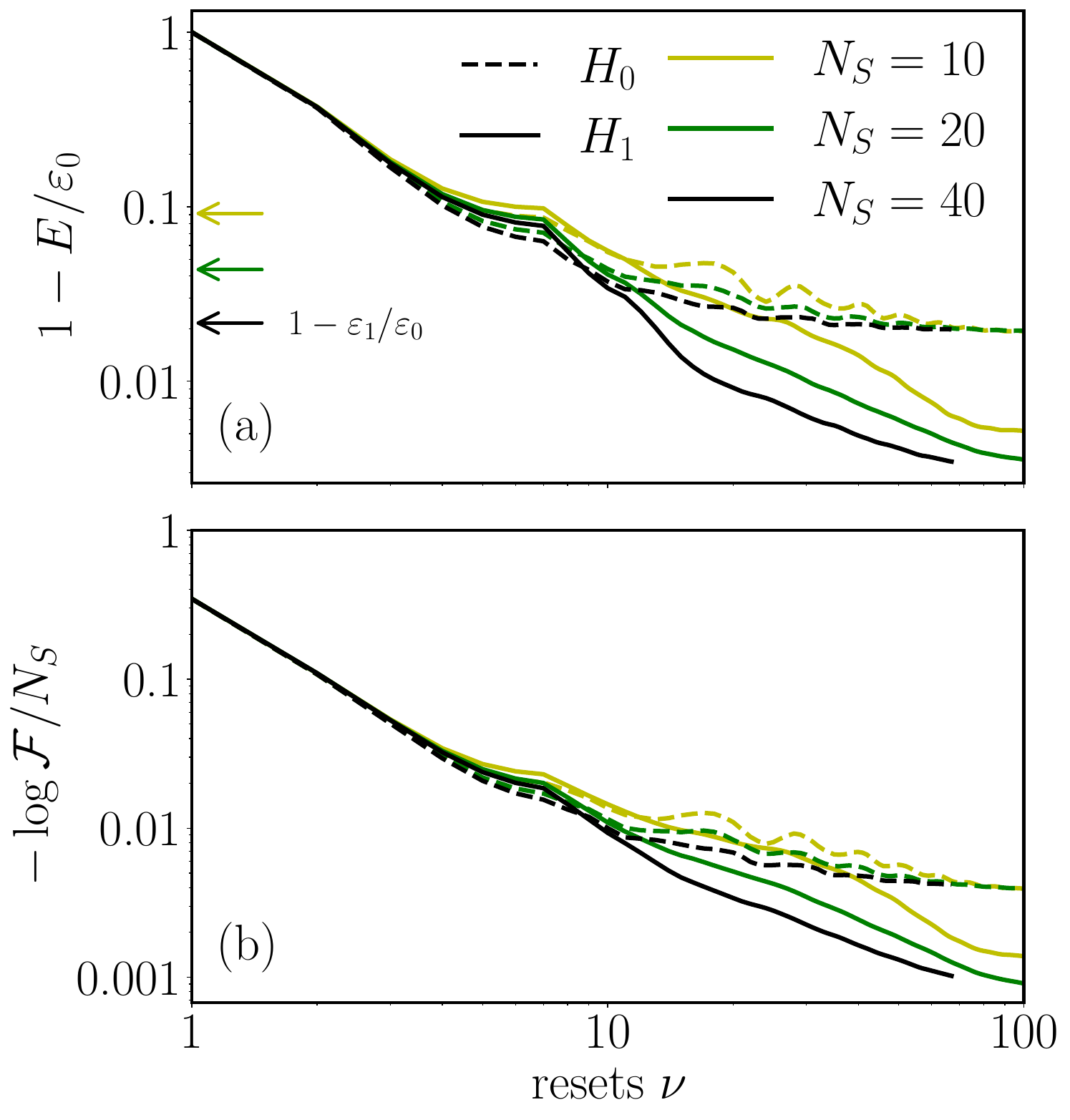}
    \caption{Cooling a non-integrable spin chain. (a) Evolution of energy, defined with respect to the ground state energy $\varepsilon_0$ of approximate effective Hamiltonian $H_0$ (dashed lines) and $H_1$ (full lines), as a function of the number of resets $\nu$, for different system sizes. Corresponding energy values for the first excited states $\varepsilon_1$ of $H_1$ are indicated by arrows, showing convergence of energy to a value well below the excitation gap. (b) Log-fidelity per qubit vs. number of resets. Model parameters are $h_x = h_z = 0.15, J = 0.1$ (PM phase), reset time $T = \beta = 20$. The bond dimension of simulations is $d = 300$.}
    \label{fig:NI}
\end{figure}

For the cooling protocol we focus on the MCP with finite density $N_A = N_S$. We take the coupling to auxiliaries to be the same as in Section \ref{Sec:CoolingSpinChain}. In contrast to the previous sections, where our aim was to develop a theory of cooling, here we generalise the protocol to cool more rapidly by using a large initial value of $\theta$, later approaching the weak coupling limit $\theta \ll 1$ where the previous analysis is expected to hold. To this end, we initialize the coupling at $\theta(0) \approx 0.4$ and progressively reduce to a final value $\theta_\infty \approx 0.083$. The value of the coupling is reduced as $\theta(t+1) = 0.8 \theta(t)$ whenever the energy of the system is reduced slower than a chosen rate,  $0<E_1(t)-E_1(t-1)<0.01$. We found this choice to be efficient for the finite time simulations. Note that the function $f_\tau$ is still modulated within a cycle, see Eq.~(\ref{eq:ftau_step}). We choose a reset time $T = \beta = 20$, for which  $\Delta\beta \approx 5.2$, where $\Delta$  is the (many-body) energy gap of $\hat H_1$. The tensor-network simulations are performed similarly to the integrable case, see\ Appendix \ref{app:numerics} for details. 

In Fig.~\ref{fig:NI}a, we show how the energies defined with respect to $\hat H_0$ and $\hat H_1$ evolve over the course of the cooling protocol, for three different system sizes $N_S = 10,\ 20,\ 40$. We observe that the system energy decays quickly toward the ground state energy of $\hat H_1$. (Note that due to the sweeping of $\theta$ the decay rate varies in time, as opposed to the clean exponential decay in the integrable case.) The expectation value of the zero-order Hamiltonian $\hat H_0$ saturates at a higher value than that of $\hat H_1$, indicating that the steady-state is well-approximated by the ground state of $\hat H_1$, as expected. Cooling efficiency appears to slightly increase with larger system sizes at later times, but larger system studies are required to provide a conclusive answer. In Fig.~\ref{fig:NI}b, we observe similar behaviour in the decay of the log-fidelity per qubit, further corroborating the ability of MCP to cool the system to the ground state of an effective prethermal Hamiltonian. 

\subsection{Antiferromagnetic Heisenberg ladder}

As a second example, we consider a two-leg ladder geometry of $N_S = 2\times L$ sites, with antiferromagnetic Heisenberg couplings between neighbouring sites within each chain and across the rungs~\cite{barnes1993excitation}, of strength $J$ and $J_\perp$ respectively. During the Floquet evolution, Heisenberg interactions are first applied to all odd-numbered bonds along the long (chain) axis of the ladder, followed by the even-numbered bonds, and finally across the rungs:

\begin{gather}
    \hat U_S^{\text{lad}} = e^{-i\hat G_\perp} e^{-i\hat G_\text{even}}e^{-i\hat G_\text{odd}}, \\ 
     \hat G_{\perp} = \frac{\pi J_\perp}{2}\sum_{\alpha, i=1}^{L}\hat S^{\alpha}_{1,i}\hat S^{\alpha}_{2,i},\;\ \hat G_{\text{even}} = \frac{\pi J}{2}\sum_{\sigma,\alpha, i=1}^{L/2-1} \hat S^{\alpha}_{\sigma,2i}\hat S^{\alpha}_{\sigma,2i+1},\nonumber \\
    \hat G_{\text{odd}} = \frac{\pi J}{2}\sum_{\sigma,\alpha, i=1}^{L/2} \hat S^{\alpha}_{\sigma,2i-1}\hat S^{\alpha}_{\sigma,2i},\nonumber
\end{gather}
where $\alpha = x,y,z$ labels the spin component, $i$ runs along the long axis of the ladder, $\sigma=1,2$ labels the two sites belonging to each rung, and $\hat S^\alpha_{\sigma,i}$ are the spin-1/2 operators. Note the convention $\hat S^x = \hat X/2$ etc, and we have assumed that $L$ is even. 

The non-perturbative regime $J\approx J_\perp$ is relevant to experiments \cite{dagotto1996surprises}. In this parameter regime, the system is gapped and adiabatically connected to the point $J/J_\perp \approx 0$. The physics can be easily understood in this limit, where the ladder decouples into isolated antiferromagnetic rungs, each hosting an $S=0$ singlet-paired ground state with energy $-3J_\perp/4$, and a triplet of $S=1$ excitations with energy $J_\perp/4$ \cite{barnes1993excitation}. The corresponding quasiparticle creation operators  (with annihilation operators identical due to Hermiticity) are specified by $\hat Q_k^\alpha = \frac{1}{\sqrt{N_S}} \sum_j e^{ikj}\hat Q_j^\alpha$, where 
\begin{equation}\label{eq:ladderQP}
    \hat Q^\alpha_i = \hat S^\alpha_{1,i}-\hat S^\alpha_{2,i},
\end{equation}
and we have chosen the operators to anticommute with the parity operator exchanging the rung index $\sigma$. At lowest order in $J/J_\perp$, the quasiparticle band develops a finite dispersion as the excitations acquire a kinetic component along the ladder. 

The adiabatic continuity argument suggests that the operators in Eq.~(\ref{eq:ladderQP}) will continue to couple effectively to quasiparticles even in the regime $J/J_\perp\sim 1$. We therefore construct the cooling operators for the protocol to match the strong-coupling quasiparticle operators, taking the system-bath coupling as $\hat V_j^\alpha = \pi\hat Q^\alpha_{\mathcal{Q}_j} (\hat \sigma^-_{\mathcal{A}_j}+\sigma^+_{\mathcal{A}_j})$, with $\mathcal{Q}_j$ ($\mathcal{A}_j$) labelling the system (auxiliary) position associated with $j$-th coupling. Each auxiliary qubit couples with a single choice of $\alpha$ during each cooling cycle (in the numerics we randomly assign the couplings at the start of each cycle). 

We test the effectiveness of the above cooling protocol, using the MCP with a finite density of auxiliary qubits. For the simulations we employ the quantum trajectory method, using Google's qsim simulator \cite{isakov2021simulations}. Each trajectory represents an exact evolution of the system's wavefunction, conditioned on a stochastic sequence of events associated with the auxiliary reset. The initial state for each trajectory is a random computational basis state, uniformly sampled from an infinite-temperature ensemble.  Similarly to the longitudinal-field TFIM, we measure the system energy with respect to the first two orders of the BCH expansion. We also track the system's average magnetization along the $\hat z$-axis, $\langle \hat S_i^z \rangle$, which is expected to be zero in the ground state due to the spin-symmetry of the Heisenberg terms. 

\begin{figure}[t]
    \centering
    \includegraphics[width=0.85\columnwidth]{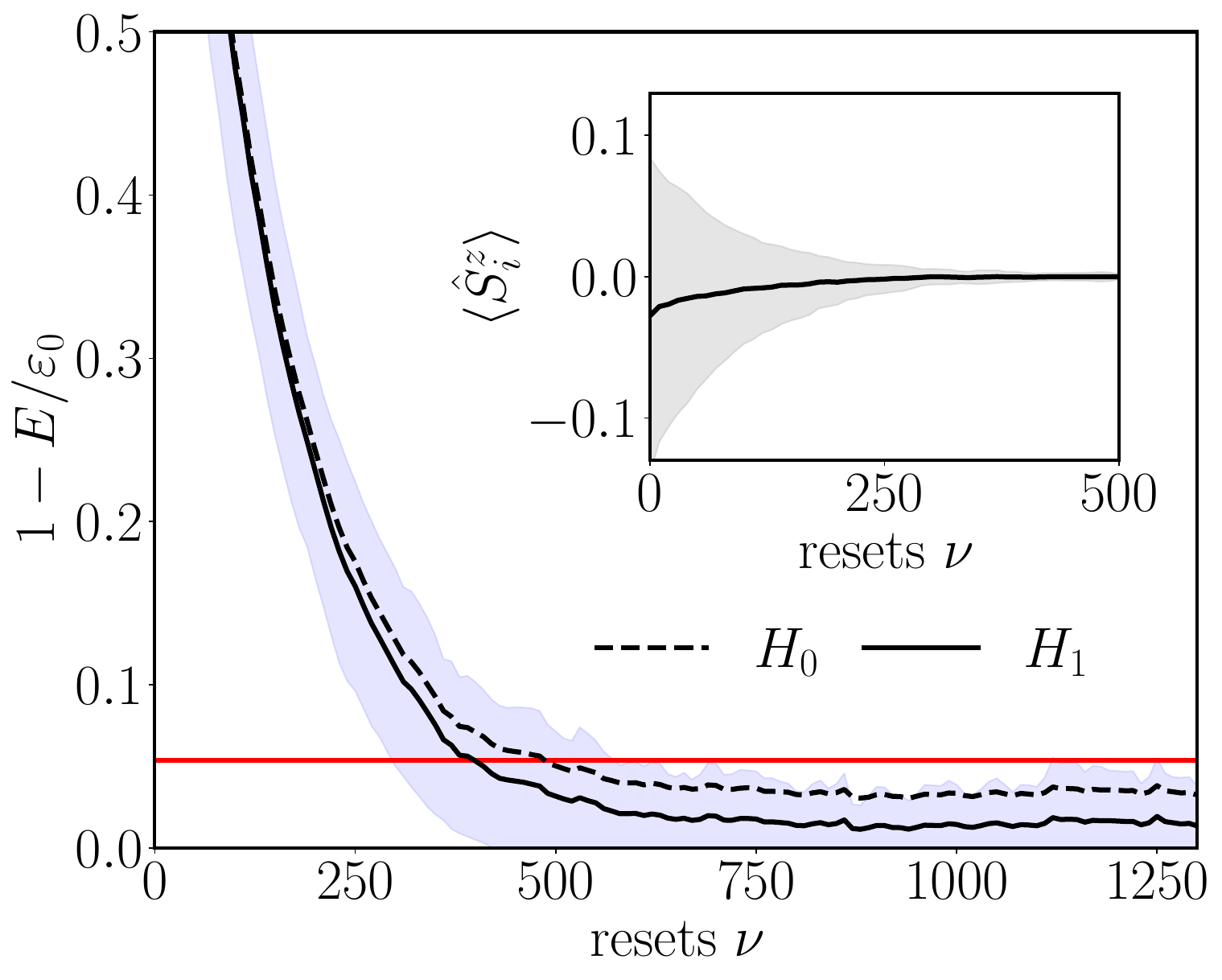}
    \caption{Cooling of a Heisenberg spin ladder. The mean energy for approximate effective Hamiltonians $H_0$ (dashed) and $H_1$ (solid) is shown, relative to the ground state energy, vs.~the number of resets $\nu$. The average is calculated over 100 trajectories, with the blue shaded area showing the standard deviation of the $H_1$ data. The excitation gap of $H_1$ is marked by the red line. \emph{Inset}: Evolution of average magnetisation vs.~number of resets. We use $J = J_\perp = 0.2$, $\theta=0.15$, and the MCP with $T=\beta=30$.}
    \label{fig:Heisenberg}
\end{figure}

The results are shown in Fig.~\ref{fig:Heisenberg}, for a ladder of $2\times 10$ qubits, coupled to 6 auxiliary qubits. The couplings $\mathcal{A}_j (\mathcal{Q}_j)$ are set randomly at the start of each cycle. We take $J = J_\perp$ = 0.2, which is in the non-perturbative regime discussed above. The coupling strength is $\theta = 0.15$ and reset time $T=\beta=30$. For these parameters the excitation gap of $\hat H_1$ is $\Delta = 0.19$ and $\Delta\beta \approx 5.8$. We observe the measured energies converging closely toward the ground state energy, with the final $H_1$ energy being slightly closer to the ground state than for $H_0$. At late times, the magnitude of the energy fluctuations falls below the excitation gap, indicative of a large overlap between the steady state and the system ground state. The measurements of the average magnetisation, displayed in the inset, show that the system approaches a magnetically disordered state on even faster timescales. We checked several other values of $J < J_\perp$ and found that the ground state convergence is robust. In Appendix \ref{app:ladder}, we provide additional data showing that the final energy density and time to reach the steady state is independent of system size, similarly to the TFIM paramagnet, assuming a finite density of auxiliary qubits. This agrees with our expectation that systems adiabatically connected to a fixed point with local excitations, can be cooled in constant time.

Together, the results for the longitudinal-field Ising model and Heisenberg spin ladder show that the MCP remains an effective tool for cooling non-integrable systems.

\section{Noise effects}\label{Sec:Noise}

The results of the previous sections demonstrate the efficiency of  our cooling protocol for preparing correlated ground states, assuming the reset time $T$ can be made large, which is the case in the absence of noise. We now turn to the effects of unwanted decoherence. Each experimental platform comes with its own sources of noise, and our goal will be to understand the limitations it imposes on the cooling efficiency. We will focus on the single-qubit decay and dephasing processes, relevant in systems of superconducting qubits. The kinetic equation can be generalized to include transitions mediated by weak noise. Using this modified kinetic theory, we study the 1d TFIM model and analyze the dependence of the steady state quasiparticle population on the noise strength.

\subsection{Rate equation in the presence of noise}
Considering continuous time evolution, single-qubit decoherence processes can be conveniently described within the Lindblad formalism \cite{breuer2002theory}. The evolution of the system's density matrix $\hat\rho_S$ is given by (for the moment, we ignore the presence of auxiliary qubits): 
\begin{gather}\label{eq:lindblad}
    \frac{d\hat\rho_S}{d t}=-i[\hat H,\hat\rho_S]+ \delta{\mathcal{D}}(\hat\rho_S; \tilde{\gamma}), \\
    \delta{\mathcal{D}}(\hat\rho_S; \tilde{\gamma}) = \sum_{j,\mu} \tilde\gamma_{ j\mu} \left( \hat L_{j \mu} \hat \rho_S  \hat L_{j \mu}^\dagger -\frac{1}{2}\{ \hat L_{j \mu}^\dagger \hat L_{j \mu}, \hat \rho_S  \} \right), \nonumber
\end{gather}
where index $\mu$ runs over possible decoherence processes, $\tilde\gamma_{j\mu}$ are their rates, and $\hat L_{j \mu}$ are the corresponding jump operators. 

In the context of Floquet evolution, we will adopt the following simplified model: we assume that decoherence can be described by a quantum channel $\mathcal{D}$ acting on the system after one period of Floquet evolution. The quantum channel will be constructed by considering only the dissipative part $\delta \tilde{\mathcal{D}}$ of the Lindblad generator in Eq.~(\ref{eq:lindblad}), acting over the Floquet cycle duration $\Delta t$. Experimentally, quantum processors operate in the regime where $\tilde\gamma_{j\mu}\Delta t\ll 1$; then, this dissipative channel can be written as (to leading order in $\tilde\gamma$ coefficients): 
\begin{equation}\label{eq:dissipative_channel1}
\mathcal{D}(\hat\rho_S)\approx \hat\rho_S+\delta\mathcal{D}(\hat\rho_S; \gamma), 
\end{equation}
where we introduced dimensionless decoherence rates:
\begin{equation}\label{eq:dimensinless_gamma}
    \gamma_{j\mu}=\tilde{\gamma}_{j\mu}\Delta t, \;\; \gamma_{j\mu}\ll 1. 
\end{equation}
While this approximate model is only expected to be quantitatively accurate in the Trotter limit $g,J\ll 1$, we expect it to correctly capture the qualitative effects of noise on cooling even away from that limit. The main motivation for introducing this simplified model is that it decouples the unitary evolution of the cooling protocol from the decoherence effects, leading to a transparent and compact form of the noisy kinetic equation. 

Within this model, the evolution of the system-bath density matrix over one cooling period is given by:

\begin{equation}\label{eq:dissipative_evolution}
       \Phi^{\gamma}(\rho_S) = \text{Tr}_B\ \mathcal{D} \mathcal{M}_{T} ...  \mathcal{D} \mathcal{M}_{1}  (\hat\rho_S\otimes \hat\rho_B^0), 
\end{equation}
where 
$$ \mathcal{M}_{\tau} (\cdot)=\hat U_{\theta,\tau} \hat U_{B}\hat U_S (\cdot) \hat U_S^\dagger \hat U^\dagger_B \hat U_{\theta,\tau}^\dagger.
$$
We follow the same steps that led to Eq.~(\ref{eq:rate_equation}) above, now keeping first-order terms in $\theta^2$ and $\gamma_{j\mu}$. The processes from cooling and noise contribute independently, since we discard next-order terms $O(\gamma\theta^2)$. The change in the quasiparticle occupations due to noise takes the standard Lindblad form
\begin{equation}
    \delta \hat n_k  = T\sum_{j,\mu} \gamma_{ j\mu} \left( \hat L^\dagger_{j \mu} \hat n_k  \hat L_{j \mu} -\frac{1}{2}\{ \hat L_{j \mu}^\dagger \hat L_{j \mu}, \hat n_k  \} \right),
\end{equation}
and following Eqs.~(\ref{eq:delta_nk}-\ref{eq:rate_equation}) we obtain the noisy rate equation, which includes cooling transitions $R_{\vec{\alpha}\vec{\beta}}$ and and an additional contribution $R_{\vec{\alpha}\vec{\beta}}^\gamma$ to transition rates from the noise terms:

\begin{align}\label{eq:rate_equation_modified}
        \delta n_k &=\sum_{\vec{\alpha},\vec{\beta}} \rho_{\vec{\alpha}\vec{\alpha}}(\beta_k-\alpha_k) [R_{\vec{\alpha}\vec{\beta}}+R_{\vec{\alpha}\vec{\beta}}^{\gamma}], \\
        R_{\vec{\alpha}\vec{\beta}} &=\theta^2 \sum_{a}|F_{h,T}(\Delta(\vec{\alpha},\vec{\beta}))|^2 |\bra{\vec{\beta}} \hat A^{a}\ket{\vec{\alpha}}|^2, \\
        R_{\vec{\alpha}\vec{\beta}}^{\gamma} &=\sum_{j=1}^{N_S}\sum_{\mu} \gamma_{j\mu} T |\langle \vec{\beta}| \hat L_{j\mu} |\vec{\alpha}\rangle|^2. 
\end{align}

\subsection{Application: cooling the transverse-field Ising model}

\begin{figure}[t]
    \centering
    \includegraphics[width=0.85\columnwidth]{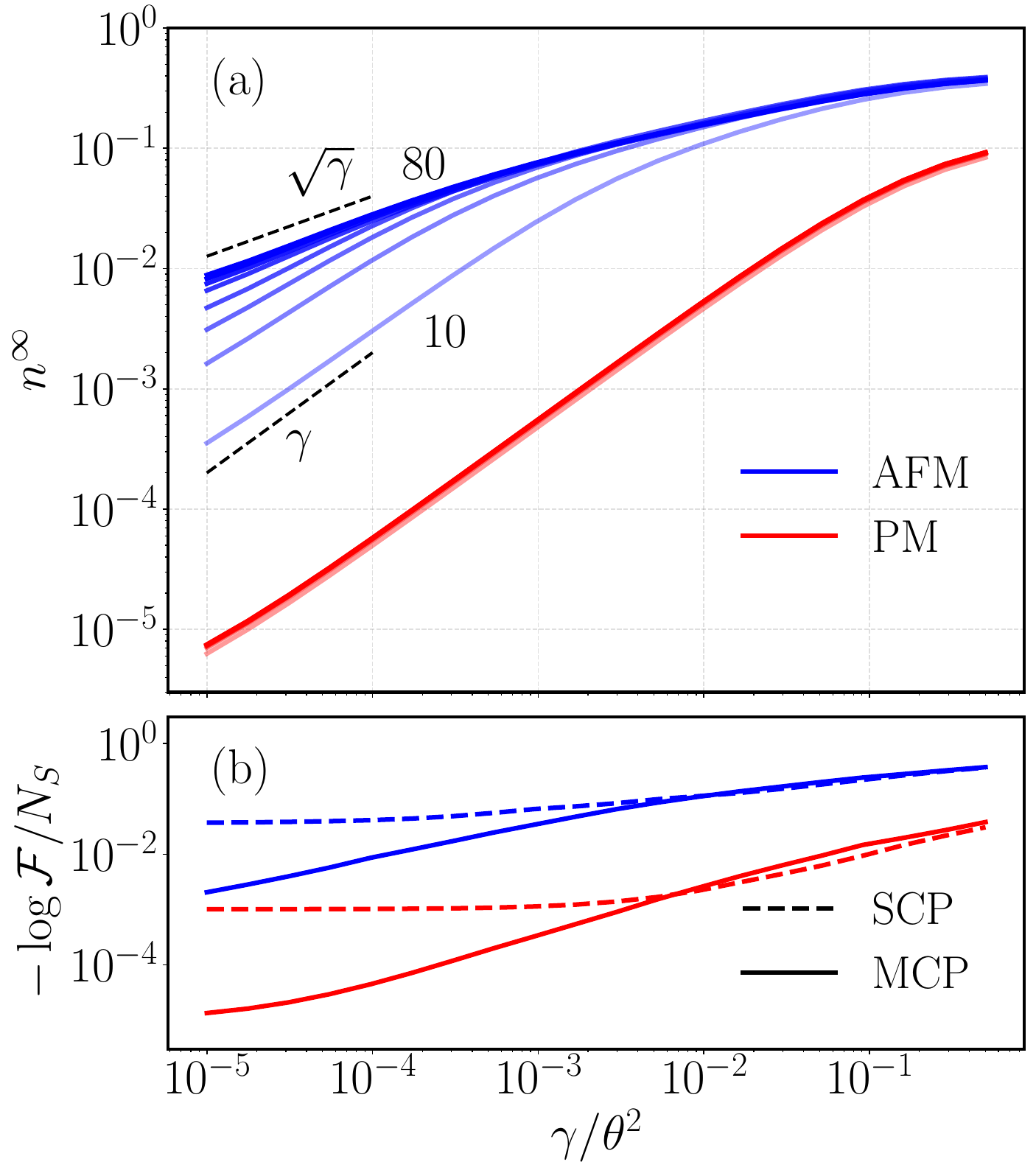}
    \caption{Performance of the cooling algorithm under decoherence. (a) Quasiparticle density in the steady state, as a function of noise ratio $\gamma/\theta^2$, in PM (red) and AFM (blue) phases, for different system sizes $N_S$ from 10 to 80. We use MCP with fixed $T=\beta = 6\pi/\Delta$. At weak noise, PM quasiparticle density scales linearly with $\gamma$, while AFM displays a crossover from $n\propto\gamma$ to $n\propto\sqrt{\gamma}$ behaviour as noise strength is reduced. (b) Log-fidelity between steady state and quasiparticle vacuum, for MCP (solid lines) and SCP (dashed lines). We fix $T=\beta$ and optimize over parameter $\beta$ for each protocol.}
    \label{fig5}
\end{figure}

Next, we apply the noisy kinetic equation to cooling of the transverse-field Ising model of Section \ref{Sec:CoolingSpinChain} in the presence of decay to the qubit ground state, and dephasing noise. This corresponds  to $\mu=d, \phi$, with jump operators 
\begin{equation}\label{eq:decoherencejump}
    \hat L_{j d}=\hat \sigma_j^+, \hspace{.5cm} \hat L_{j\phi}=\hat{Z}_j. 
\end{equation}
We assume homogeneous decoherence rates for simplicity, $\gamma_{j\phi} = \gamma_\phi$, $\gamma_{jd} = \gamma_d$, and rescale the dephasing rate, $\gamma_\phi\to \gamma_\phi/2$, following standard conventions in the literature on open quantum system. Non-uniform decoherence rates can be further included in the rate equation without complication (we refer to \cite{mi2022noise} for a characterisation of site-dependent noise rates for the Google Quantum Processor).

Similar to our analysis in Sec.~\ref{Sec:CoolingSpinChain}, we restrict to processes involving at most two quasiparticle levels. (Processes involving a higher number of quasiparticles can be systematically accounted for within our formalism.) This approximation yields the same kinetic equation derived in Sec.~\ref{Sec:CoolingSpinChain}, Eq.~(\ref{eq:bulk_rate_equation}), but with the rates modified by noise, $W \to \tilde{W}$, $V\to \tilde{V}$:
\begin{align}
    &\tilde{W}_k^{\mp} = {W}_k^{\mp} + \sum_{j=1}^{N_S} \gamma_{d} T |\sigma^\pm_{j;0;k}|^2+\frac{\gamma_{\phi}T}{2} |Z_{j;0;k}|^2, \label{eq:Wtilde1} \\ 
    &\tilde{W}_{k,q}^{\mp} = W_{k,q}^{\mp}+
    \sum_{j=1}^{N_S} \gamma_{d} T |\sigma^\pm_{j;0;k,q}|^2 + \frac{\gamma_{\phi} T}{2}  |Z_{j;0;k,q}|^2,\label{eq:Wtilde2} \\  
    &\tilde{V}_{k,q}^{\mp} =V_{k,q}^\mp + \sum_{j=1}^{N_S} \gamma_{d}T |\sigma^\pm_{j;q;k}|^2 + \frac{\gamma_{\phi}T}{2} |Z_{j;q;k}|^2. \label{eq:Vtilde} 
\end{align}
The matrix elements satisfy the selection rules discussed in Sec.~\ref{Sec:CoolingSpinChain}. Note that while the matrix elements for cooling and decay processes are identical here, the spectral filtering provided by the filter function distinguishes the two processes. We now summarise the main behaviour of the noisy rate equation in the paramagnetic and antiferromagnetic phases of the TFIM, with a more detailed discussion of matrix elements for $\hat Z_j$ and $\hat \sigma^\pm_j$ operators given in Appendix \ref{app:matrixelements}. In the PM phase, single particle loss and gain processes occur throughout the chain (cooling and decay contributions); quasiparticle pair creation/annihilation and number-conserving scattering events occur due to dephasing. In the AFM phase, on account of the domain-wall nature of excitations, only two-particle scattering and pair creation/annihilation processes can occur in the bulk, and they are induced by cooling, decay and dephasing. Single particle transitions require an interaction with the Majorana edge mode, and are localized to the chain edge.

We have compared the predictions of the noisy kinetic equation and MPS simulations in the presence of noise, finding again a good agreement in the parameter regimes accessible to MPS (moderate value of noise strength, or moderate times at a very weak noise). We refer the reader to Appendix \ref{app:noise} for a comparison of the steady state occupations, also for different coupling strengths. The rate equation, however, allows us to study system sizes and parameter regimes far beyond the reach of matrix product state simulations~\footnote{The steady state can be efficiently found by root-finding algorithms, after the cooling and noise-induced rates have been calculated via Wick's theorem. In practice, the system reaches its steady state quickly and we evolve the rate equation in time until the quasiparticle occupations have effectively converged.}.

In Fig.~\ref{fig5}a, we illustrate the quasiparticle density in the steady state, obtained from the kinetic equation, as a function of the noise strength. For simplicity, we choose $\gamma_{d}=\gamma_{\phi}=\gamma$ and fix auxiliary density $N_A = N_S$. We focus on the MCP with $T=\beta = 6\pi/\Delta$. We observe a qualitatively different behavior in the two phases, in line with above results on noiseless cooling. In the PM phase, the steady state quasiparticle density decays as a power-law function of noise, $n^\infty_{\rm PM}\propto \gamma^a$, $a\approx 1$, and is independent of the system size for $N_S \gtrsim 10$. In the AFM phase, in contrast, the density displays a power-law scaling  which depends on system size, $n^{\infty}_{\rm AFM} \propto \gamma^{b(N_S)}$. For small systems and weak noise, we observe similar scaling to the PM, with  $b\approx 1$; as either the system size or noise strength is increased, we find a crossover to a different power $b\approx 1/2$. 

The utility of the kinetic theory is demonstrated with the following simplified versions of the kinetic equations, which capture the observed features in Fig.~\ref{fig5}a:
In the PM phase, where single-quasiparticle processes can occur throughout the system (bulk and edge),  the evolution of the occupation number $n_k(t)$ is given by
\begin{equation}\label{eq:dn/dt-PM}
    \frac{dn_k(t)}{dt}\approx  -C_1 \theta^2 n_k+C_\gamma \gamma T, 
\end{equation}
where $C_1, C_\gamma$ are constants of order one. We assumed that MCP is in the regime where heating processes due to coupling to auxiliaries are strongly suppressed, and neglected the dependence of $C_1, C_\gamma$ on $k$. For the noise we neglect terms $O(\gamma T n)$, since we are interested in the regime of weak noise and low density. Then, the expected steady state quasiparticle density is given by
\begin{equation}\label{eq:SS_PM}
    n^{\infty}_{\rm PM}\approx \frac{C_\gamma \gamma T}{C_1 \theta^2}, 
\end{equation}
where we averaged over $k$, which gives the linear dependence on $\gamma T/\theta^2$ in Fig.~\ref{fig5}a, for weak noise.

In the AFM phase, as discussed above, the single-quasiparticle cooling processes can only occur at the edge, and their rates are suppressed by 
\begin{equation}
    \theta^2 \Gamma_{k} = \theta^2 C_e |u_{1,k}|^2 \approx \theta^2 C_e \frac{k^2}{N_S}
\end{equation} (strictly speaking, this expression holds for small quasimomenta, $k\ll 1$, but we will see below that the steady state quasiparticle density is determined mostly by this quasimomentum range). The processes removing quasiparticle pairs can be modeled by a term $-C_2\theta^2 n_k n$, neglecting the $k,q$ dependence of the rates $W_{k,q}^\mp$. The dominant heating mechanism is from pair creation in the bulk. Then,  we arrive at  
\begin{gather}\label{eq:dn/dt-AFM}
    \frac{dn_k(t)}{dt}\approx -\theta^2 \Gamma_{k} n_k - C_2\theta^2 n_k n \nonumber \\  + C_\gamma' \gamma T (1-n_k) (1-n).  
\end{gather}
The steady state occupation numbers are given by
\begin{equation}\label{eq:nk_SS_AFM}
    n_k^{\infty}\approx \frac{C_\gamma' \gamma T}{\theta^2\Gamma_{k}+C_2 \theta^2 n^{\infty}_{\rm AFM} +C_\gamma'\gamma T},
\end{equation}
supplemented by a self-consistency condition
\begin{equation}\label{eq:self-consistency}
    n^{\infty}_{\rm AFM} =N^{-1}_S\sum_{m=1}^{N_S} n^\infty_{k_m},  
\end{equation}
where we again neglected terms $O(\gamma T n)$ for the regime of interest. The solution $n^{\infty}_{\rm AFM}(\gamma)$ exhibits two regimes, depending on the noise strength. At stronger noise values, where $n^{\infty}N_S\gg 1$ (such that the total number of quasiparticles in the steady state of the system is much greater than 1), two-particle cooling processes dominate. Neglecting the sub-dominant single-particle cooling contribution $\Gamma_k$ yields
\begin{equation}\label{eq:strong_noiseAFM}
    n^{\infty}_{\rm AFM}\approx \left( \frac{C_\gamma'\gamma T}{C_2\theta^2} \right)^{1/2}. 
\end{equation}
In the opposite limit, $n^{\infty}N_S\ll 1$, the single-particle cooling processes determine the steady state and we can neglect the $C_2$ term in Eq.~(\ref{eq:nk_SS_AFM}). Then, 
\begin{gather}\label{eq:weak_noise_AFM}
    n^{\infty}_{\rm AFM}\approx \sum _{m=1}^{N_S} 
    \frac{C_\gamma' \gamma T}{C_e\theta^2 m^2 N_S^{-2} + C_\gamma'\gamma TN_S}\approx C \frac{ \gamma T N_S^2}{\theta^2},
\end{gather}
with $C = \pi^2 C_\gamma'/ (6 C_e)$. This expression holds in the weak-noise limit, $n^{\infty} N_S \sim {\gamma TN_S^3}/{\theta^2}\ll 1$, and we made use of the approximate quantization condition $k_m \approx \pi(1-m)/N_S$, which holds for $k_m \ll \pi$. These approximate rate equations, Eqs.~(\ref{eq:strong_noiseAFM},\ref{eq:weak_noise_AFM}), capture the crossover behaviour displayed in Fig.~\ref{fig5}a. 

In Sec.~\ref{Sec:CoolingSpinChain} we argued that the fidelity of the steady state with the ground state approaches 1 as the MCP parameter $T \propto \beta$ is made large. In the presence of noise, there will be an optimal $T_*$ and $\beta_*$ that maximise the fidelity, since the noise strength also scales with $T$. We compare the log-fidelity per qubit for the MCP and the SCP vs.~the noise ratio $\gamma/\theta^2$, in Fig.~\ref{fig5}b. For each point, we optimize the parameter $\beta$, fixing $T=\beta$ for simplicity (for the SCP this simply corresponds to optimizing $T$). We observe that, while at moderate noise values $\gamma/\theta^2 \gtrsim 10^{-2}$, the SCP marginally outperforms MCP, for weak noise the MCP reaches higher fidelities in both phases, approaching fidelity 1 as the noiseless limit is approached. We conclude that the MCP is able to prepare target states with high fidelity even in the presence of weak noise, but noise is more destructive in systems with topological excitations such as the 1d AFM example.

\section{Discussion and outlook}\label{sec:conclusions}


In summary, in this work we investigated cooling protocols for preparing ground states of many-body systems on quantum processors. Our approach is reminiscent of cooling strategies in systems of ultracold atoms, including sympathetic cooling~\cite{Mogudno2001Science}, but our protocol primarily aims to remove quasiparticles -- emergent excitations above a many-body ground state. We developed a phenomenological kinetic theory framework, which accounts for competing cooling and heating processes induced by auxiliaries and by noise, discussed how to tailor the protocol parameters to remove quasiparticles from the system, and compared the efficiency of different cooling protocols. 

Our results are directly applicable to studying correlated states with current digital quantum processors, as well as in analog processors with the fast reset capability: indeed, SCP was recently implemented on the Google Quantum Processor~\cite{mi2024stable}. While reaching the ground state in that experiment was challenging due to the level of noise $\gamma\approx 0.01$, our work shows that, with e.g.~$\theta=0.1$, noise values of $\gamma\approx 10^{-3}$ will be sufficient for preparing steady states with quasiparticle density below $n\approx 10^{-2}$ in the paramagnetic phase (see Fig.~\ref{fig5}b). In systems of $N_S=100$ qubits, this translates to approximately one quasiparticle in the sample. Cooling the antiferromagnetic phase is more challenging, because the quasiparticles cannot be removed individually by local operators; nevertheless, noise level $\gamma\approx 10^{-6}$ allows to reach comparable densities. We expect that such low-energy states can be further efficiently purified to extract properties of the ground state and low-energy dynamics~\cite{mi2024stable}. 

The quasiparticle cooling algorithm is a promising alternative to existing state-preparation methods, including variational unitary circuits~\cite{cerezo2021variational, TILLY2022PR, fedorov2022vqe}, thanks to a degree of robustness with respect to noise it offers, as well as its generality. First, dissipation removes excitations created by noise, and the steady state is independent of the initial state and the history of errors. 
Second, our results for the non-integrable models studied here suggest that dissipative cooling should be suitable for preparing a broad class of correlated states, not restricted to integrable systems. Further, we note that the dissipative protocol can be adapted to complement existing state-preparation methods, for example by using the output of variational algorithms for the protocol initial state, or to aid in avoiding local minima in the variational landscape \cite{chen2024local}. 

One area where we expect our cooling protocol to prove instrumental is in the study of 2d correlated systems: the ground states of such systems are challenging to probe numerically, but should be amenable to state preparation on quantum devices. The study of the Heisenberg ladder (a quasi-1d system) carried out in Section \ref{Sec:Non-Integrable} suggests that the cooling protocol will remain effective also in 2d, in particular for the cases where the system's excitations can be adiabatically connected to local operators. It will be interesting to understand the advantages or limitations of dissipative cooling for these systems. 

As pointed out in Ref.~\cite{RudnerCoolingArxiv2022}, and further highlighted by our analysis, systems with topological, non-local quasiparticles are more challenging to cool. On the fundamental level, this stems from lower bounds~\cite{bravyi2006lieb, hastings2011topological} on the preparation times of topologically ordered states (e.g., toric code). Exploring strategies of optimal cooling of such systems, for example by creating quasiparticle traps or making use of dynamical gauge fields \cite{kishony2023gauged}, is a promising direction of future research.

The quasiparticle theory typically breaks down when crossing a quantum phase transition. It would be instructive to apply our analysis to quantum-critical points with gapless quasiparticles, and to study the possibility of cooling strongly interacting systems without long-lived quasiparticles. Although local operators are no longer expected to couple strongly to single-particle excitations, we expect that our cooling protocol should drive the system into a low energy steady state, as long as heating processes are efficiently suppressed (e.g.~by the filter function).
Generally, we may expect~\cite{chen2021fast} that algorithms similar to ours will be able to efficiently prepare low energy states of \emph{thermalizing systems} (which rules out examples such as spin glasses, where the cooling procedure will drive the system to one of the many metastable minima). We note that, despite the generality of this approach, we expect certain states will yet prove hard to prepare. One reason is that preparing ground states with high fidelity in general would allow one to solve QMA-complete problems~\cite{kempe2006complexity}, which would contradict a widely believed conjecture that QMA-complete problems cannot be solved efficiently on quantum computers. Still, we do not expect the extreme hard cases to arise in the absence of spin-glass order.

Our work suggests several other avenues for future research. First, we note that our protocol may be modified to prepare approximate finite-temperature Gibbs states~\cite{TehralDiVincenzoPRA2000,chen2021fast,shtanko2023preparing,chen2023quantum}, by using the MCP filter function, Eq.~(\ref{eq:ftau_step}), with effective temperature $T_{\rm eff}=1/\beta$. This is due to the fact that the MCP already satisfies the conditions for detailed balance, Eq.~(\ref{eq:detailedbalance}). We demonstrate cooling of finite temperature states in Appendix \ref{app:thermal}, leaving a more detailed investigation for future works. Weak noise processes in quantum processors generally violate detailed balance, and will make the steady state slightly non-thermal. In contrast, intrinsic thermalization mechanisms in interacting systems are expected to favor the thermal state. This competition, and the resulting steady state may be studied using digital~\cite{mi2024stable} and analog quantum processors. 

Finally, our protocol can be modified to engineer non-equilibrium quasiparticle distribution in the steady state \cite{ulvcakar2024generalized}. The properties of the steady state, and evolution towards it can give useful insights into interactions between quasipaticles, their dynamics, and lifetimes -- properties that play a central role in our understanding of correlated materials. 
We note that the accuracy of the kinetic theory we used to model the cooling protocol suggests that the dominant thermalization mechanisms are well approximated by few-particle processes. 
Applications of suitably modified protocols in quantum simulator experiments would unlock new avenues to study underlying dynamics and thermalization mechanisms.

\acknowledgements 
We thank Zala Lenar\v{c}i\v{c}, Iris Ul\v{c}akar and Guifre Vidal for insightful discussions, and Rolando Somma and Guifre Vidal for useful comments on the manuscript. This work was partially supported by the European Research Council via the grant agreements
TANQ 864597 (A.M., D.A.).

\appendix
\begin{widetext}
    
\end{widetext}
\section{Details of modulated coupling pulse}\label{app:MCP}

In this Appendix, we provide additional details regarding the modulated coupling protocol (MCP) defined in Eq.~(\ref{eq:ftau_step}).
As explained in Section \ref{Sec:protocol}, this pulse is advantageous in that the corresponding filter function,  $F^{\rm MCP}_{h,T}(\epsilon)$, approximately satisfies the thermal detailed balance condition, Eq.~(\ref{eq:detailedbalance}), with  the inverse `temperature' set by the parameter $\beta$. First we derive the form of the MCP, starting from the $2\pi$-periodic function given by 
\begin{equation}\label{eq:symmFermi2}
    F_\beta(\epsilon) = \frac{\pi}{2A}\sum_{m=-\infty}^{\infty}  e^{im\pi}\tanh([\epsilon - m\pi]\beta/4).
\end{equation}
The summation limit above should be taken as $\lim_{k\to\infty} \sum_{m=-2k}^{2k+1}$, although for $\beta \gtrsim 1$ is is sufficient to restrict to a small number of terms when considering the inverval $-\pi < \epsilon \leq \pi$. By inserting a resolution of the delta-function we arrive to the following Fourier series
\begin{gather}\label{eq:MCPfourier}
    F_\beta(\epsilon) =  \sum_{m=-\infty}^{\infty} e^{im\epsilon} g(m), \\
    g(m) = \frac{\pi}{2A}\int_{-\infty}^\infty \frac{d\epsilon'}{2\pi} e^{-im\epsilon'}[\tanh(\epsilon'\beta/4) - \tanh([\epsilon'-\pi]\beta/4)]. \nonumber
\end{gather}
\normalsize
The components can be evaluated via residues,
\begin{gather}
    g(m) = \frac{\pi}{iA\beta} \frac{(1-e^{-i\pi m})}{\sinh(2\pi m/\beta)},
\end{gather}
 and then, shifting the summation index to $\tau = m +\tau_0$ ($\tau_0 = T/2$) we have (up to an unimportant phase factor)
\begin{equation}
    F_\beta(\epsilon) =  \frac{2\pi}{A\beta} \sum_{\tau=-\infty}^\infty\frac{\sin [\pi (\tau-\tau_0)/2]}{\sinh[2\pi(\tau-\tau_0)/\beta]}e^{i\tau(\epsilon-\pi/2)}.
\end{equation}
Truncating the series at a finite number of terms $1 \leq \tau \leq T$,  we arrive at the MCP given in Eq.~(\ref{eq:ftau_step}) of the main text. The truncation error is bounded by
\begin{equation}
    |F_\beta - F^{\rm MCP}_{h,T}| \leq 2\sum_{m=\tau_0+1}^\infty|g(m)|  \lesssim A^{-1} e^{-2\pi T/\beta},
\end{equation}
so is small in the limit $\beta \ll T$. By design, in the prethermal regime $|\epsilon| \ll \pi$ we have 
\begin{equation}
    F_\beta(\epsilon) \approx \frac{\pi}{2A}(1+\tanh(\epsilon\beta/4)),
\end{equation}
which satisfies the detailed balance condition Eq.~(\ref{eq:detailedbalance}), and therefore so does the MCP filter function, up to the errors given above. 

\begin{figure}
    \vspace{0.5cm}
    \centering
    \includegraphics[width=0.85\columnwidth]{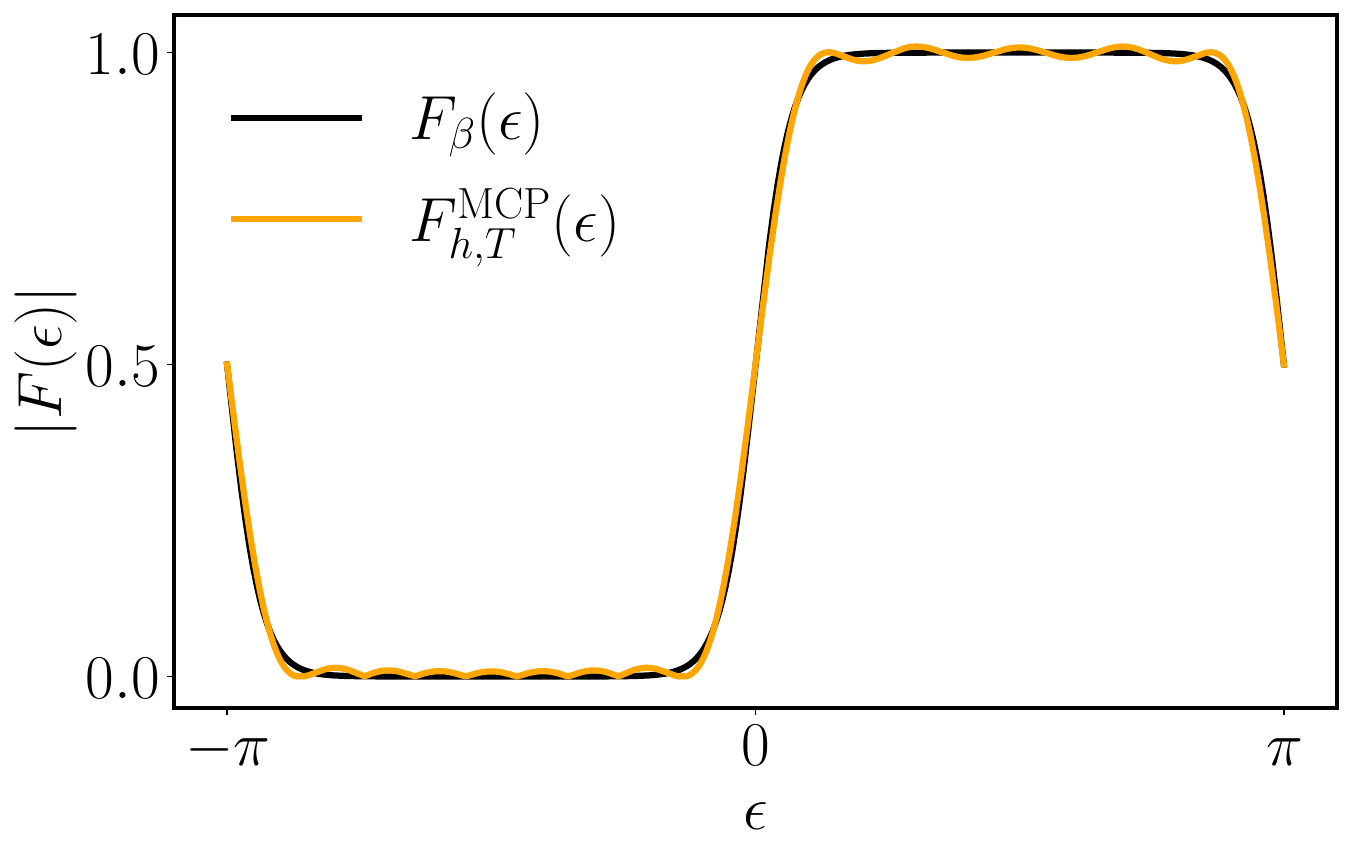}
    \caption{Comparison of $F^{\rm MCP}_{h,T}$ (Eq.~(\ref{eq:ftau_step}), orange) and $F_\beta$ (Eq.~(\ref{eq:symmFermi2}), black), for parameters $T=\beta=20$. }
    \label{fig:MCP}
\end{figure}

In Fig.~\ref{fig:MCP} we compare $F^{\rm MCP}_{h,T}(\epsilon)$ as obtained from the MCP, with the Fermi function Eq.~(\ref{eq:symmFermi2}). (For the latter we truncate the series after a few terms with negligible error.) We take $\beta = 20$ and show the summation result for $T=\beta$. Even for this large value of $\beta/T$, the truncation (`ringing') error is small and $F^{\rm MCP}_{h,T}$ clearly captures the step-like form of the Fermi function. In realistic systems the errors due to noise will typically be larger than the above truncation errors, and we can obtain efficient cooling with values of $T$, $\beta$, considered in the main text.

%
\section{Detailed balance and thermal state preparation with MCP}
\label{app:thermal}
%

\begin{figure}[t!]
    \centering
    \includegraphics[width=0.98\columnwidth]{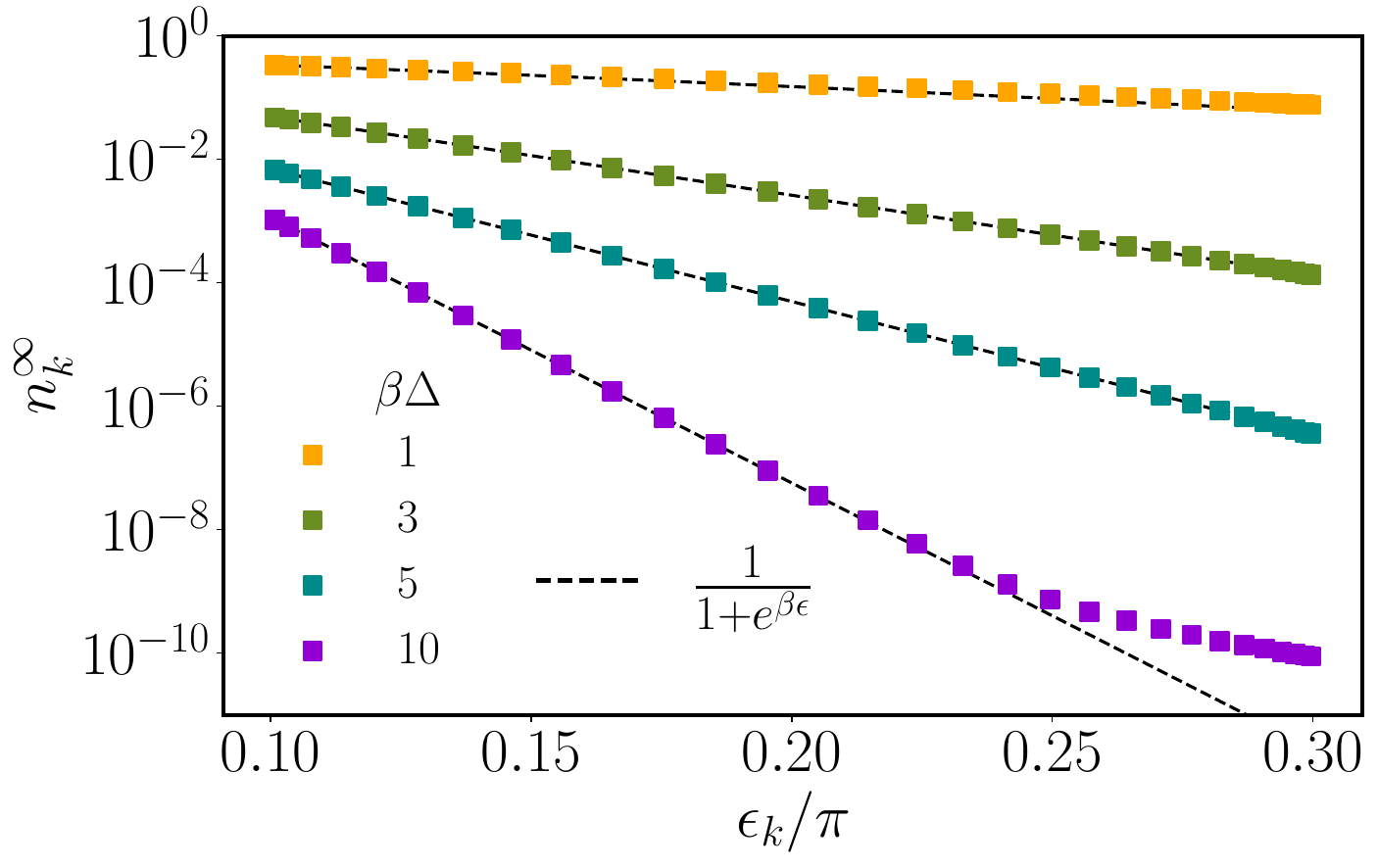}
    \caption{Steady state quasiparticle populations for cooling with thermal MCP, for effective temperature $T_{\rm eff} = 1/\beta$. Results shown are for PM phase with one auxiliary per site, with system size $N_S=30$ and reset time $T=5\beta/2$. The dashed black lines are the expected thermal (Fermi) distributions.}
    \label{fig:thermal}
\end{figure}

Above, we focused on preparation of ground states, which corresponds to the limit $\beta\Delta \to \infty$, where $\Delta$ is the excitation gap. However, our MCP protocol also provides a way to prepare \emph{thermal states} (finite temperature Gibbs states) of the same effective Hamiltonian, a subject which has been the focus of several recent studies ~\cite{TehralDiVincenzoPRA2000,chen2021fast,Rall2023thermalstate,chen2023quantum}. The thermal Gibbs state at an effective temperature $T_{\rm eff}$ is defined by 
\begin{equation}\label{eq:gibbsstate}
    \frac{\rho_{\vec{\alpha}\vec{\alpha}}}{\rho_{\vec{\beta}\vec{\beta}}} = \exp\bigg(-\frac{\varepsilon(\vec{\alpha})-\varepsilon(\vec{\beta})}{T_{\rm eff}}\bigg),
\end{equation}
where $\rho_{\vec{\alpha}\vec{\alpha}}$ are the many-body populations in Eq.~(\ref{eq:GGE}), $\varepsilon$ is the total quasienergy of the many-body eigenstate. 

This can be achieved by cooling the system using the MCP with the parameter $\beta = 1/T_{\rm eff}$, and the additional constraint that the cooling operators $\hat A^a$ appearing in Eq.~(\ref{eq:AB}) are Hermitian. Then, the general rate equation, Eq.~(\ref{eq:rate_equation}), with the Hermitian operators $\hat A^a$, can be written as
\begin{equation}
    \delta n_k = \frac{1}{2}\sum_{\vec{\alpha}\vec{\beta}}(\beta_k - \alpha_k)w_{\vec{\alpha}\vec{\beta}}\bigg(\rho_{\vec{\alpha}\vec{\alpha}}-\rho_{\vec{\beta}\vec{\beta}} \frac{w_{\vec{\beta},\vec{\alpha}}}{w_{\vec{\alpha}\vec{\beta}}}\bigg)
\end{equation}
where 
\begin{equation}
   w_{\vec{\alpha}\vec{\beta}} = \sum_a |F_{h,T}(\Delta(\vec{\alpha}, \vec{\beta}))|^2|\bra{\vec{\beta}} \hat A^a \ket{\vec{\alpha}}|^2,
\end{equation}
and $\Delta(\vec{\alpha}, \vec{\beta}) = \varepsilon(\vec{\alpha})-\varepsilon(\vec{\beta})$ are the Bohr frequencies. If the \emph{detailed balance} condition holds, that is, when the terms inside the bracket above vanish identically,
\begin{equation}
    \frac{\rho_{\vec{\alpha}\vec{\alpha}}}{\rho_{\vec{\beta}\vec{\beta}}} = \frac{|F_{h,T}(\Delta(\vec{\beta}, \vec{\alpha}))|^2\sum_a|\bra{\vec{\alpha}} \hat A^a \ket{\vec{\beta}}|^2}{|F_{h,T}(\Delta(\vec{\alpha}, \vec{\beta}))|^2\sum_b |\bra{\vec{\beta}} \hat A^{b} \ket{\vec{\alpha}}|^2}, 
\end{equation}
then this provides a solution to the steady state $\delta n_k = 0$. With the Hermitian constraint and choice $\beta = 1/T_{\rm eff}$, using Eq.~(\ref{eq:detailedbalance}) for the MCP gives the required Gibbs state, Eq.~(\ref{eq:gibbsstate}).

As a demonstration, we calculate the steady state quasiparticle populations from the rate equation in the PM phase of the Floquet TFIM, with one auxiliary per site and keeping single-particle processes as in Section \ref{Sec:CoolingSpinChain}. We choose hermitian operators $\hat A^a = \hat Z_j$. The results are shown in Fig.~\ref{fig:thermal}, for several values of the inverse temperature $\beta$. We take system size $N_S=30$ and $T=5\beta/2$ to reduce ringing artifacts. The results are compared to the expected Fermi distributions which hold for fermionic quasiparticles  when Eq.~(\ref{eq:gibbsstate}) is satisfied; the agreement is generally excellent, suggesting that thermal Gibbs states can be efficiently prepared by the quasiparticle cooling algorithm (where, as in the ground state case, errors due to noise etc.~will result in slightly non-equilibrium steady state populations).

%
\section{Quasiparticles in the TFIM}
\label{app:TFIM}
%

The Floquet TFIM with open boundary conditions, considered as our cooling example throughout the main text, is defined by the Floquet unitary
\small
\begin{gather}\label{eq:Utfim}
    \hat U_S =  \exp\bigg({-\frac{i\pi J}{2}\sum_{i=1}^{N_S-1} \hat X_i\hat X_{i+1}}\bigg)\exp\bigg(\frac{i\pi g}{2}\sum_{i=1}^{N_S} \hat Z_i\bigg).
\end{gather}
\normalsize
This model can be mapped onto a quadratic fermionic chain by the Jordan-Wigner mapping, Eq.~(\ref{eq:JW}). For this purpose it is convenient to introduce a basis of Hermitian Majorana fermion operators, defined as
\begin{equation}\label{eq:majoranas}
    \hat a_{2i-1} = \hat c_i+\hat c^\dagger_i, \hspace{.5cm} \hat a_{2i} = i(\hat c^\dagger_i-\hat c_i).
\end{equation}
Then the unitary $\hat U_S$ acts on Majorana operators as a linear transformation: 
\begin{equation}\label{eq:eigK}
   \hat U_S^\dagger \hat a_i \hat U_S = \sum_{j=1}^{2N_S} (K_S)_{ij} \hat a_j.
\end{equation}
From Eq.~(\ref{eq:Utfim}) we find the $K_S$ matrix given by
\begin{widetext}
\begin{gather} 
    K_S = 
\begin{pmatrix} 1 & 0 & 0 & \ldots & 0\\
                0 & \cos \pi J & -\sin\pi J  & \ldots &0 \\
                0 & \sin \pi J & \cos\pi J  & \ldots &0\\
                \vdots & \vdots & \vdots & \ddots & 0 \\
                0 & 0 & 0 & 0 & 1
\end{pmatrix}
\begin{pmatrix}
                \cos \pi g & \sin\pi g & 0 & 0 & \ldots \\
                -\sin \pi g & \cos\pi g & 0 & 0 & \ldots \\
                 0 & 0 & \cos \pi g & \sin\pi g & \ldots \\
                0 & 0 & -\sin \pi g & \cos\pi g & \ldots \\
                \vdots & \vdots & \vdots & \vdots & \ddots
\end{pmatrix}.\label{eq:Kmatrix}
\end{gather}
\end{widetext}
The eigenmodes of $\hat U_S$ are Dirac fermions satisfying 
\begin{equation}
    \hat U_S^\dagger \hat \eta_k \hat U_S = e^{-i\epsilon_k} \hat \eta_k,
\end{equation}
where $k$ is a quantum number labeling the eigenmodes (the quasimomentum). Defining the eigenvectors of $K_S$ (for $\epsilon_k \geq 0$) according to
\begin{equation}
    K_S\psi_k = e^{-i\epsilon_k} \psi_k,
\end{equation}
we have the expansion
\begin{equation}\label{eq:eta_expansion}
    \hat\eta_k = \frac{1}{\sqrt{2}}\sum_{j=1}^{N_S} \psi^*_{k,2j-1} \hat a_{2j-1}+\psi^*_{k,2j} \hat a_{2j}. 
\end{equation}
Due to particle-hole symmetry, the eigenvectors of $K_S$ with negative quasienergy are related to those of positive quasienergy by the conjugation $\varphi_k = \psi_{k}^*$, $K_S \varphi_k = e^{i\epsilon_k}\varphi_k$. The eigenvectors $\psi_{k}$ and quasienergies $\epsilon_k$ can be found by exact diagonalisation of the $K_S$ matrix; see e.g.~Ref.~\cite{lerose2021scaling} or Appendix A of Ref.~\cite{mi2024stable} for an analytical derivation. Note that the boundary condition defines the quantization condition for the $N_S$ quasimomenta $k_m$, in terms of which the quasienergies are given by the exact formula 
\begin{equation}\label{eq:quasienergy}
  \cos \epsilon_k= \cos(\pi J) \cos(\pi g)-\sin(\pi J) \sin (\pi g) \cos k. 
\end{equation}

Along with the definition of the Bogoliubov coefficients, 
\begin{equation}
    \hat c_j=\sum_k u_{jk}\hat \eta_k +v_{jk} \hat \eta_k^\dagger,
\end{equation}
this gives the relations
\begin{gather}
    u_{jk} = \frac{\psi_{k,2j-1} + i\psi_{k,2j}}{\sqrt{2}}, \\ 
    v_{jk} = \frac{\psi^*_{k,2j-1} + i\psi^*_{k,2j}}{\sqrt{2}},
\end{gather}
specifying the standing-wave form of the eigenmodes. 

\begin{figure}
    \centering
    \vspace{.2cm}
    \includegraphics[width=0.85\columnwidth]{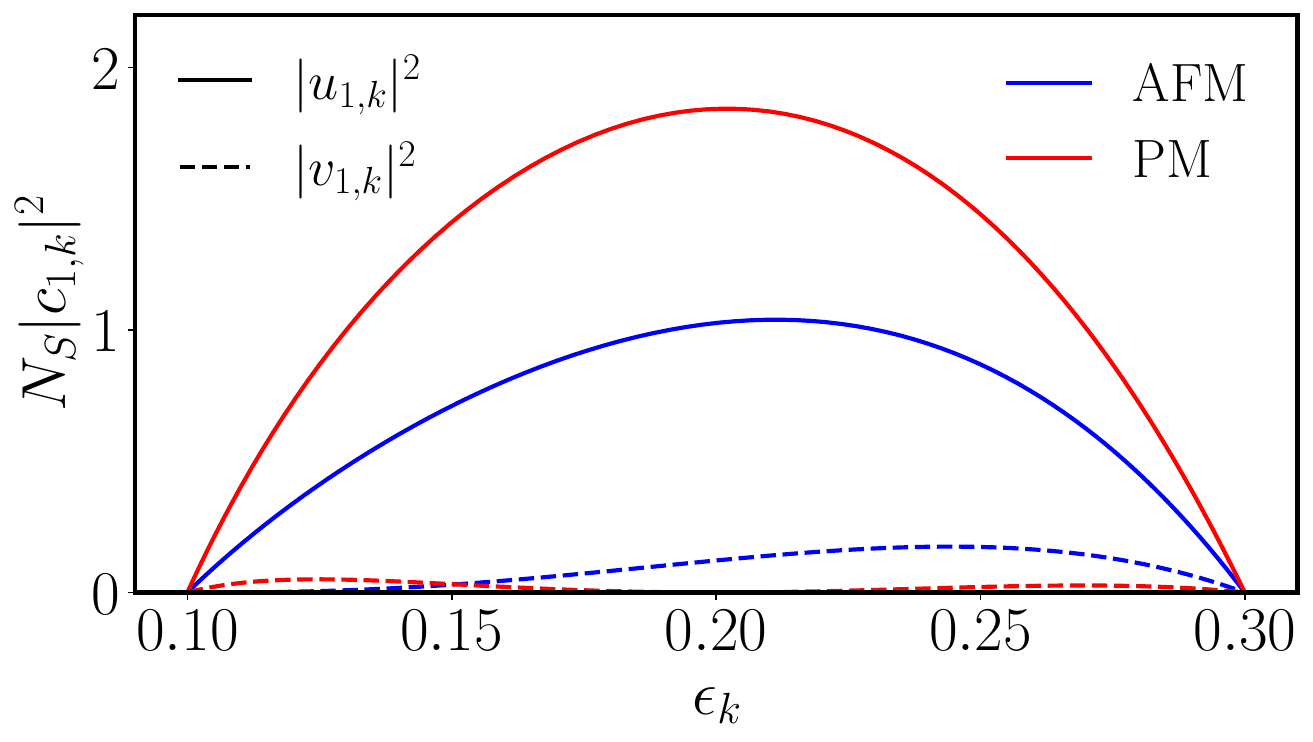}
    \caption{Profile of edge Bogoliubov coefficients $|u_{1k}|^2$, (solid) $|v_{1k}|^2$ (dashed), plotted vs quasienergy $\epsilon_k$. We plot the coefficients rescaled by chain length $N_S$: squared coefficients are $O(1/N_S)$ in middle of quasiparticle band, but vanish as $O(1/N_S^3)$ at band edges.}
    \label{fig:bogo}
\end{figure}

Since the Bogoliubov coefficients at the boundary play a particularly important role in the cooling rates entering the kinetic equation for edge cooling, we plot the values of $|u_{1k}|^2$, $|v_{1k}^2|$ in Fig.~\ref{fig:bogo}, vs.~the quasienergy $\epsilon_k$. We observe that quasiparticle in the middle of the band typically have boundary overlap $|u_{1k}|^2 = O(N_S^{-1})$, while the overlap vanishes for quasiparticles with quasienergy near the two band edges. The overlap for these quasiparticles is expected to go as $O(N_S^{-3})$ \cite{vznidarivc2015relaxation}.

\vspace{.5cm}
\section{Matrix elements for bulk cooling}\label{app:matrixelements}

In Sections \ref{Sec:CoolingSpinChain}, \ref{Sec:Noise}, we considered cooling in the example of the transverse-field Ising model, with two types of matrix elements appearing in the derivation of the rate equations: the $\hat Z_j$ matrix elements, which we denote as
\begin{equation}
    Z_{j;\vec{\alpha};\vec{\beta}} = \bra{\vec{\alpha}} \hat{Z}_j \ket{\vec{\beta}},
\end{equation}
and the $\hat\sigma^\pm_j$ matrix elements,
\begin{equation}
    \sigma^\pm_{j;\vec{\alpha};\vec{\beta}} = \bra{\vec{\alpha}} \hat{\sigma}^\pm_j \ket{\vec{\beta}}.
\end{equation}
Here $\ket{\vec{\alpha}}$ are the many-body eigenstates of the TFIM, which are completely defined by their quasiparticle occupation numbers. 

Although the TFIM defines an integrable model, calculation of spin matrix elements can be challenging \cite{Iorgov_2011}, due to the fact that the Jordan-Wigner transformation between spin and fermion operators is non-local. We will see that this is the case for the $\hat\sigma^\pm_j$ matrix elements below. First, we focus on the simpler case of $\hat Z_j$: 

\subsection{$Z_{j;\vec{\alpha};\vec{\beta}}$ matrix elements}
In the fermion basis $\hat Z$ is a quadratic operator, so only matrix elements between eigenstates differing in the occupation of at most two quasiparticle modes survive:

\begin{widetext}
\begin{gather}
    Z_{j;\vec{\alpha};\vec{\beta}} = \bra{\vec{\alpha}} (1-2\hat c^\dagger_j \hat c_j) \ket{\vec{\beta}} \nonumber \\
    = \delta_{\vec{\alpha}\vec{\beta}} -2\sum_{kq}\Big( u^*_{jk}u_{jq} [\hat\eta_k^\dagger\hat\eta_q]_{\vec{\alpha}\vec{\beta}} +u^*_{jk}v_{jq}[\hat\eta_k^\dagger\hat\eta^\dagger_q]_{\vec{\alpha}\vec{\beta}}
    + v^*_{jk}u_{jq}[\hat\eta_k\hat\eta_q]_{\vec{\alpha}\vec{\beta}}+v^*_{jk}v_{jq}[\hat\eta_k\hat\eta^\dagger_q]_{\vec{\alpha}\vec{\beta}}\Big)
\end{gather}
\end{widetext}
where e.g. $[\hat\eta^\dagger_k \hat\eta_q]_{\vec{\alpha}\vec{\beta}} \equiv \bra{\vec{\alpha}} \hat\eta^\dagger_k \hat\eta_q \ket{\vec{\beta}} = 0,1$ is only non-zero if $\vec{\alpha}$ and $\vec{\beta}$ differ only by the occupations of the $k$ and $q$ quasiparticle levels (including the case $k=q$). Since the Bogoliubov coefficients $u_{jk}$, $v_{jk}$ can be obtained by diagonalisation of the single-particle Hamiltonian, the matrix elements are easily obtained. 

In deriving the rate equation, we focused on the case of transitions to and from the ground state, and scattering between single quasiparticle levels. Thus we have 
\begin{equation}
    |Z_{j;0;k,q}|^2 = |\bra{\Omega} \hat{Z}_j \ket{k,q}|^2 = 4|v_{jk}^*u_{jq}|^2
\end{equation}
as the probability for two quasiparticles with quasimomenta $k,\ q$, to annihilate, and 
\begin{equation}
    |Z_{j; k ;q}|^2 = |\bra{k} \hat{Z}_j \ket{q}|^2 = 4|u_{jk}^*u_{jq}|^2
\end{equation}
as the scattering probability $q \to k$. Since $\hat Z$ is hermitian, the pair-creation probability is equal to the pair-annihilation, i.e.~$|Z_{j;0;k,q}|^2 = |Z_{j;k,q;0}|^2$. Note that, in the AFM phase there is also an amplitude for single-quasiparticle processes, due to the selection rules discusses in the main text i.e.~one of $k$, $q$ can be taken as the Majorana zero mode. 

\subsection{$\sigma^\pm_{j;\alpha;\beta}$ matrix elements}

We now turn to quasiparticle processes mediated by the $\hat\sigma^\pm_j$ matrix elements. We focus on $\hat\sigma^+_j$ for illustration. In the fermion basis, $\hat\sigma^+_j$ comes attached with the Jordan-Wigner string operator,
\begin{equation}
    \sigma^+_{j;\vec{\alpha};\vec{\beta}} = \bra{\vec{\alpha}}\Big( \prod_{i=1}^{j-1} e^{i\pi \hat c^\dagger_i \hat c_i} \Big)\hat c_j \ket{\vec{\beta}}.
\end{equation}
Due to this non-locality in the fermionic eigenbasis, multi-particle processes are allowed with $\hat \sigma_j^+$ changing the occupation numbers of up to $2j-1$ quasiparticle levels. Amazingly, the exact finite-volume matrix elements have been worked out for the case of the $\hat X_j$ operators in the TFIM with periodic boundary conditions \cite{Iorgov_2011}. Here, we focus on singe- and two-particle transitions near the ground state, which capture the dominant cooling processes at late times, as described in the main text. Our theory can be straightforwardly generalised to include higher-order processes.

Focusing on few-particle processes alows us to use Wick's theorem to evaluate the necessary matrix elements for systems up to several hundred sites: For free (more generally, Gaussian) theories, Wick's theorem states that
\begin{gather}
    \langle \hat a_{i_1}\hat a_{i_2}\ldots \hat a_{i_{2N}} \rangle_{\Omega} = \sum_{P} \zeta_P \prod_{k=1}^N \langle \hat a_{P(i_{2k-1})} \hat a_{P(i_{2k})}\rangle \\ 
    = \text{Pf}\ \begin{pmatrix}
        0 & \langle \hat a_{i_1} \hat a_{i_2} \rangle & \langle \hat a_{i_1} \hat a_{i_3} \rangle & \ldots & \langle \hat a_{i_1} \hat a_{i_{2N}} \rangle \\
        {} & 0 & \langle \hat a_{i_2} \hat a_{i_3} \rangle & \ldots & \langle \hat a_{i_2} \hat a_{i_{2N}} \rangle \\ 
        {} & {} & 0 & \ldots &  \langle \hat a_{i_3} \hat a_{i_{2N}} \rangle \\ 
        {} & {} & {} & \ddots & \vdots \\ 
        {} & {} & {} & {} & 0
    \end{pmatrix}.\label{eq:Wicks}
\end{gather}
The sum runs over all possible permutations of the Majorana indices, and $\zeta_P = \pm 1$ is the sign of the permutation, carrying the fermion statistics. All expectation values are taken with respect to the ground state of the free theory, and for notational simplicity we introduced the Majorana fermion operators (Eq.~(\ref{eq:majoranas})). In the second line we introduced the Pfaffian representation of Wick's theorem in terms of a skew-symmetric matrix $\Sigma$. In our case, since only the modulus squared of the correlation functions enters the rate equation, we may use 
\begin{equation}
    |\langle a_{i_1}a_{i_2}\ldots a_{i_{2N}} \rangle_{\Omega}|^2 = |\text{Pf}(\Sigma)|^2 = |\text{det}(\Sigma)|.
\end{equation}
which can be computed in time $O(N^3)$. For further reference, see e.g.~\cite{barouch1971statistical, IsingModelReviewSantoro2020}.

\begin{figure}
    \centering
    \vspace{0.3cm}
    \includegraphics[width=0.9\columnwidth]{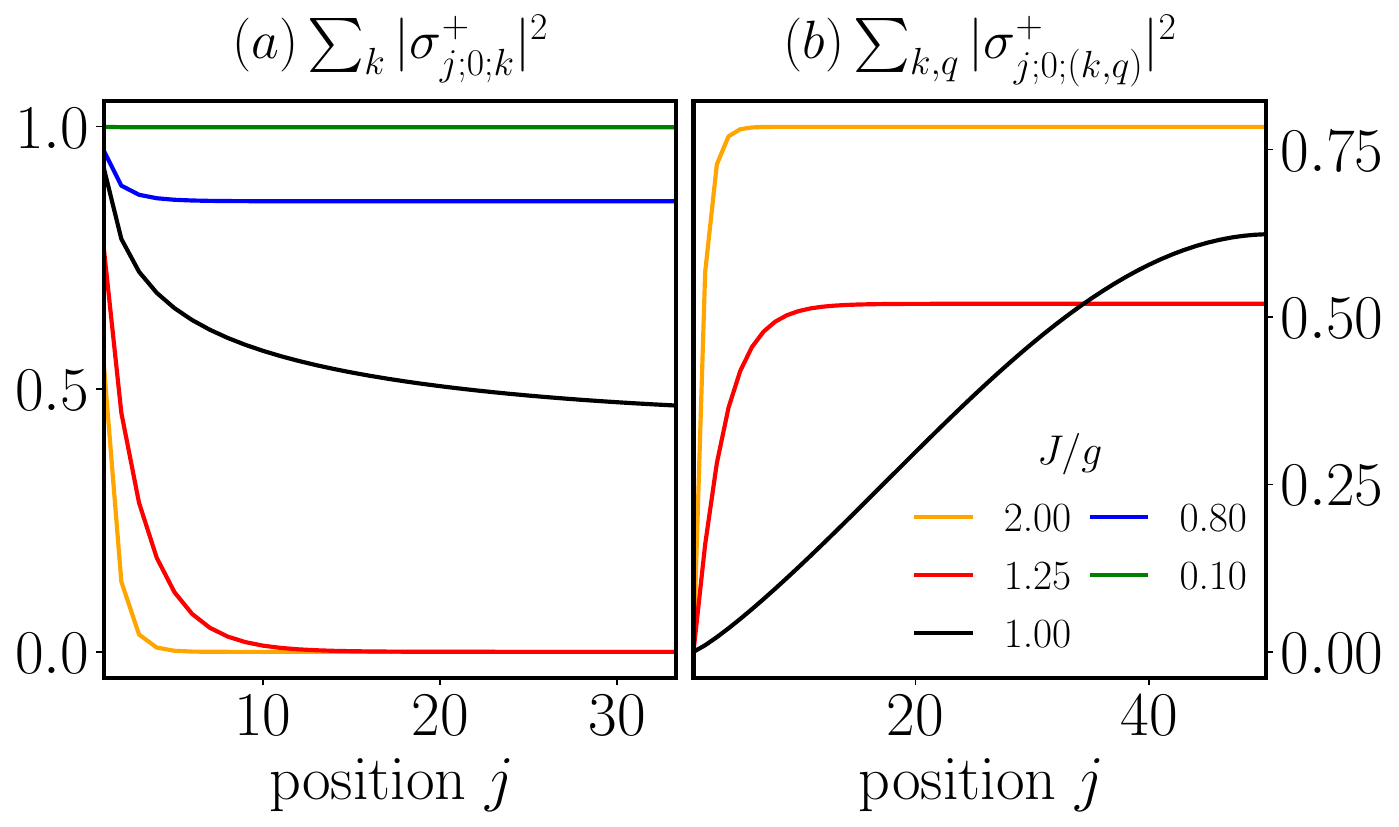}
    \caption{(a) Single-quasiparticle matrix elements $|\sigma^+_{j;0;k}|^2$ summed over quasimomenta $k$ and plotted against operator position $j$, for different values of $J/g$. (b) Two-particle matrix elements $|\sigma^+_{j;0;(k,q)}|^2$ summed over $k$, $q$, and plotted against operator position $j$. Note that the two-particle matrix element is non-zero only in the AFM phase.}
    \label{fig:AD1}
\end{figure}

\vspace{.5cm}
\textit{Single quasiparticle processes:} We first consider single-quasiparticle transitions above the ground state, 
\begin{equation}
    \sigma^+_{j;0;k} = \bra{\Omega}\hat\sigma^+_j \ket{k}
\end{equation}
being the amplitude to directly annihilate a quasiparticle of quasimomentum $k$. We rewrite the single-particle element in fermion basis 
\begin{equation}\label{eq:1P}
    |\sigma^+_{j;0;k}|^2 = |\langle\prod_{i=1}^{j-1} \hat a_{2i-1}\hat a_{2i}\ \hat c_j \hat \eta^\dagger_k \rangle_\Omega |^2.
\end{equation}
The corresponding Wick's matrix $\Sigma$ had dimension $2j\times 2j$. In Fig.~\ref{fig:AD1}a we show how the single-particle matrix elements in  Eq.~\ref{eq:1P} (summed over $k$), depend on the TFIM phase parameter $J/g$ and on the position of the operator $j$. The system size is $N_S = 100$. The matrix elements display a clear change in behaviour between the two phases: in the PM phase, single-particle transitions occur almost uniformly throughout the system bulk; in the AFM phase, transitions are localized within a few correlation lengths of the edge, with a correlation length $\xi_{J} \propto J/g$. This is as expected, based on the physical picture of domain-wall-like quasiparticles in the AFM (removing a single domain wall requires flipping a macroscopic number of spins).

\vspace{.5cm}
\textit{Two-quasiparticle processes:} Based on the physical picture, we expect that in the AFM phase, a pair of domain walls may collide in the bulk and annihilate, as this requires only local reconfigurations of spins. However, there is a subtlety, from the fact that $\hat \sigma^+_j$ has non-zero matrix elements only between eigenstates of different fermion parity. This rules out two-particle transitions in the PM phase, but as discussed in the main text, the issue is sidestepped in the AFM by interacting with the Majorana zero mode, whose action maps between the two nearly-degenerate ground states. We therefore write 
\begin{equation}\label{eq:2P}
    |\sigma^+_{j;0;k,q}|^2 = |\bra{\Omega_+}\prod_{i=1}^{j-1} \hat a_{2i-1}\hat a_{2i}\ \hat c_j \hat \eta^\dagger_k \hat \eta^\dagger_q \ket{\Omega_-}|^2,
\end{equation}
with $\ket{\Omega_-} = \hat\eta^\dagger_0 \ket{\Omega_-}$, and $ \hat\eta^\dagger_0$ the Majorana mode. These matrix elements may be evaluated with Wick's theorem on the $(2j+2)\times (2j+2)$ dimensional $\Sigma$-matrix. In Fig.~\ref{fig:AD1}b, we show how the summed matrix elements $\sum_{k,q}|\sigma^+_{j;0;k,q}|^2$ depend on the position $j$, for different values of $J/g$ in the AFM phase. The pair matrix element is almost homogeneous in the bulk, vanishing in the edge region.

For completeness we also give the two-particle scattering elements, which are again non-zero only for the AFM phase:
\begin{equation}\label{eq:2Pscat}
    |\sigma^+_{j;k;q}|^2 = |\bra{\Omega_+} \hat\eta_k \prod_{i=1}^{j-1} \hat a_{2i-1}\hat a_{2i}\ \hat c_j  \hat \eta^\dagger_q \ket{\Omega_-}|^2.
\end{equation}
The scattering does not directly lead to cooling/heating of the system (the total number of quasiparticles is conserved by the scattering processes), however it can lead to a redistribution of quasiparticle density, which itself affects the cooling rate.

\section{Details of numerical simulations}\label{app:numerics}

\subsection{Free fermion simulations} 

For the edge cooling protocol analysed in Section \ref{Sec:CoolingSpinChain} for the TFIM, the non-unitary dynamics described by Eq.~(\ref{eq:cooling_channel}) defines a Gaussian channel --- a channel mapping Gaussian states to Gaussian states. A Gaussian state obeys Wick's theorem (see Eq.~\ref{eq:Wicks}), meaning that the expectation values of all observables can be factorised purely in terms of $2$-point fermionic correlation functions. Under the action of the cooling channel, Eq.~(\ref{eq:cooling_channel}), the matrix of 2-point functions can be efficiently updated, as we show below, and hence we have access to the arbitrary correlators of the time-dependent many-body state. 

We introduce the antisymmetrised covariance matrix of Majorana correlators as
\begin{equation}
    \Gamma_{ij} = \frac{i}{4} \langle [\hat a_i, \hat a_j] \rangle = \begin{cases}i\langle \hat a_i \hat a_j \rangle/2, & i \neq j \\ 0, & i=j \end{cases},
\end{equation}
where the Majorana fermions were defined in Eq.~(\ref{eq:majoranas}). The matrix includes both system and bath degrees of freedom, and has dimension $(2N_S+2)\times(2N_S+2)$ (we limit our discussion to the case of a single edge auxiliary for simplicity). We block the covariance matrix according to 
\begin{equation}
    \Gamma = \begin{pmatrix}
        \Gamma_{BB} & \Gamma_{SB} \\ \Gamma_{BS} & \Gamma_{SS}
    \end{pmatrix}.
\end{equation}
where $\Gamma_{SS}$ is the $2N_S\times 2N_S$ matrix containing system-system correlations, etc. For the initial condition, representing the system in the maximally mixed state and the auxiliary in the reset state,  Eq.~(\ref{eq:initialbath}), we have
\begin{equation}
    \Gamma_{SS} = 0
\end{equation}
\begin{equation}\label{eq:Gamma1}
    \Gamma_{BB} = \frac{1}{2}\begin{pmatrix}
        0 & -1 \\ 1 & 0
    \end{pmatrix}, \hspace{.5cm} \Gamma_{SB} = \Gamma_{BS} = 0.
\end{equation}

Due to the integrability of the TFIM, unitary evolution of the Majoranas in the Heisenberg picture acts as a linear map:
\begin{equation}
    (\hat U^\dagger_S \hat U^\dagger_B \hat U^\dagger_\theta)\  \hat a_i\ (\hat U_\theta \hat U_B \hat U_S) = \sum_j (K_\theta K_B K_S)_{ij} \hat a_j,
\end{equation}
where $K_S$ is defined in (\ref{eq:Kmatrix}), and $K_B = \exp(\pi h_B)$, $K_\theta = \exp(\pi h_\theta)$, with
\small
\begin{equation}
h_B =
\begin{pmatrix} 0 & h & 0 & \ldots \\
                -h & 0 & 0 & \ldots \\
                0 & 0 & 0 & \ldots \\
               \vdots & \vdots & \vdots & \ddots \end{pmatrix}, \hspace{.3cm}
h_\theta = 
\begin{pmatrix} 0 & 0 & 0 & \theta & \ldots \\
                0 & 0 & -\theta & 0 & \ldots \\
                0 & \theta & 0 & 0 & \ldots \\
                -\theta & 0 & 0 & 0 & \ldots \\
               \vdots & \vdots & \vdots & \vdots & \ddots \end{pmatrix},
\end{equation}
\normalsize
all other elements being zero. Then, defining $\hat U = \hat U_\theta \hat U_B \hat U_S$ and $K = K_\theta K_B K_S$, the covariance matrix evolves under the unitary part of the cooling channel as, 
\begin{equation} \langle \hat U^\dagger\hat \Gamma_{ij}\hat U \rangle = K_{il}K_{jm} \langle [ \hat a_l, \hat a_m] \rangle = (K \Gamma K^T)_{ij}.
\end{equation}

\begin{figure}[t!]
    \centering
    \vspace{.5cm}
    \includegraphics[width=0.85\columnwidth]{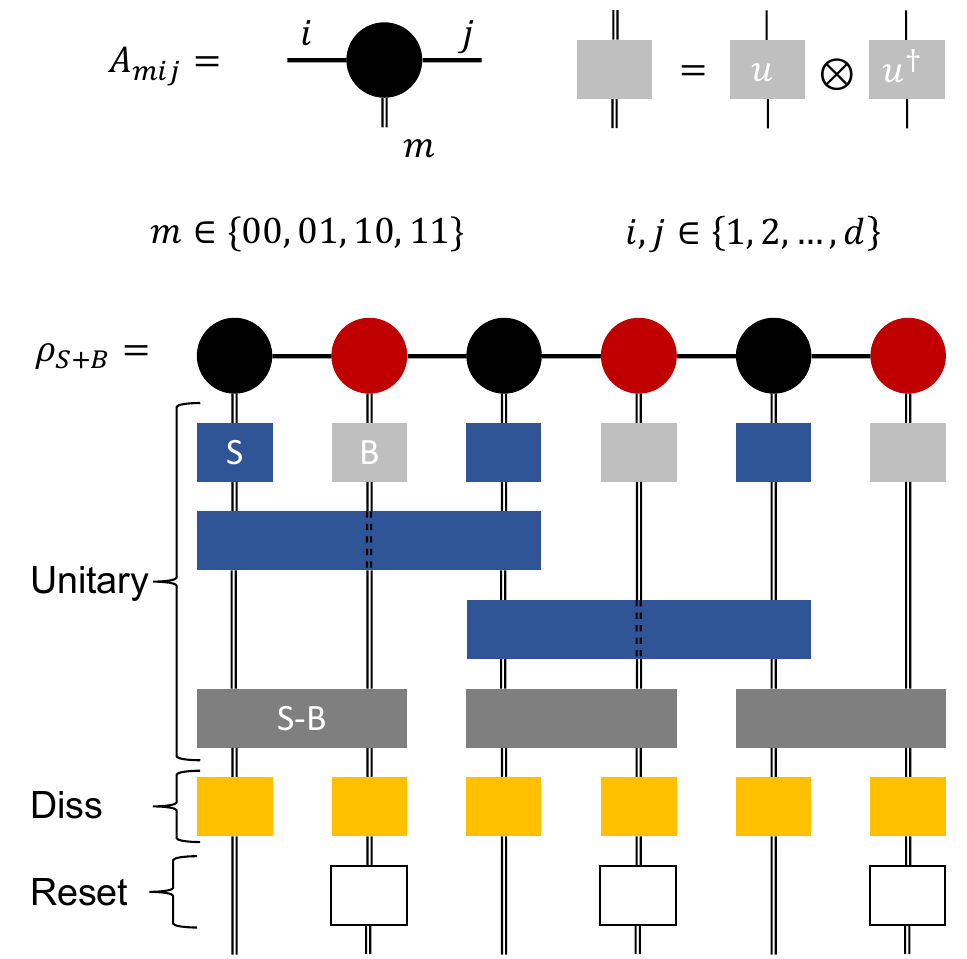}
    \caption{Tensor network implementation for one cooling cycle. System (auxiliary) qubits are denoted by black (red) tensors respectively. Different channels correspond to different blocks: bath evolution (light grey); system evolution (dark blue); system-bath coupling (dark grey); decoherence (yellow); auxiliary reset (white). The network is composed of rank-3 tensors $A$, has a bond dimension $d$ and a 4-dimensional physical bond corresponding to the four (vectorised) states of the qubit's density matrix. The system is encoded in a one-dimensional geometry at the price of introducing next-nearest neighbor channels.}
    \label{figMPS}
\end{figure}

The reset of the auxiliaries can be done by reinstating condition (\ref{eq:Gamma1}). If we denote the corresponding map by $\mathcal{R}$, then the cooling channel (\ref{eq:cooling_channel}) acts on the covariance matrix according to:
\begin{equation}
    [\Phi (\Gamma)]_{ij} = [\mathcal{R}(K\Gamma K^T)]_{ij}.
\end{equation}
Thus the matrix $\Gamma$ can be efficiently updated to obtain the dynamics, or (by vectorising the matrix $\Gamma$) the linear map $\Phi$ diagonalised to yield the steady state. Note that qubit dephasing can also be efficiently implemented in this scheme \cite{barthel2022solving}, but bulk auxiliaries or qubit decay break the necessary integrability.

\subsection{Tensor network simulations}

Here, we summarize the tensor network algorithm that was employed to simulate the dissipative dynamics in Sections \ref{Sec:CoolingSpinChain}-\ref{Sec:Non-Integrable}. We give an overview of the relevant details, assuming the reader is familiar with the basics of matrix product algorithms; introductions can be found in e.g.~\cite{orus2014practical, bridgeman2017hand}.

We assume a finite density of auxiliaries, one per system site. The total density matrix of the system plus auxiliaries is $\rho_{SB}$. We work with the vector unfolding of the density matrix product operator, which is written as a matrix product state (MPS),
\small
\begin{equation}
    \rho_{SB} =\sum_{\textbf{m},\textbf{s}}A^{1}_{m_1,s_1}A^{2}_{m_2,s_1,s_2}\ldots A^{N}_{m_N,s_{N-1}}|m_1\ldots m_N\rangle,
\end{equation}
\normalsize 
where $A^{i}$ is a rank-3 tensor graphically defined in Fig.~\ref{figMPS}. Virtual indices $\textbf{s} = \{s_1,\ldots, s_{N-1}\}$ are traced while physical indices  $\textbf{m} = \{m_1,\ldots, m_N\}$ denote the different states of the system+bath, $N=2N_S$. The system geometry in the presence of auxiliaries is quasi-1d, see Fig.~\ref{fig1}d, and we adopt a 1-dimensional ordering that alternates between system and auxiliary qubits i.e.~$\mathcal{Q}_1\mathcal{A}_1\ldots \mathcal{Q}_{N_S}\mathcal{A}_{N_S}$, as illustrated in Fig.~\ref{figMPS}. The dimension of the virtual indices of the MPS (bond dimension) is denoted $d$, and controls the amount of non-local correlations that can be represented by the state. 

\begin{figure}[t!]
    \centering
    \vspace{.5cm}
    \includegraphics[width=0.9\columnwidth]{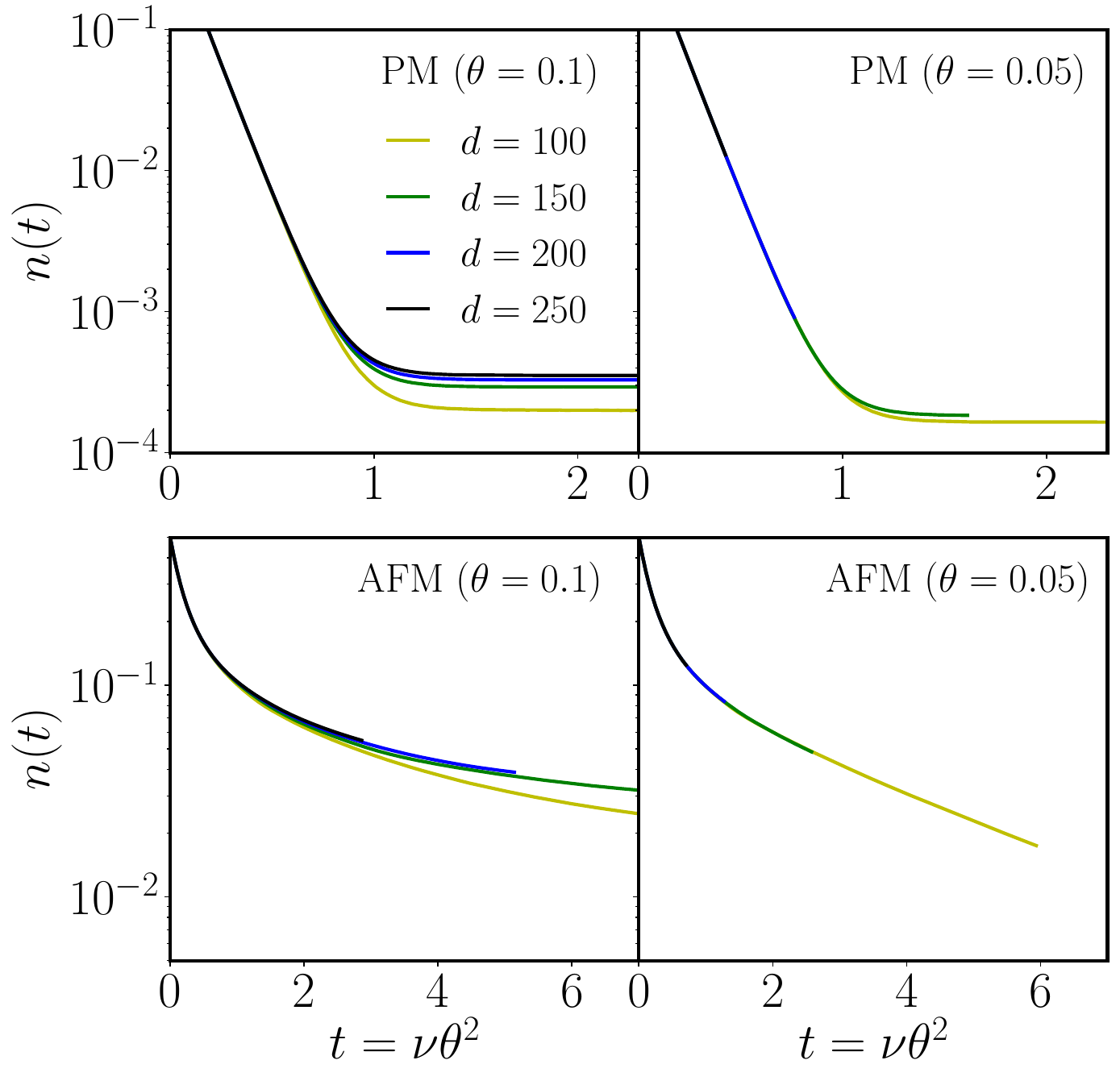}
    \caption{Convergence of ground state cooling simulations as a function of the MPS's bond dimension, for the integrable TFIM with the parameters used in Figure~\ref{fig3}. We show data for both PM and AFM phases at two different bath couplings $\theta = 0.1$ $(0.05)$, and for bond dimensions $d=100,150,200,250$.}
    \label{figBondDim1}
\end{figure}

\begin{figure}[t!]
    \centering
    \vspace{.5cm}
    \includegraphics[width=0.9\columnwidth]{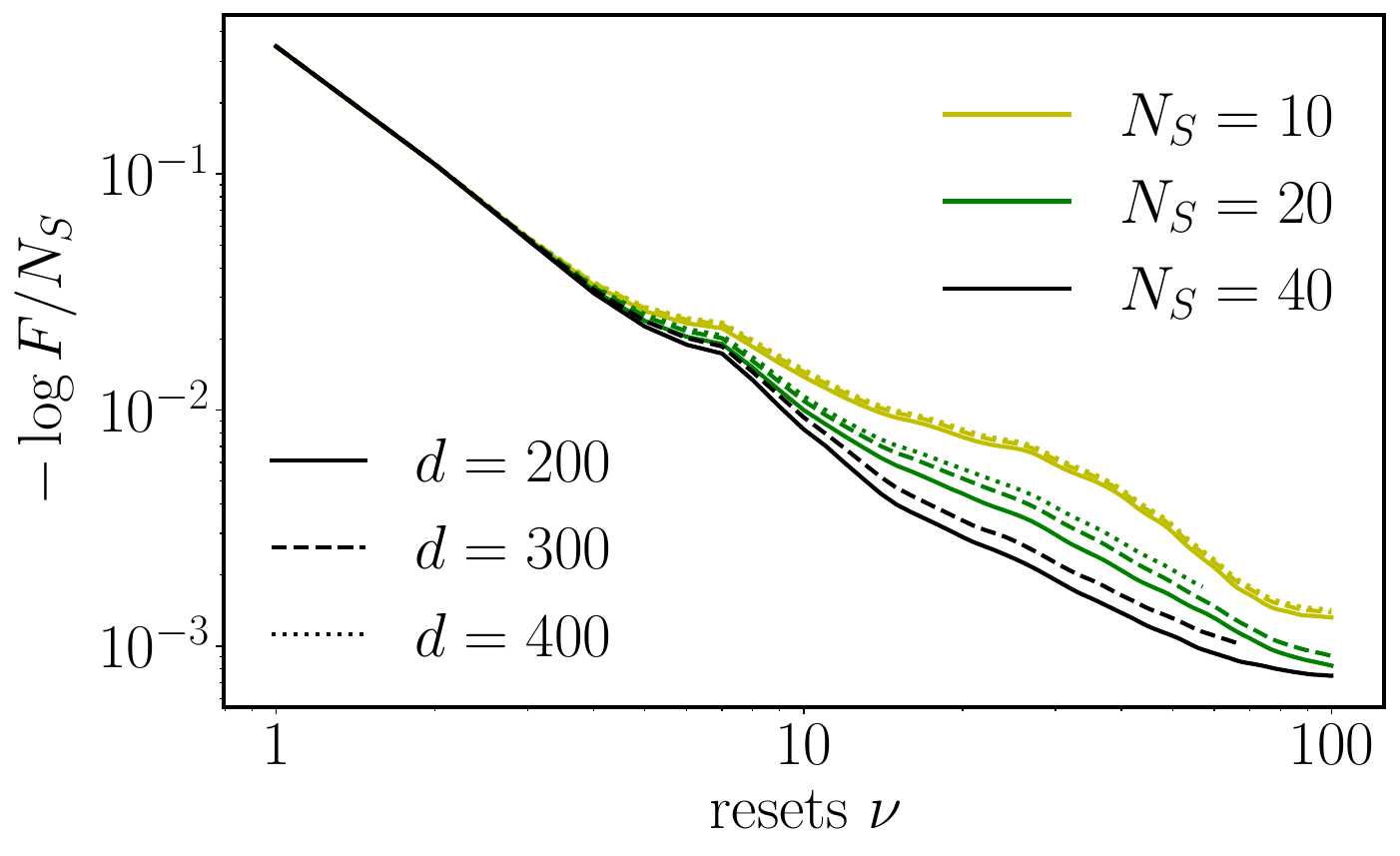}
    \caption{Convergence of ground state cooling simulations as a function of the MPS's bond dimension, for the non-integrable Ising model with the parameters used in Figure~\ref{fig:NI}. Full/dashed/dotted bond dimensions $d=200,300,400$ respectively. For increasing system sizes, there is a small drift with $d$ at $\nu>10$, but the monotonic increase of fidelity is observed for all bond dimensions.}
    \label{figBondDim2}
\end{figure}

The time evolution is performed via the standard TEBD algorithm \cite{vidal2003efficient, vidal2004efficient}, which updates the MPS in real time, followed by the reset channel acting on the bath, which can be efficiently implemented as discussed below. In Fig.~\ref{figMPS} we illustrate the evolution for the example of the Floquet TFIM (\ref{eq:TFIM}) with bath evolution and interaction, Eqs.~(\ref{eq:UB}-\ref{eq:SWAP}). We first simultaneously perform the single-site unitary channels for the system and bath, then the two-body system channels, which correspond to next-nearest-neighbor gates, followed by the nearest-neighbor system-bath channel. Note that the next-nearest-neighbour gate arises from the quasi-1d geometry. This is the most expensive step of the algorithm as it requires two singular value decompositions and scales as $O(\text{dim}(m)^4d^3)$.  

\begin{figure}
    \centering
    \includegraphics[width=0.9\columnwidth]{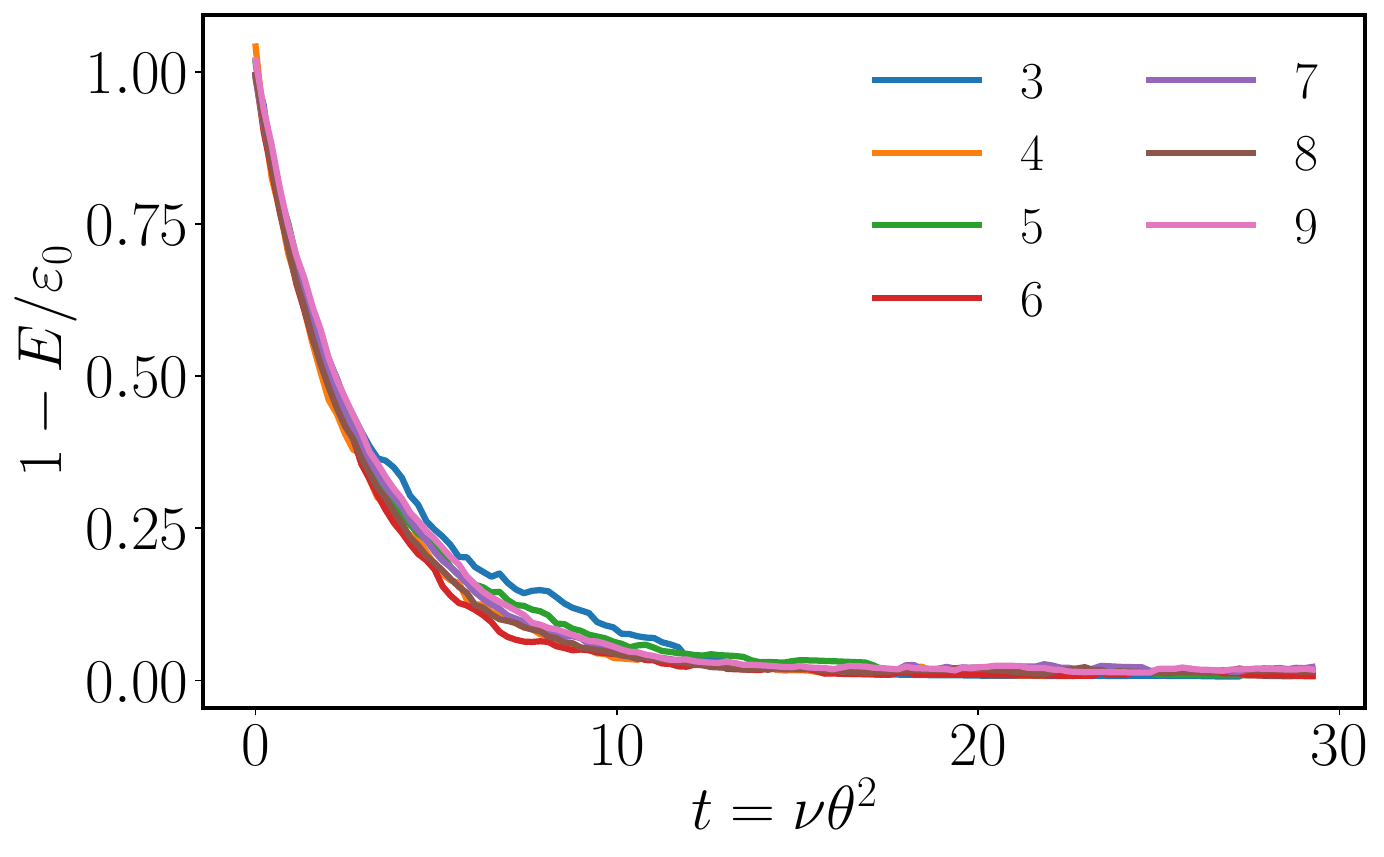} 
    \caption{Energy of Heisenberg spin ladder vs.~rescaled time $t=\nu\theta^2$, where $\nu$ is the number of resets and $\theta=0.15$ is the coupling strength. Parameters are the same as in Fig.~\ref{fig:Heisenberg}, except here we vary the system size, with ladders of $2\times L$ qubits, with $3 \leq L \leq 9$. The data shows that cooling dynamics is independent of system size for the model considered.}
    \label{fig:ladderscaling}
\end{figure}

For the simulations with decoherence, following the unitary channel we apply the decoherence channel Eq.~(\ref{eq:dissipative_evolution}) with jump operators in Eq.~(\ref{eq:decoherencejump}). Finally the reset of auxiliaries is performed by simply modifying the matrix elements of each auxiliary qubit tensor as,
\begin{gather}
A^i_{00 s_{i-1} s_{i}} \rightarrow A^i_{00 s_{i-1} s_{i}} +A^i_{11 s_{i-1} s_{i}}, \\
A^i_{m\neq 00 s_{i-1} s_{i}} \rightarrow 0.\notag
\end{gather}
In the non-integrable TFIM in the presence of a longitudinal field, the local unitary operations are defined according to Eq.~(\ref{eq:Ug}), and the rest of the algorithm is unaffected. 

Next, we study the effects of  truncation due to finite bond dimension during the cooling protocol in Figures~\ref{figBondDim1} and \ref{figBondDim2}. In  Fig.~\ref{figBondDim1} we consider the TFIM cooling for the parameters shown in Fig.~\ref{fig3} of the main text. We only show data for two values of the coupling, $\theta = 0.1, 0.05$ for the MCP, since  the SCP protocol is well converged for all bond dimensions. We observe that the convergence is slower for the larger coupling, suggesting higher bond dimensions are needed to capture non-perturbative effects.  Phenomenologically, such behavior is expected since higher-order in $\theta$ corrections are expected to lead processes that are less spatially local, thus requiring a larger bond dimension to be accurately captured.

The comparison of data obtained with different bond dimension $d=200,300,400$ for the non-integrable model is shown in Fig.~\ref{figBondDim2}. We note that the bond dimension necessary for convergence depends on the correlation length of the ground state. For the parameters of the TFIM with longitudinal field used in the main text ($h_x = h_z = 0.15, J = 0.1$), the effects of truncation are almost negligible for bond dimension values considered. 

\section{System-size dependence for Heisenberg ladder cooling}
\label{app:ladder}

Here we provide data for system-size dependence of cooling in the Heisenberg spin ladder, discussed in Section \ref{Sec:Non-Integrable}. We take the same parameters as in Section \ref{Sec:Non-Integrable}, i.e.~$J=J_\perp = 0.2$, $\theta=0.15$, and $T=\beta=30$.  Since the quasiparticle excitations are quasi-local (adiabatically connected to local operators), we expect that both the cooling time and the final energy density scale independently of system size. In Fig.~\ref{fig:ladderscaling} we show how the mean energy (averaged over 60 quantum trajectories) of the $H_1$ Floquet Hamiltonian decays, for systems with sizes of $2\times L$ qubits, with $3 \leq L \leq 9$. We fix the number of auxiliary qubits at $N_A = \text{floor}(L/2)+1$. The data clearly shows that the cooling behaviour is independent of system size, and cooling is achieved in time $O(\theta^{-2})$.

\begin{figure}[t]
    \centering
    \includegraphics[width=0.9\columnwidth]{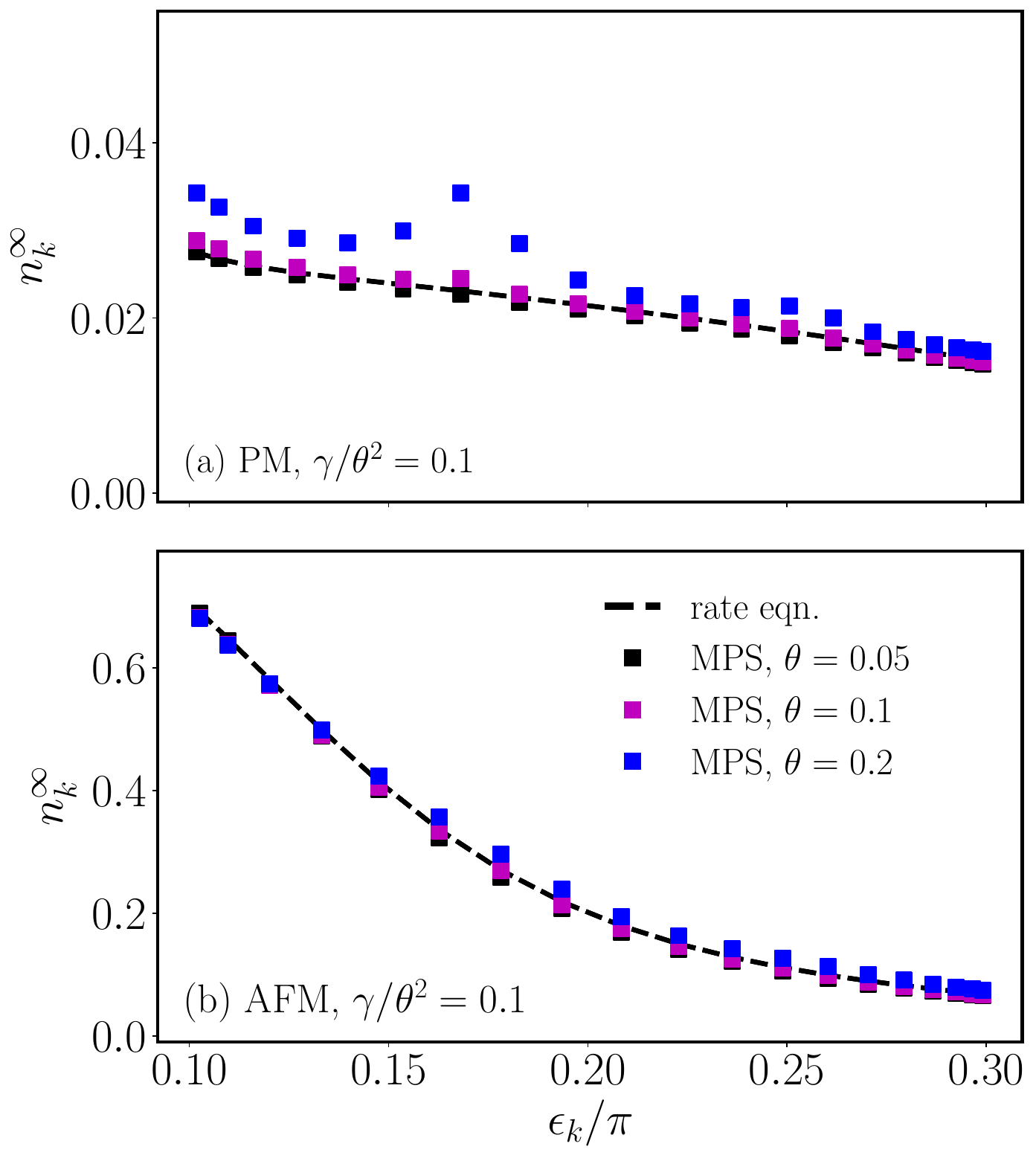}
    \caption{Quasiparticle occupation numbers $n^\infty_k$ in steady state, as a function of quasienergy $\epsilon_k$, for noise strength $\gamma_d = \gamma_\phi = 0.1\theta^2$. (a) PM and (b) AFM phases. Rate equation prediction (dashed line) vs.~MPS simulations (points), for three values of the coupling $\theta$. Protocol parameters are as in Fig.~\ref{fig4}. The MPS simulations use a maximum bond dimension of $d= 100$.}
    \label{figAE}
\end{figure}

\section{Comparison of noisy kinetic equation and MPS simulations}\label{app:noise}

In Section \ref{Sec:Noise} we derived the kinetic equation in the presence of noise, Eq.~(\ref{eq:rate_equation_modified}), which includes transitions induced by qubit dephasing and decay. Here we compare the accuracy of the rate equation predictions against MPS simulations. We focus on the MCP with the same parameters as in Fig.~\ref{fig4}, namely finite density of auxiliaries, $N_A=N_S=20$, and $T=\beta=12$. In Fig.~\ref{figAE} we plot the quasiparticle densities in the steady state, obtained from the kinetic equation prediction (black dashed line) and MPS simulations (scatter points). We use a maximum MPS bond dimension of $d=100$ and run the simulations starting from the exact ground state, as described in Section~\ref{Sec:CoolingSpinChain}. We fix the noise strength $\gamma_d = \gamma_\phi = 0.1\theta^2$, and show results for three different coupling strengths, $\theta = 0.2, 0.1, 0.05$, finding good qualitative agreement for the smaller coupling values $\theta = 0.1, 0.05$.

\bibliography{cooling.bib}

\begin{thebibliography}{93}
\bibitem [{a()}]{a}%
  \BibitemOpen
  \bibinfo {note} {We note that the normalization factor is chosen according to the condition $\DOTSB \sum@ \slimits@ _{\tau =1}^T f_\tau =1$. The modulated coupling in Eq.~(\ref {eq:ftau_step}) is sign-changing, however, $\DOTSB \sum@ \slimits@ _{\tau =1}^T |f_\tau |$ is of order 1 for all cases considered below.}\BibitemShut {Stop}%
\end{thebibliography}
\end{document}